\documentclass[nohyperref]{article}

\PassOptionsToPackage{capitalise,noabbrev}{cleveref}

\usepackage[accepted]{include/icml/icml2024}

\usepackage{microtype}
\usepackage{graphicx}
\usepackage{subfig}
\usepackage{booktabs} 
\usepackage{caption}
\usepackage{subcaption}
\usepackage{subfig}
\usepackage{siunitx}
\usepackage[inline]{enumitem}

\usepackage{titletoc}
\usepackage{amsmath}
\usepackage{amsthm}
\usepackage{amssymb}
\usepackage{xfrac}
\usepackage{hyperref}
\usepackage{setspace}
\usepackage{lipsum}

\usepackage{subfig}
\usepackage{multirow}
\usepackage{array}
\usepackage{multicol}
\usepackage{tabularx}
\usepackage[inline]{enumitem}
\usepackage{minitoc}

\usepackage[utf8]{inputenc}
\usepackage{microtype}
\usepackage{graphicx}
\usepackage{booktabs} 
\usepackage{titletoc}
\usepackage{hyperref}

\usepackage{amsmath}
\usepackage{amssymb}
\usepackage{bbm} 
\usepackage{bm}
\usepackage{verbatim}
\usepackage{float}
\usepackage{color,soul}
\usepackage{enumitem}
\usepackage{mathtools}
\usepackage{hhline}
\usepackage[title]{appendix}
\usepackage[nameinlink]{cleveref} 
\usepackage{tabularx}
\usepackage{tikz}
\usepackage{wrapfig,booktabs}
\usepackage{xspace}
\usepackage{cancel}
\usepackage{sidecap}

\graphicspath{ {../figures/} }

\DeclarePairedDelimiterX{\infdivx}[2]{(}{)}{%
  #1\;\delimsize\|\;#2%
}

\crefname{appsec}{appendix}{appendices}
\Crefname{appsec}{Appendix}{Appendices}

\definecolor{mydarkblue}{rgb}{0,0.08,0.45}
\hypersetup{ %
    pdftitle={},
    pdfauthor={},
    pdfsubject={Proceedings of the International Conference on Machine Learning 2020},
    pdfkeywords={},
    pdfborder=0 0 0,
    pdfpagemode=UseNone,
    colorlinks=true,
    linkcolor=mydarkblue,
    citecolor=mydarkblue,
    filecolor=mydarkblue,
    urlcolor=mydarkblue,
    pdfview=FitH
}

\usetikzlibrary{shapes.geometric, arrows, bayesnet, calc, positioning}

\DeclareMathOperator{\xbf}{\mathbf{x}}
\DeclareMathOperator{\Xbf}{\mathbf{X}}
\DeclareMathOperator{\ybf}{\mathbf{y}}
\DeclareMathOperator{\Ybf}{\mathbf{Y}}

\DeclareMathOperator{\mbf}{\mathbf{m}}

\newcommand{\by}{\mathbf y}
\newcommand{\bx}{\mathbf x}

\newcommand{\calN}{\mathcal{N}}

\newcommand{\calL}{\mathcal{L}}

\newcommand{\y}{\mathbf{y}}

\newcommand{\R}{\mathbb{R}}

\newcommand{\bmu}{{\boldsymbol{\mu}}}

\newcommand{\closer}[3]{{\kern-#1ex{#2}\kern-#3ex}}

\mathchardef\mhyphen="2D

\DeclareMathOperator{\E}{\mathbb{E}}

\usepackage{titletoc}
\usepackage{algorithm}
\usepackage{algorithmic}
\usepackage{colortbl}

\definecolor{azure}{rgb}{0.0, 0.5, 1.0}
\definecolor{airforceblue}{rgb}{0.36, 0.54, 0.66}
\definecolor{darkgreen}{rgb}{0.0, 0.2, 0.13}

\newcommand\defines{\,\dot{=}\,}

\newcommand{\vbar}{\,|\,}

\newcommand{\calX}{\mathcal{X}}
\newcommand{\calY}{\mathcal{Y}}
\newcommand{\calD}{\mathcal{D}}

\newcommand{\calC}{\mathcal{C}}

\newcommand{\pms}[1]{\ensuremath{{\scriptstyle\pm #1}}}

\usepackage{standalone}
\usepackage{tikz}
\usepackage{pgfplots}
\usepackage{pgf}
\usetikzlibrary{calc}
\usetikzlibrary{positioning}
\usetikzlibrary{angles,quotes}
\usetikzlibrary{backgrounds}
\usetikzlibrary{fit}
\usetikzlibrary{arrows}
\usetikzlibrary{arrows.meta}
\usetikzlibrary{shapes.symbols}
\usetikzlibrary{shadings}
\usetikzlibrary{shapes}
\usetikzlibrary{fadings}
\usetikzlibrary{bayesnet}
\usetikzlibrary{matrix}
\usetikzlibrary{plotmarks}
\usetikzlibrary{intersections}
\usetikzlibrary{pgfplots.fillbetween}
\pgfplotsset{compat=1.14}
\usepackage{siunitx}

\usepackage{colortbl}
\definecolor{mediumgray}{gray}{0.7}
\definecolor{lightgray}{gray}{0.85}
\definecolor{lightlightgray}{gray}{0.9}
\definecolor{C1}{HTML}{1F77B4}
\definecolor{C2}{HTML}{FF7F0E}
\definecolor{C3}{HTML}{2CA02C}
\definecolor{C4}{HTML}{D62728}
\definecolor{C5}{HTML}{9467BD}
\colorlet{C1light}{C1!70!white}
\colorlet{C2light}{C2!70!white}
\colorlet{C3light}{C3!70!white}
\colorlet{C4light}{C4!70!white}
\colorlet{C5light}{C5!70!white}
\colorlet{C1lighter}{C1!50!white}
\colorlet{C2lighter}{C2!50!white}
\colorlet{C3lighter}{C3!50!white}
\colorlet{C4lighter}{C4!50!white}
\colorlet{C5lighter}{C5!50!white}
\colorlet{C1vlight}{C1!20!white}
\colorlet{C2vlight}{C2!20!white}
\colorlet{C3vlight}{C3!20!white}
\colorlet{C4vlight}{C4!20!white}
\colorlet{C5vlight}{C5!20!white}
\colorlet{linkcolor}{violet}

\usepackage{tcolorbox}

\newcommand{\codebox}[2]{%
\begin{tcolorbox}[colback=blue!3!white,leftrule=2.5mm,size=title]
#1: #2
\end{tcolorbox}
\vspace{-0.1cm}%
}
\usepackage{fancyhdr}

\usepackage[algo2e]{algorithm2e}
\usepackage{eqparbox}
\renewcommand{\algorithmiccomment}[1]{\hfill\eqparbox{COMMENT}{#1}}

\icmltitlerunning{Context-Guided Diffusion for Out-of-Distribution Molecular and Protein Design}

\begin{document}

\twocolumn[
\icmltitle{Context-Guided Diffusion for\\Out-of-Distribution Molecular and Protein Design}

\begin{icmlauthorlist}
\icmlauthor{Leo Klarner}{ox}
\icmlauthor{Tim G. J. Rudner}{nyu}
\icmlauthor{Garrett M. Morris}{ox}
\icmlauthor{Charlotte M. Deane}{ox}
\icmlauthor{Yee Whye Teh}{ox}
\end{icmlauthorlist}

\icmlaffiliation{ox}{University of
Oxford, UK}
\icmlaffiliation{nyu}{New York University, New York, USA}

\icmlcorrespondingauthor{Leo Klarner}{leo.klarner@stats.ox.ac.uk}

\icmlkeywords{Machine Learning, ICML}

\vskip 0.3in
]



\printAffiliationsAndNotice{}  

\begin{abstract}
Generative models have the potential to accelerate key steps in the discovery of novel molecular therapeutics and materials. Diffusion models have recently emerged as a powerful approach, excelling at unconditional sample generation and, with data-driven guidance, conditional generation within their training domain. Reliably sampling from high-value regions beyond the training data, however, remains an open challenge---with current methods predominantly focusing on modifying the diffusion process itself. In this paper, we develop context-guided diffusion (CGD), a simple plug-and-play method that leverages unlabeled data and smoothness constraints to improve the out-of-distribution generalization of guided diffusion models. We demonstrate that this approach leads to substantial performance gains across various settings, including continuous, discrete, and graph-structured diffusion processes with applications across drug discovery, materials science, and protein design.
\end{abstract}

\vspace{-2em}
\section{Introduction}
\label{sec:intro}

The central goal of molecular discovery is to identify novel compounds
with desirable functional properties.
The immense size of the underlying search spaces---up to $10^{60}$ for drug-like small molecules~\citep{bohacek1996art} and $20^{N}$ for $N$-length protein sequences~\citep{maynard1970natural}---renders this a challenging combinatorial optimization problem.
Considering the substantial cost of synthesizing and validating candidate compounds experimentally, the question of how to efficiently navigate these search spaces to locate high-value subsets lies at the core of modern molecular design~\citep{gomez2018automatic, sanchez2018inverse, bilodeau2022generative}.

\begin{figure}[t!]
\vspace{-2pt}
    \centering
    \includegraphics[width=0.905\linewidth]{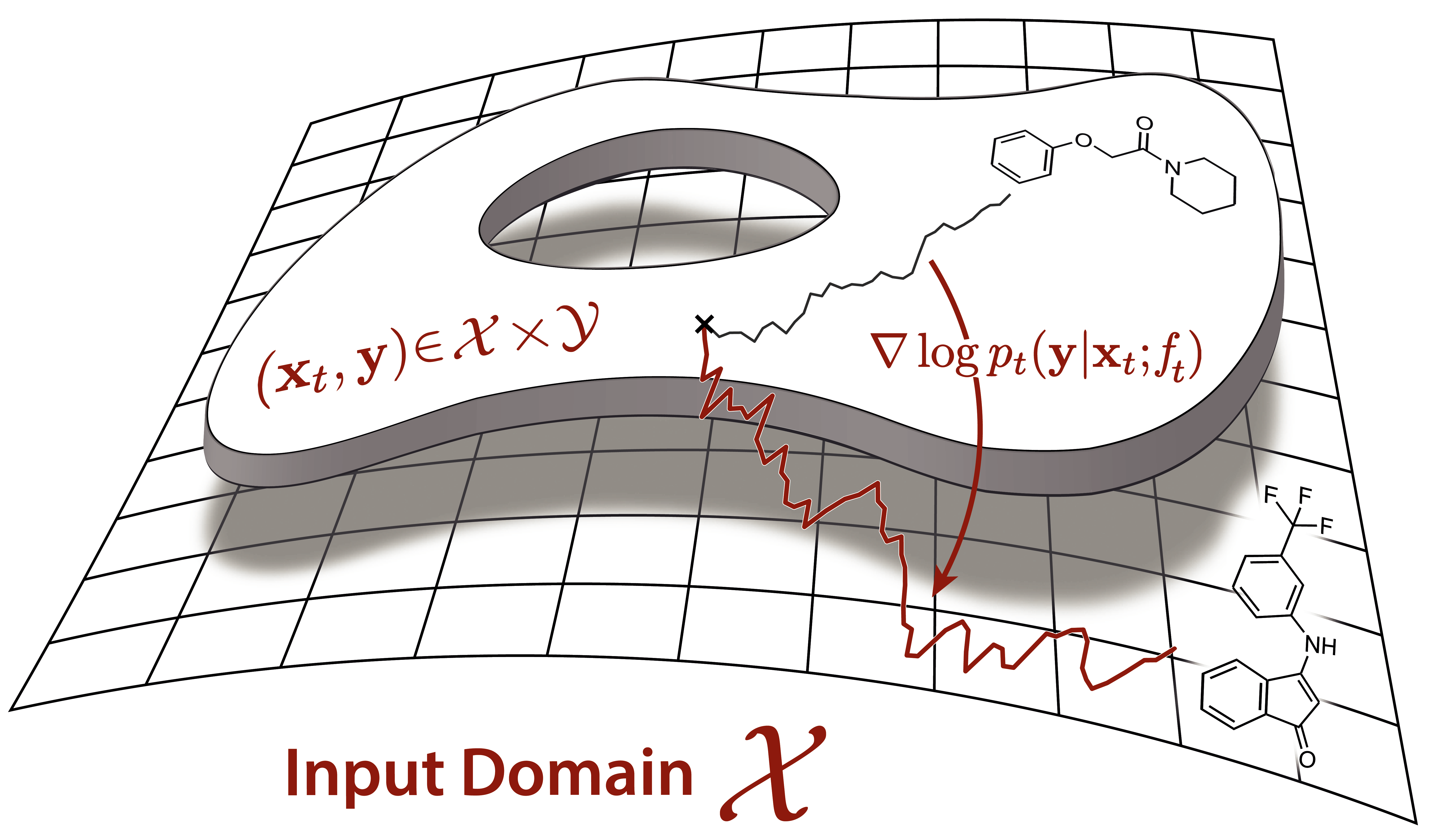}
    \vspace{-9pt}
    \caption{
    Guidance models that generalize poorly under distribution shifts can be a major performance bottleneck for property-guided diffusion models.
    We introduce a guidance model regularizer that improves generalization under distribution shifts and enables {\em context-guided diffusion} (CGD).
    We show that CGD leads to conditional sampling processes that consistently generate novel, high-value molecules {\color{red!50!black}(\textbf{red})}.
    }
    \label{fig:graphical abstract}
    \vspace{-9pt}
\end{figure}

Deep generative models have the potential to accelerate this process by capturing and abstracting key structural properties of their input domain~\citep{sanchez2018inverse, bilodeau2022generative}.
In recent years, denoising diffusion models~\citep{sohl2015deep, ho2020denoising, song2020score, karras2022elucidating} have emerged as the method of choice, displaying impressive performance in conditional and unconditional image generation tasks~\citep{dhariwal2021diffusion, saharia2022photorealistic, rombach2022high, zhang2023adding}, as well as applications across chemistry \citep{jing2022torsional,corso2023diffdock} and biology \citep{watson2023novo,abramson2024accurate}.

\begin{figure*}[t!]
\centering\begin{minipage}{0.4\textwidth}%
  \centering
  \includegraphics[width=\linewidth]{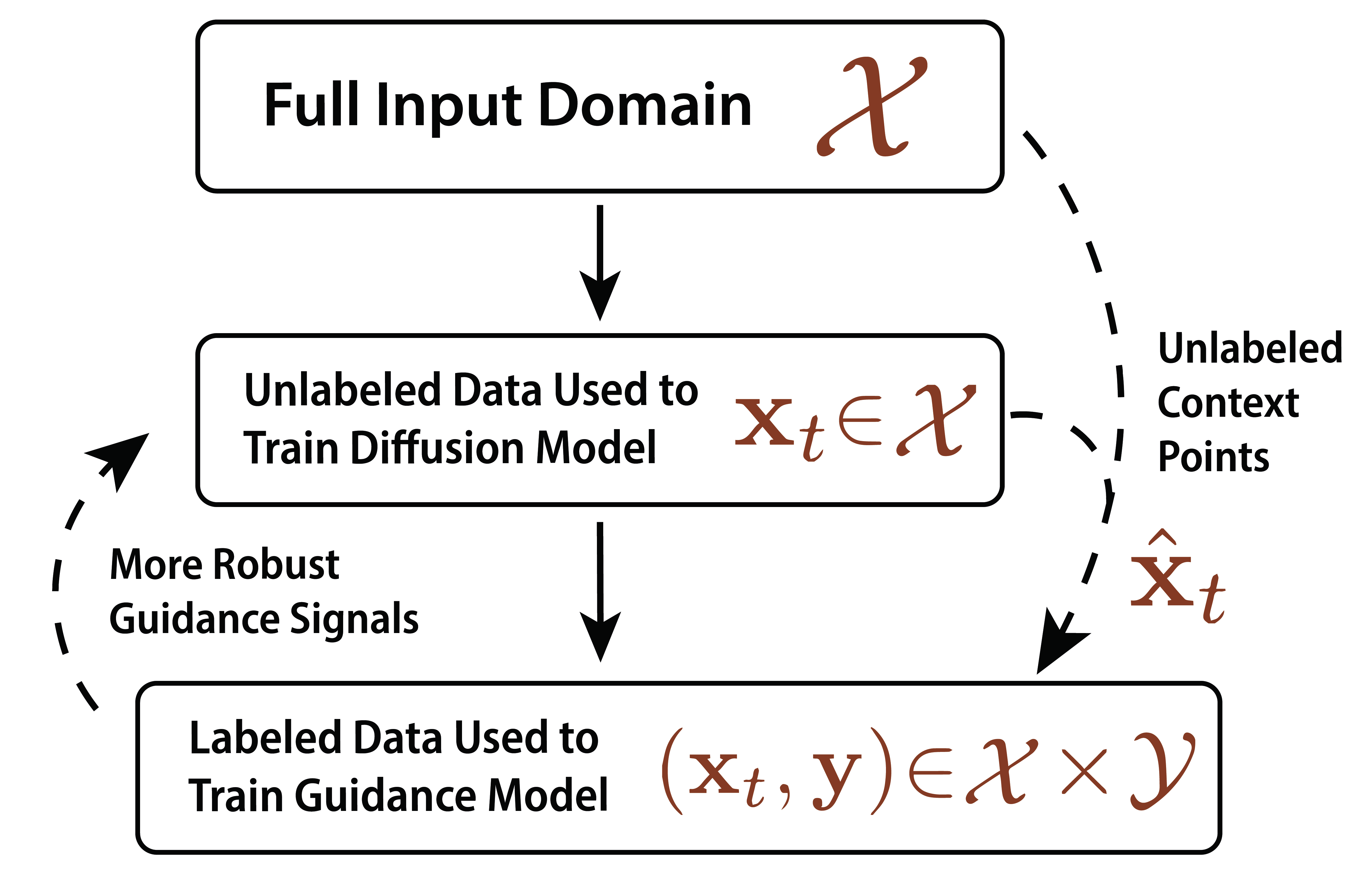}%
\end{minipage}%
\begin{minipage}{0.55\textwidth}%
  \centering
  \includegraphics[width=\linewidth]{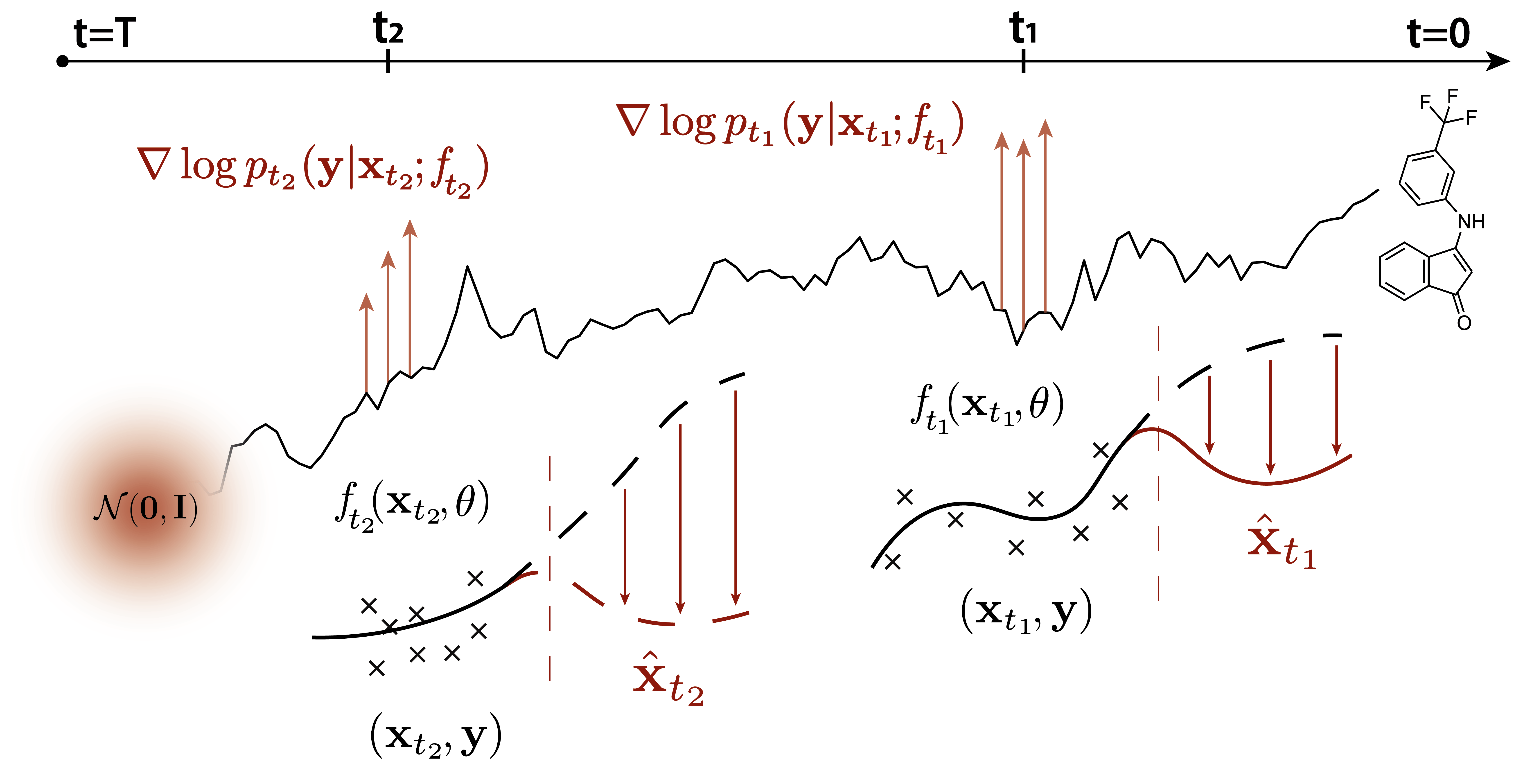}%
\end{minipage}
\vspace*{-12pt}
\caption{
    {\em Context-guided diffusion} leverages unlabeled context data to combine signals from labeled training data with structural information of the broader input domain \textbf{(left)}.
    Specifically, we construct a data- and noise scale-dependent guidance model regularizer that encourages smooth gradients, mean reversion, and high predictive uncertainty on out-of-distribution (OOD) inputs, allowing the conditional denoising process under a context-guided diffusion model to focus on promising near-OOD subsets of chemical and protein sequence space \textbf{(right)}.
    }
\label{fig:overview_plot}
\vspace*{-6pt}
\end{figure*}



While diffusion models excel at capturing complex, multi-modal densities \citep{kadkhodaie2023generalization}, accurately modeling the training distribution is only the first step.
To fully leverage diffusion models for applied molecular optimization, we require carefully designed guidance functions that steer the sample generation process toward compounds with desirable properties~\citep{weiss2023guided,gruver2023protein,lee2023exploring}.
However, when labeled data for training these guidance functions is scarce and only available for a biased and unrepresentative subset of the input domain, as is often the case in practice, overconfident guidance signals risk steering the generative process toward false-positive regions of chemical or protein sequence space (\Cref{fig:graphical abstract}).


In this paper, we propose \emph{context-guided diffusion} (CGD)---a simple plug-and-play method that leverages unlabeled data and smoothness constraints to improve the out-of-distribution generalization of property-guided diffusion models.
Specifically, we construct a guidance model regularizer that leverages unlabeled context data to train guidance models that both (i) fit the training data well and (ii) exhibit high uncertainty and smooth gradients in out-of-distribution (OOD) regions of the input domain (\Cref{fig:overview_plot}).
We show that using the resulting context-aware guidance model to perform context-guided diffusion leads to conditional sampling processes that consistently generate novel, near-OOD molecules with desirable properties.

CGD is straightforward to implement, requires no changes to the architecture of the guidance model, and adds no computational overhead at sampling time.
Crucially, it is agnostic to the underlying generative process and thus compatible with latent \cite{rombach2022high, xu2023geometric}, equivariant \citep{hoogeboom2022equivariant}, Riemannian \citep{de2022riemannian,huang2022riemannian}, and constrained \citep{pmlr-v202-lou23a,fishman2023diffusion,fishman2023metropolis} diffusion models.

We demonstrate the versatility of our approach by applying it to the design of small molecules with graph-structured diffusion processes \citep{lee2023exploring}, the generation of novel materials with equivariant diffusion models \citep{weiss2023guided}, and the optimization of discrete protein sequences with categorical diffusion \citep{gruver2023protein}.
We find that CGD consistently outperforms standard guidance models, as well as more sophisticated pre-training and domain adaptation techniques, and that it enables more reliable generation of novel samples from high-value subsets of chemical and protein sequence space.


\vspace*{-3pt}\codebox{\centering The code for our experiments can be accessed at}{\begin{center}
    \footnotesize
    \vspace{-0.4em}
    \href{https://github.com/leojklarner/context-guided-diffusion}{\centering\texttt{
    https://github.com/leojklarner/\\context-guided-diffusion}
    }
\end{center}}

\section{Guided Diffusion Models}
\label{sec:background}

We start by providing a brief overview of standard unconditional diffusion models, following~\citet{song2020score}, and then discuss conditional sampling schemes, focusing on classifier guidance as introduced in~\citet{sohl2015deep}, \citet{song2020score} and~\citet{dhariwal2021diffusion}.

\paragraph{Unconditional Diffusion Models.} Denoising diffusion models progressively add noise to a data distribution $p_0$ until it approaches a tractable reference distribution $p_{T}$.
This forward noising process is governed by a stochastic differential equation (\textsc{SDE}) \citep{song2020score}, defined as
\begin{align}
\label{eq:forward_noising_process}
    \mathrm{d}\mathbf{X}_{t} = \mathbf{f}(\Xbf_{t}, t)+g(t)\mathrm{d}\mathbf{B}_{t},
\end{align}
where $\Xbf_0 \sim p_0$ is sampled from the data distribution and $\mathbf{B}_{t}$ represents the $d$-dimensional Brownian motion.
A common choice is the VP-SDE \citep{sohl2015deep,ho2020denoising}, with drift coefficient \mbox{${\mathbf{f}(\Xbf_{t}, t)=-\frac{1}{2}\beta_{t}\Xbf_{t}}$} and diffusion coefficient \mbox{$g(t)=\sqrt{\beta_{t}}$}, where the noise scale \mbox{$\beta_{t}>0$} determines the amount of corruption.
Over time steps \mbox{$t\in[0,T]$}, this \textsc{SDE} gradually noises $\Xbf_{t}\sim p_{t}$ until it converges to the reference distribution $p_T=\calN(\mathbf{0}, \mathbf{I})$.



To generate new samples from $p_0$, diffusion models leverage the fact that the forward noising process defined by \Cref{eq:forward_noising_process} admits a time-reversal \citep{anderson1982reverse,haussmann1986time}.
For the VP-SDE, this process is given by
\begin{align}
\label{eq:reverse_denoising_process}
\begin{split}
\mathrm{d} \mathbf{X}_{t} = -\beta_{t} \left\{\frac{1}{2} \mathbf{X}_{t} +\nabla \log p_{t}\left(\mathbf{X}_{t}\right)\right\} \mathrm{d} t + \sqrt{\beta_{t}} \mathrm{d} \mathbf{B}_{t},
\end{split}
\end{align}
which flows backward in time. \mbox{$\mathbf{X}_T \sim p_T$} is sampled from the reference distribution and incrementally denoised until it reflects $p_0$.
We have used the shorthand $\nabla \log p_{t}\left(\mathbf{X}_{t}\right)$ to denote $\nabla_{\xbf} \log p_{t}\left(\xbf\right)|_{\xbf=\Xbf_{t}}$ and will continue using this shorthand in the remainder of the text.







The gradient of the log-density $\nabla \log p_{t}(\bx_{t})$ in \Cref{eq:reverse_denoising_process} is estimated by a time-dependent score network $s_{\psi}(\bx_{t}, t)$ using techniques from score matching \citep{hyvarinen2005estimation}.
Once trained, $s_{\psi}$ can be used to generate samples from the data distribution $p_0$ by simulating the reverse-time \textsc{SDE} with a range of numerical solvers \citep{song2020score}.

\paragraph{Guided Diffusion Models.}
In many applications, we may wish to condition the generative process of diffusion models to produce samples with specific properties $\ybf$, such as images of a certain class or molecules with a desirable pharmacological effect.
That is, we want to modify the denoising process in \Cref{eq:reverse_denoising_process} to incorporate the conditioning information $\ybf$, and sample from the conditional distribution $p_0(\bx_0 \vbar \mathbf{y})$ rather than the unconditional data distribution $p_0(\bx_0)$.
The corresponding reverse-time \textsc{SDE} is given by
\begin{align*}
\begin{split}
    \mathrm{d} \mathbf{X}_{t} = -\beta_{t} \left\{\frac{1}{2} \mathbf{X}_{t} + \nabla \log p_{t}\left(\mathbf{X}_{t}\vbar \ybf\right)\right\} \mathrm{d} t + \sqrt{\beta_{t}} \mathrm{d} \mathbf{B}_{t},
    \end{split}
\end{align*}
differing from \Cref{eq:reverse_denoising_process} only in the need for a conditional score function $\nabla \log p_{t}(\Xbf_{t} \vbar \ybf)$, which can be expressed as
\begin{align}
\begin{split}
\label{eq:conditional_score}
    \hspace*{-5pt}\nabla \log p_{t}(\Xbf_{t} \vbar \ybf)
    &=
    \nabla \log \frac{p_{t}(\ybf \vbar \Xbf_{t})p_{t}(\Xbf_{t})}{Z}\\
    &=
    \nabla \log p_{t}(\ybf \vbar \Xbf_{t}) + \nabla \log p_{t}(\Xbf_{t}) ,
\end{split}
\end{align}
since $Z$ is a normalization constant independent of $\Xbf_{t}$.

To approximate $\nabla \log p_{t}(\ybf \vbar \Xbf_{t})$, a time-dependent discriminative {\em guidance model} $f_t(\xbf_{t} ; \theta)$ is trained on a labeled dataset \mbox{$\smash{\calD = \cup_{t=1}^{T} \{(\bx_{t}^{(n)}, \Bar{\ybf}^{(n)})\}_{n=1}^{N}}$} with inputs corrupted by the same forward noising process as the diffusion model.
Once trained, the guidance model can be incorporated into a {\em guidance function} $p(\ybf \vbar \Xbf_{t} ; f(\cdot ; \theta))$ that is used to compute the \emph{conditional scores} $\nabla \log p_{t}(\Xbf_{t} \vbar \ybf; f(\cdot ; \theta))$, which in turn steer the reverse-time SDE toward samples with a high predicted likelihood of observed label $\by$ via
\begin{align*}
\begin{split}
    \mathrm{d} \mathbf{X}_{t}
    \hspace*{-2pt}=\hspace*{-2pt}
    -\beta_{t} \left\{\frac{1}{2} \mathbf{X}_{t} \hspace*{-2pt}+\hspace*{-2pt} \nabla \log p_{t}(\mathbf{X}_{t}\vbar \ybf; f_t(\cdot ; \theta) )\hspace*{-2pt}\right\}\hspace*{-2pt} \mathrm{d} t \hspace*{-2pt}+\hspace*{-2pt} \sqrt{\beta_{t}} \mathrm{d} \mathbf{B}_{t} .
    \end{split}
\end{align*}

Alternative approaches such as classifier-free guidance \citep{ho2022classifier} estimate the conditional scores by passing the conditioning information to the score network
\begin{align*}
    \nabla \log p_{t}(\Xbf_{t} \vbar \ybf)
    \approx
    s_{\psi}(\bx_{t}, \mathbf{y}, t),
\end{align*}
either in the form of class labels \citep{ho2022classifier} or pre-trained embeddings \citep{nichol2021glide, ramesh2022hierarchical,saharia2022photorealistic}.
However, the requirement to provide $\ybf$ as an explicit input to the score network limits the applicability of these methods to regression-based optimization problems frequently encountered in molecular and protein design \citep{lee2023exploring, weiss2023guided, gruver2023protein}, where the objective is to maximize or minimize a property for which the optimal value is often unknown and likely to lie well outside the training distribution of the conditional score network.

\section{Context-Guided Diffusion Models}
\label{sec:method}

    


\begin{figure*}[!t]
  \begin{tabular}{c}
    \subfloat[An algorithmic overview of the proposed context-guided regularization scheme.]{
    \begin{minipage}{.375\linewidth}

    \vspace{-5pt}
    \begin{algorithm}[H]
    \setstretch{1.22}
    \caption{One guidance model training epoch with the context-guided optimization objective from~\Cref{eq:fseb_loss}.}
    \vspace{0.2em}
    
        




    \For{$(\xbf_0, \Bar{\ybf})\in\calD_0$}{

        \vspace{0.5em}

        $t\sim\mathcal{U}(0, T)$ \;

        $\xbf_{t}\gets\operatorname{NoisingProcess}(\xbf_0, t)$ \algorithmiccomment{\labelcref{eq:forward_noising_process}}\;

        $\mathcal{L}\gets - \log p_{t}(\Bar{\ybf}|\xbf_{t}; f^{j}_{t}(\xbf_{t};\theta )) $ \algorithmiccomment{\labelcref{eq:supervised_loss}}\;

        $\Bar{\mathcal{L}} \gets \mathcal{L} + \frac{1}{2 \lambda} ||\theta||_{2}^{2} $  \algorithmiccomment{\labelcref{eq:standard_loss}}\;

        \vspace{0.5em}

        $\Hat{\xbf}_0\sim p_{\hat{\Xbf}_{0}}$\;

        $s\sim\mathcal{U}(0, T)$ \;
        
        $\Hat{\xbf}_{s}\gets\operatorname{NoisingProcess}(\Hat{\xbf}_0, s)$ \algorithmiccomment{\labelcref{eq:forward_noising_process}}\;

        
        $R \gets \sum\limits_{j} D_{\mathcal{M}}(f^{j}_{s}(\Hat{\xbf}_{s};\theta), m_{t}^{j}, {K}_{t})^2$ \algorithmiccomment{\labelcref{eq:regularizer}}\;
        
        \vspace{0.5em}
    
        $\theta\gets\operatorname{Update}(\theta, \Bar{\mathcal{L}} + R)$\;
        
        \vspace{0.5em}
    }

    \vspace{0.2em}

    \end{algorithm}
\end{minipage}%
\hspace*{-10pt}}
  \end{tabular}
  \begin{tabular}{c}
    \subfloat[Predictions and samples from an $L_{2}$-regularized guidance model.]{\includegraphics[width=0.6\linewidth]{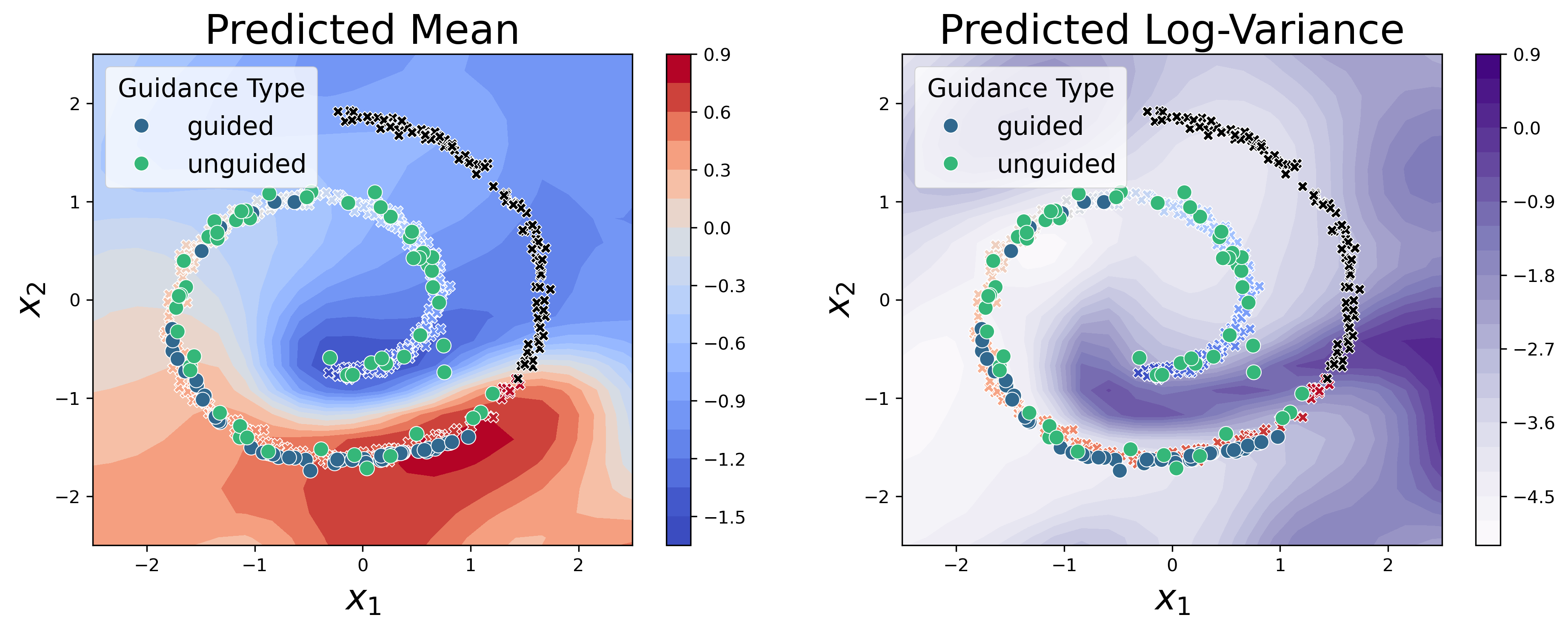}} \\
    \subfloat[Predictions and samples from a context-aware guidance model.]{\includegraphics[width=0.6\linewidth]{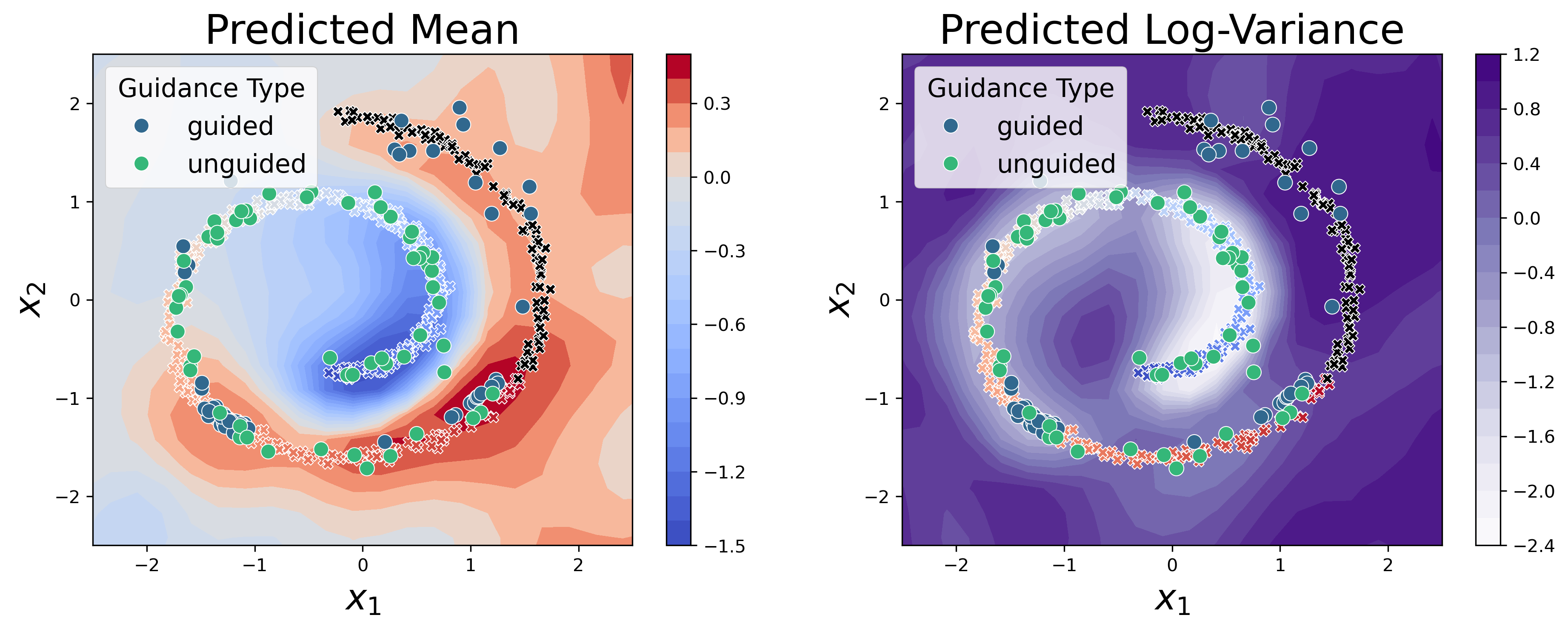}}
  \end{tabular}
  \vspace*{1pt}
  \caption{Panel (a) presents an algorithmic perspective on how the proposed regularization scheme fits into the training loop of a standard guidance model. Panels (b) and (c) illustrate how the regularization scheme affects the functions a guidance model learns. Two regression models with different regularizers are trained on a low-label subset of an illustrative dataset (color-coded crosses). However, without additional information about the input domain, the predictions of the $L_{2}$-regularized model (b) exhibit poor generalization and miscalibrated uncertainty estimates when
    evaluated in new, unseen regions. In contrast, the context-aware model (c) is able to leverage unlabeled data sampled uniformly from $[-2.5, 2.5]^2\subset\R^{2}$ to generate more accurate and better-calibrated predictions. The corresponding context-guided denoising process generates samples that recover a held-out, high-label test set (black crosses). Full experimental details are presented in~\Cref{app:toy_example}.}
  \label{fig:reg_illustration}
\end{figure*}

In this section, we present \emph{context-guided diffusion} (CGD)---a method that produces more robust conditional gradient estimates that consistently generate novel samples with improved properties.
We first introduce the prevailing approach to training guidance models and highlight its limitations.
We then present a domain-informed regularization term designed to bias guidance model training toward functions that both (i) fit the training data well and (ii) revert to domain-appropriate behavior in out-of-distribution settings.
Finally, we describe how to incorporate context-aware guidance models into guided diffusion to generate novel molecules and protein sequences with desirable properties.

\subsection{Standard Guidance Model Training}
\label{sec:standard_guidance_model_training}

We consider a supervised learning setting with diffusion timesteps \mbox{$t=0,...,T$} and $N$ i.i.d. data realizations \mbox{$\smash{{\calD_{t} = \{(\xbf_{t}^{(n)}, \Bar{\ybf}^{(n)})\}_{n=1}^{N}}}$} of the noised inputs \mbox{$\xbf_{t} \in \calX$} and targets \mbox{$\Bar{\ybf} \in \calY \subseteq \mathbb{R}^{k}$}.
Furthermore, we define the guidance model \mbox{$\smash{f_{t}^{j}(\cdot \,; \theta) \defines g^{j}(h_t(\cdot \,; \theta_{h}) ; \theta_{L}^{j} )}$} with \mbox{$j=1,2$} and $\smash{\theta \defines \{ \theta_{h}, \theta_{L}^{1}, \theta_{L}^{2} \}}$.
Here, \mbox{$h_t(\cdot \,; \theta_{h}):\calX\to\R^d$} is an embedding model, and $g^{1}(h_t(\cdot \,; \theta_{h}) ; \theta_{L}^{1} ) \defines h_t(\cdot \,; \theta_{h}) \theta_L^{1}$ and $g^{2}(h_t(\cdot \,; \theta_{h}) ; \theta_{L}^{2} ) \defines \exp{(h_t(\cdot \,; \theta_{h}) \theta_L^{2})}$ are $k$-dimensional mean and variance heads, respectively.

Conventional supervised learning techniques for training guidance models define the diffusion timestep-dependent negative log-likelihood objective\vspace*{-6pt}
\begin{align}
\label{eq:supervised_loss}
    \calL(\theta, \calD_{\mathcal{T}}, \mathcal{T})
    \defines
    -\sum_{n=1}^{N} \log{p_{\mathcal{T}}(\Bar{\ybf}^{(n)} \vbar \xbf^{(n)}_{\mathcal{T}} ; f^{j}_{\mathcal{T}}(\cdot ; \theta))}
\end{align}\\[-10pt]
where \mbox{$\mathcal{T} \sim p_{\mathcal{T}}(t)$} is a randomly sampled timestep.
The guidance model parameters $\theta$ are optimized by minimizing the \mbox{$L_2$-regularized} objective function\vspace*{-5pt}
\begin{align}
\label{eq:standard_loss}
    \Bar{\calL}(\theta, \calD_{\mathcal{T}})
    \defines
    \E_{p_{\mathcal{T}}}\left[ \calL(\theta, \calD_{\mathcal{T}}, \mathcal{T}) \right]
    + 
    \frac{1}{2\lambda} ||\theta||_{2}^{2},
\end{align}\\[-12pt]
where $\lambda\in\R_{+}$ is chosen as a hyperparameter.

Unfortunately, this approach is only able to operate over subsets of the input domain $\calX$ for which labeled data is available. 
When labels are scarce, as is the case in many practical settings, models trained with standard supervised learning objectives are prone to generating incorrect and overconfident predictions when exposed to distribution shifts \citep{ovadia2019can, liu2020simple, van2020uncertainty, koh2021wilds, band2021benchmarking, rudner2021pathologies, tran2022plex, rudner2022tractable, klarner2023drug}.
This issue is exacerbated by the fact that guidance models are trained on noised data points:
Since neural networks are both highly sensitive to noise perturbations \citep{szegedy2013intriguing,goodfellow2014explaining,ho2022classifier} and able to overfit to fully corrupted inputs \citep{zhang2021understanding}, the errors of systematically overconfident gradient signals 
may accumulate over hundreds of denoising steps.

\subsection{Context-Aware Guidance Models}
\label{sec:regularization_objective}

To improve the performance and reliability of guidance models in this setting, we propose a family of diffusion timestep-dependent regularizers designed to favor parameters $\theta$ that encode desirable model behaviors on out-of-distribution data.
While ideal model behaviors under distribution shifts are often application-dependent, we design a general-purpose regularizer---applicable across domains---that encourages uncertain predictions in areas of the input domain with little or no signal from the labeled data.
This regularizer maximizes sensitivity at the expense of specificity and enables the conditional denoising process to focus on regions in data space that are near the training data and have the highest likelihood of containing molecules with improved properties.

To encode this behavior, we construct a timestep-dependent distribution over guidance function outputs that is explicitly designed to exhibit high uncertainty on a problem-specific, out-of-distribution context set \mbox{$\smash{\Hat{\xbf}_{\calC} \defines \{ \xbf_{\calC}^{(i)} \}_{i=1}^{N_\calC}}$}.

More specifically, we consider regularizers of the form
\begin{align}
\begin{split}
\label{eq:regularizer}
    &
    R(\theta, f_{t}, t, p_{\hat{\Xbf}_{t}})
    \\
    &
    ~
    \defines
    \mathbb{E}_{p_{\hat{\Xbf}_{t}}} \left[\sum_{j=1}^{2} 
    D_{\mathcal{M}}(f_{t}^{j}(\Hat{\xbf}_{t} ; \theta), m_{t}^{j}(\Hat{\xbf}_{t}), {K}_{t}(\Hat{\xbf}_{t}))^2 
    \right] ,
\end{split}
\end{align}\\[-5pt]
%
%
where
\mbox{$\smash{p_{\hat{\Xbf}_{t}}}$} is a uniform distribution over noised context batches $\Hat{\xbf}_{t}$ of size $M$ (with \mbox{$M \ll N_\calC$}), and
\begin{align}
\label{eq:mahalanobis_distance}
    &
    D_{\mathcal{M}}(f_{t}^{j}(\Hat{\xbf}_{t} ; \theta), m_{t}^{j}(\Hat{\xbf}_{t}), {K}_{t}(\Hat{\xbf}_{t}))^2
    \\
    &
    \defines
    (f_{t}^{j}(\Hat{\xbf}_{t} ; \theta) - m_{t}^{j}(\Hat{\xbf}_{t}))^\top {K}_{t}(\Hat{\xbf}_{t})^{-1} (f_{t}^{j}(\Hat{\xbf}_{t} ; \theta) - m_{t}^{j}(\Hat{\xbf}_{t}))
    \nonumber
\end{align}
is the squared Mahalanobis distance between noised model predictions $f_{t}^{j}(\Hat{\xbf}_{t} ; \theta)$ and a data- and timestep-dependent distribution over random vectors with mean $m_{t}^{j}(\Hat{\xbf}_{t})$ and covariance ${K}_{t}(\Hat{\xbf}_{t})$.
Here, $m_{t}^{j}(\Hat{\xbf}_{t})$ specifies the desired out-of-distribution behavior for each guidance model head, and ${K}_{t}(\Hat{\xbf}_{t})\in\R^{M\times M}$ allows us to encode additional information about structural properties of the input domain.

Specifically, ${K}_{t}(\Hat{\xbf}_{t})\neq\mathbf{I}$ controls how strongly the guidance model predictions $f_{t}^{j}(\Hat{\xbf}_{t} ; \theta)$ can vary between similar context points, allowing us to enforce smoother gradient estimates and more stable guidance signals in regions of data space with limited signal from the labeled training data.





This covariance function can be defined in terms of any valid, domain-appropriate similarity metric.
In the following, we present a general approach to constructing ${K}_{t}$ from model embeddings.
Specifically, we consider a separate set of fixed parameters $\phi$ to generate the network embeddings $h_t(\Hat{\xbf}_{t}, \phi) \in \R^{M\times d}$ of all molecules in the context batch $\Hat{\xbf}_{t}$. We then use their inner product to construct
\begin{align}
\label{eq:covariance_matrix}
\begin{split}
    {K}_{t}(\Hat{\xbf}_{t})
    \defines
    \sigma_{t} h_t(\Hat{\xbf}_{t}, \phi)h_t(\Hat{\xbf}_{t}, \phi)^T + \tau_{t}\mathbf{I} ,
\end{split}
\end{align}
where $\sigma_{t}$ is a covariance scale parameter and $\tau_{t}$ is a diagonal offset parameter.
These hyperparameters allow us to calibrate the behavior of our regularizer, with $\tau_{t}$ determining how closely the predictions have to match the mean function $m_{t}^{j}(\Hat{\xbf}_{t})$ and $\sigma_{t}$ determining the strength of the smoothness constraints placed on $f_{t}^{j}(\Hat{\xbf}_{t};\theta)$ for both \mbox{$j=1,2$}.

This approach is agnostic to how the embeddings $h_t(\Hat{\xbf}_{t}, \phi)$ are derived, so long as they encode relevant structural properties of the input domain.
We find that a relatively simple approach is already sufficient to produce substantial performance gains in the settings we consider in~\Cref{sec:experiments}. Specifically, we use a fixed set of randomly initialized parameters $\phi$ to obtain $h_t(\Hat{\xbf}_{t}, \phi)$ from the model's embedding trunk. 
This is applicable across different network types and domains, and motivated by the observation that neural networks with suitable inductive biases, such as permutation- or roto-translational equivariance in~\Cref{sec:exp_small_molecules,sec:exp_materials}, can already produce robust and informative representations in a randomly initialized state~\citep{kipf2016semi}.

Having constructed the covariance matrix ${K}_{t}(\Hat{\xbf}_{t})$, we now specify the mean function $m_{t}^{j}(\Hat{\xbf}_{t})$ for $j=1,2$.
In particular, letting $f_{t}(\cdot ; \theta)$ be as defined above, with a learned mean function $f_{t}^{1}(\Hat{\xbf}_{t} ; \theta)$ and a learned variance function $f_{t}^{2}(\Hat{\xbf}_{t} ; \theta)$, we wish for the mean predictions to revert to the mean of the training labels and for the variance to be high when the model is evaluated on points that are meaningfully different from the training data.
To encode this desired behavior into the regularizer defined in \Cref{eq:regularizer}, we define $\smash{m^{1}_{t}(\Hat{\xbf}_{t}) \defines \frac{1}{N}\sum_{n=1}^{N} \Bar{\ybf}^{(n)}}$ for the training data labels $\Bar{\ybf}^{(n)}$, $\smash{m^{2}_{t}(\Hat{\xbf}_{t}) \defines \boldsymbol{\sigma}_{0}^{2}}$, for a target variance parameter $\boldsymbol{\sigma}_{0}^{2}$.
%

Adding this regularizer to the supervised learning loss from \Cref{eq:standard_loss}, we arrive at the modified training objective
\begin{align}
\label{eq:fseb_loss}
\begin{split}
    \mathcal{L}^\ast(\theta, \calD_{\mathcal{T}})
    =
    \E_{p_{\mathcal{T}}}[ \calL(\theta, \calD_{\mathcal{T}}, \mathcal{T}) + R(\theta, f_{\mathcal{T}}, \mathcal{T}, p_{\hat{\Xbf}_{\mathcal{T}}}) ],
\end{split}
\end{align}\\[-10pt]
where both expectations (over $p_{\mathcal{T}}$ and $p_{\hat{\Xbf}_{t}}$) can be estimated via simple Monte Carlo estimation.

\pagebreak


\paragraph{Implementation.}
The proposed regularizer $R$ does not require any changes to the guidance model itself and can be easily integrated into standard guidance model training pipelines:
At each iteration, we sample a mini-batch $\Hat{\xbf}_{t}$ of size $M$ from the context set $\Hat{\xbf}_{\calC}$ and perturb it with the same noising process as the unconditional diffusion model in~\Cref{eq:forward_noising_process}.
We then perform a gradient-free forward pass of the embedding model to compute $h_t(\Hat{\xbf}_{t}, \phi)$ and construct the covariance matrix ${K}_{t}(\Hat{\xbf}_{t})$.
Finally, we use the predictions of the guidance model $f_{t}(\Hat{\xbf}_{t} ; \theta)$ on the context batch to compute the full regularization term in~\Cref{eq:regularizer}.
An algorithmic outline of this process is shown in~\Cref{fig:reg_illustration}.

As the risk of miscalibrated gradient estimates is particularly high in the earlier stages of the sample generation process---where highly diffused input points are associated with the labels of the corresponding un-noised points---it is important to ensure that our regularizer is able to take the noise level $\beta_t$ into account.
Specifically, we increase $\tau_{t}$ with the same schedule as the noise levels $\beta_t$ to penalize overconfident predictions as context points approach the reference distribution $p_T=\calN(\mathbf{0},\mathbf{I})$.
Similarly, as the distinction between in-distribution and out-of-distribution data points becomes less meaningful at larger noise levels, we decrease $\sigma_{t}$ with an inverted schedule.

\subsection{Generation via Context-Guided Diffusion}

Once the context-aware guidance model $f_{t}^{j}(\cdot ; \theta)$ has been trained, we can use it to steer the denoising process towards out-of-distribution samples that are most likely to exhibit desired properties. 
To do so, we use the context-aware guidance model to specify a context-aware guidance function $p(\ybf \vbar \xbf_{t} ; f_{t}(\cdot ; \theta))$.
This function defines a distribution, conditioned on both the training data and the context set, that reflects the probability of obtaining the desired properties given $\xbf_{t}$. This allows us to integrate signals from labeled data points with structural knowledge about the broader input domain into the sampling process.
Details about guidance function specification are provided in \Cref{app:guidance}, and an illustrative example that showcases the effect of the proposed regularizer on the guidance model (and the resulting guided diffusion process) is shown in~\Cref{fig:reg_illustration}.

Finally, the context-aware guidance function can be seamlessly integrated into existing guided diffusion model frameworks by replacing standard guidance functions at sampling time.
This modular, plug-and-play approach ensures that our method is independent of the specific diffusion model architecture and generative process and requires minimal implementational overhead.
Importantly, like conventional classifier-guided diffusion, our method does not necessitate any \mbox{(re-)training} or fine-tuning of the diffusion model itself, making it a convenient out-of-the-box tool for conditioning the generation process of a pre-trained model.

\clearpage

\section{Related Work}
\label{sec:related_work}

\paragraph{Guided Diffusion for Molecular Design.}
Guiding the denoising process of diffusion models with gradients from a classifier or conditional score network is an effective approach that has been applied to a broad range of scientific problems, such as the generation of desirable protein folds \citep{watson2023novo,ingraham2023illuminating} and molecular conformations \citep{hoogeboom2022equivariant,pmlr-v202-peng23b,pmlr-v202-guan23a}.
Similarly to our approach, recent work has explored property-guided diffusion models for molecular optimization with a particular focus on out-of-distribution sample generation.
For instance, \citet{lee2023exploring} proposed a modification of the unconditional denoising process that, in combination with standard supervised guidance, generated more diverse small molecules with improved properties.
Likewise, \citet{weiss2023guided} used a guided diffusion model to optimize the electronic properties of polycyclic aromatic systems, demonstrating generalization to unseen heterocycles.
Furthermore, \citet{gruver2023protein} explored property-guided discrete diffusion for multi-objective protein sequence optimization, balancing naturalness with improved properties through a grid of sampling hyperparameters.
However, all of these approaches primarily focus on modifying the unconditional sampling process itself and rely on standard supervised learning techniques for guidance model training. Our method is orthogonal to these contributions and aims to further improve the performance of property-guided diffusion models across application domains and generative processes.

\paragraph{Regularization with Unlabeled Data.} Using unlabeled data to improve the performance of predictive models is a well-established paradigm that underpins many successful semi-supervised \citep{kingma2014semi,laine2016temporal,berthelot2019mixmatch} and self-supervised \citep{brown2020language,bommasani2021opportunities} algorithms. This includes a range of unsupervised domain adaptation and generalization techniques that use unlabeled data to mitigate the adverse effect of distribution shifts \citep{tzeng2015simultaneous,sun2016deep,ganin2016domain,kang2019contrastive,li2020model}. However, these methods predominantly focus on aligning model embeddings and are, contrary to our approach, unable to directly regularize model predictions.
The method most closely related to our approach is the regularization scheme proposed in \citet{rudner2023function}.
However, in contrast to~\citet{rudner2023function}, who only consider supervised image classification tasks, our approach is specifically designed for the conditional generation of molecular structures.
Key methodological developments in this regard include the introduction of a factorized regularization term that is suitable for use with regression models, as well as an explicit timestep-dependence that accounts for the noising process of the underlying diffusion model.

\newpage
\section{Empirical Evaluation}
\label{sec:experiments}

We compare our method to both standard guidance models and more sophisticated pre-training and domain adaptation techniques across a range of experimental settings. Specifically, we demonstrate the versatility of context-guided diffusion by applying it to the design of small molecules with graph-structured diffusion processes (\Cref{sec:exp_small_molecules}), to the generation of novel materials with equivariant diffusion models (\Cref{sec:exp_materials}), and to the optimization of discrete protein sequences with categorical diffusion models  (\Cref{sec:exp_proteins}).
We observe that context-guided diffusion leads to improved conditional sampling processes that consistently and substantially outperform existing methods across different diffusion model types and application domains.

\subsection{Graph-Structured Diffusion for Small Molecules}
\label{sec:exp_small_molecules}

To evaluate our method in the context of graph-structured diffusion processes for small molecule generation, we follow the framework of \citet{lee2023exploring} and consider a discrete diffusion process over the space of molecular graphs. 
Defining a graph $\mathbf{g}_t$ as a tuple of node feature and adjacency matrices $\mathbf{g}_t=(\xbf_t, \mathbf{a}_t)$, the forward and reverse processes are given by a joint system of \textsc{SDE}s \citep{jo2022score}. 
To encourage the exploration of novel chemical space, \citet{lee2023exploring} condition the denoising process on a hyperparameter $\gamma\in[0,1)$ that shrinks the unconditional gradients
\begin{align}
\label{eq:exp_mood_lambda}
\begin{split}
    \nabla\log p_t(\mathbf{g}_t \vert\gamma) 
    &=
    \nabla \log p_t(\mathbf{g}_t)+\nabla\log p_t(\gamma \vert \mathbf{g}_t)\\
    &= \left(1-\sqrt{\gamma}\right)\nabla \log p_t(\mathbf{g}_t).
\end{split}
\end{align}
In addition, the denoising process is conditioned on desirable molecular properties with a standard supervised guidance model as outlined in~\Cref{sec:background,sec:standard_guidance_model_training}.
This model is trained to predict the labels $\ybf\in[0, 1]$, defined as a composite of the synthetic accessibility~\citep[\textsc{SA};][]{ertl2009estimation}, drug-likeness~\citep[\textsc{QED};][]{bickerton2012quantifying}, and QuickVina docking scores \citep{alhossary2015fast}, described in further detail in~\Cref{app:small_molecules}. In this setting, larger labels denote more desirable properties.

\begin{figure*}[!t]
    \centering
    \includegraphics[width=\textwidth]{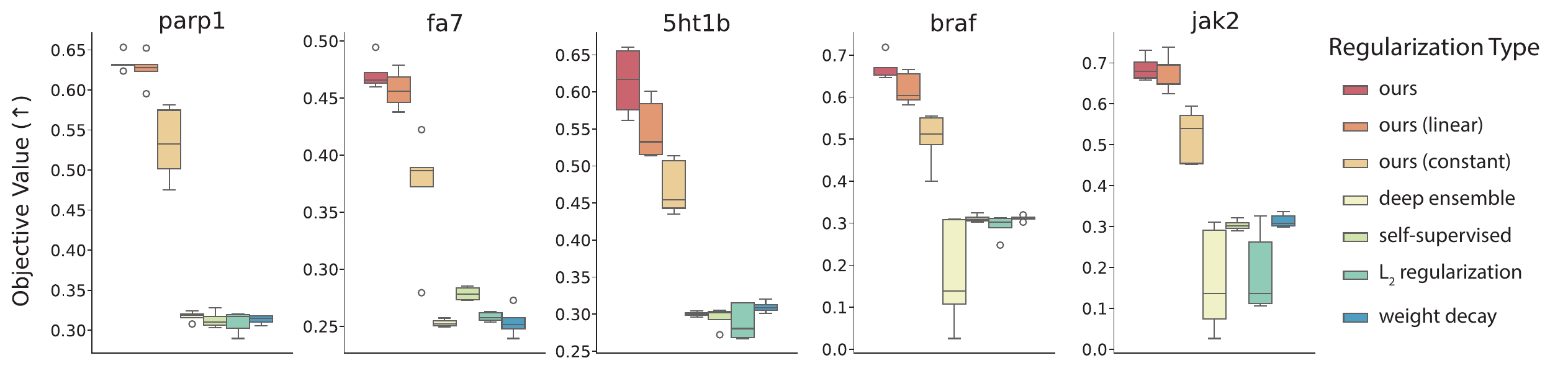}
    \caption{Comparison of the small molecules generated with different guided diffusion models across five distinct protein targets. Objective values ($\uparrow$) are normalized with respect to the highest score of the held-out high-property validation set and averaged across five independent training and sampling runs with different random seeds.}
    \label{fig:small_molecules_results}
\end{figure*}

As the docking scores are protein pocket-dependent, we repeat all experiments across the five different targets chosen by \citet{lee2023exploring}, namely \textsc{PARP1}, \textsc{FA7}, \textsc{5HT1B}, \textsc{BRAF} and \textsc{JAK2}.
We closely follow their experimental setup and train guidance models on a subset of \num{250000} small molecules from the \textsc{ZINC} database \citep{irwin2012zinc}. Specifically, we pre-compute the labels $\ybf$ for every target and use them to split the data into a low-property training and a high-property validation set. This label split allows us to select regularization hyperparameters that maximize the ability of the guidance models to generalize well to novel, high-value regions of chemical space, serving as a proxy for the desired behavior at sampling time.

\newpage

To construct an appropriate context set for use with our regularizer, we sample an additional \num{500000} unlabeled small molecules from the \textsc{ZINC} database.
Following ~\Cref{sec:regularization_objective}, we specify a predictive distribution that reverts to the training set mean and high uncertainty on out-of-distribution context points. We compare this approach to a range of alternative regularization schemes, including weight decay (used in \citet{lee2023exploring}), explicit $L_2$ regularization, and an ensemble of independently trained networks \citep{lakshminarayanan2017simple}. 
To facilitate a fair comparison of these methods, we ensure that all training parameters, except regularization type and strength, are kept constant across experiments. After selecting the optimal hyperparameters using the high-property validation set, all models are independently retrained with five different random seeds. We refer to~\Cref{app:small_molecules} for full implementational details, hyperparameter ranges, additional results, and sanity checks.

\begin{figure*}[!t]
\begin{minipage}[c]{0.65\textwidth}%
  \centering
  \includegraphics[width=\textwidth]{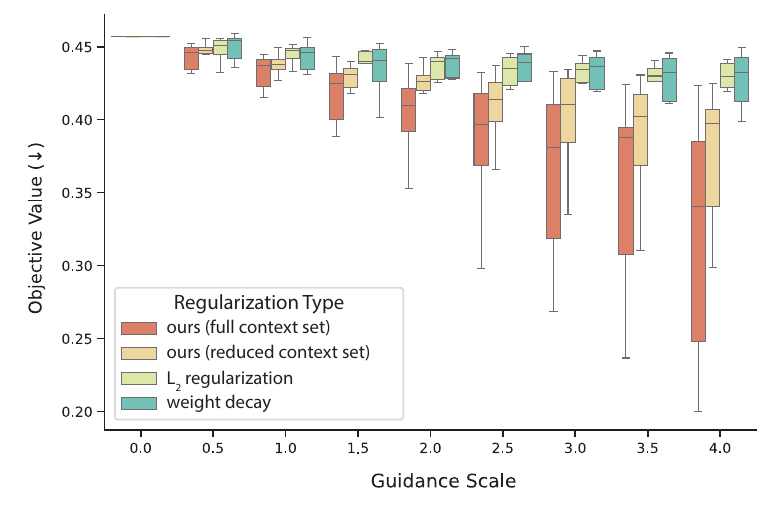}%
\end{minipage}%
\hfill%
\begin{minipage}[c]{0.33\textwidth}%
  \vspace{0pt}
  \centering
  \includegraphics[width=\textwidth]{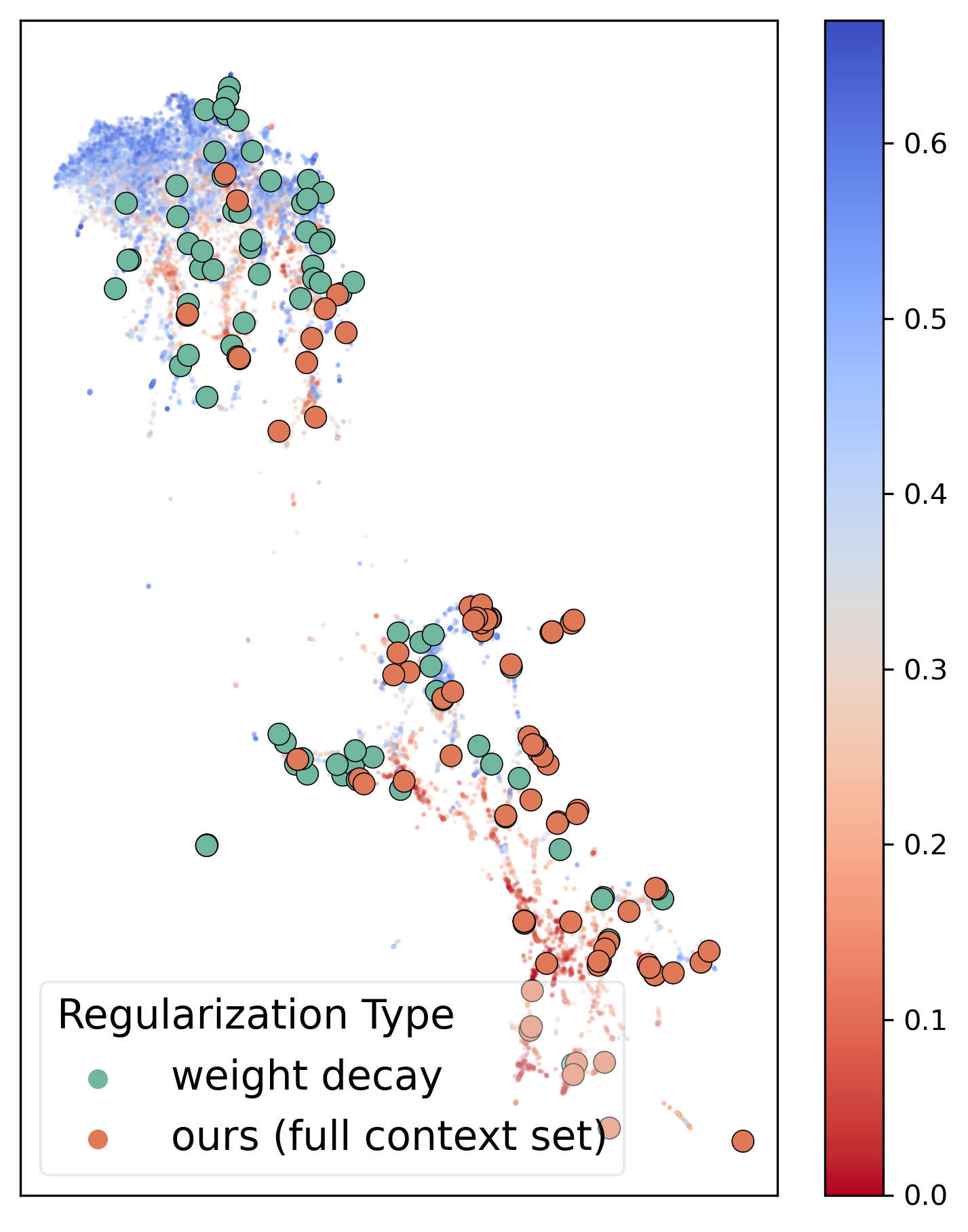}%
\end{minipage}%
\caption{Comparison of polycyclic aromatic systems generated with different guidance models across ten independent training and sampling runs. \textbf{Left:} Full distribution of generated objective values ($\downarrow$) ablated over different context sets and guidance scales. \textbf{Right:} \textsc{UMAP} plot \citep{mcinnes2018umap} of training (upper left) and test set (lower right), as well as samples from guided diffusion models. Validity and novelty are analyzed in~\Cref{app:materials} and show similar trends.}
\label{fig:materials_results}
\end{figure*}

We additionally evaluate the performance of other approaches that are able to make use of unlabeled data. Specifically, we pre-train a guidance model on all molecules from our context set with a self-supervised denoising objective \citep{zaidi2023pretraining} and then fine-tune it on the labeled data, using weight decay. 
Furthermore, we consider a range of unsupervised domain adaptation and generalization techniques, namely DeepCoral \citep{sun2016deep}, domain-adversarial neural networks \citep{ganin2016domain}, and domain confusion \citep{tzeng2015simultaneous}, and present the results for these methods in~\Cref{app:small_molecules}.

Adhering to the evaluation protocol of \citet{lee2023exploring}, we use the retrained guidance models to generate \num{3000} molecules each and derive the corresponding labels with the provided docking and evaluation pipeline.
As in \citet{lee2023exploring}, we only retain compounds that pass synthetic accessibility and drug-likeness thresholds and display a maximum ECFP4-based Tanimoto similarity \citep{rogers2010extended} of less than 0.4 to any molecule in the training set, before reporting the top 5\% of the remaining docking scores. 

To put these absolute values into perspective, we normalize them so that \num{1} corresponds to the highest docking score in the high-property validation set. 
The results are reported in~\Cref{fig:small_molecules_results} and demonstrate that our method consistently improves the performance of property-guided diffusion models across different protein targets. 
The full distributions of all components of the objective function, the results of models trained with label noise, as well as versions normalized by heavy atom count, are presented in \Cref{app:small_molecules} and display identical trends.
While ensembling and pre-training techniques perform best on in-distribution tasks (see \Cref{app:small_molecules}), we find that they alone are not sufficient to overcome the limitations of standard supervised training/fine-tuning techniques. However, we note that context guidance is an orthogonal approach that is straightforward to use in combination with these methods.

To investigate how important the timestep-dependence of our method is to its empirical performance, we carry out a series of ablations in which the regularization hyperparameters $\sigma_t$ and $\tau_t$ either \begin{enumerate*}[label=(\roman*)]
    \item mimic the $\beta_t$ schedule of the noising process,
    \item change linearly over time, or 
    \item stay constant across all noise scales.
\end{enumerate*}
We observe a strict performance drop across these settings, indicating that adapting the regularizer to the noising process is highly beneficial (\Cref{fig:small_molecules_results}). 

Furthermore, we investigate the impact that the size and composition of our context set have on the performance of our method by comparing models trained with the full context set to models trained with randomly sampled subsets of size 10\% and 1\%, respectively.
As expected, we observe a strong decrease in performance, with the results of models trained with the smallest context set (1\%) reverting to that of standard weight decay.
Additionally, we select the 10\% of compounds in the original context set that are either most or least similar to the labeled training data and find that using a much smaller set of more similar, near-OOD molecules is able to match the performance of the full context set.
Full details and experimental results are presented in~\Cref{tab:app_context_set_ablation}.

\subsection{Equivariant Diffusion For Materials}
\label{sec:exp_materials}

We additionally evaluate our method by applying it to equivariant diffusion models for materials design. Specifically, we follow the experimental setup of \citet{weiss2023guided} and train an E(3)-equivariant diffusion model \citep{hoogeboom2022equivariant} on all polycyclic aromatic systems consisting of up to 11 cata-condensed benzene rings \citep{wahab2022compas}. As in \citet{weiss2023guided}, we guide the sampling process towards molecules with desirable electronic properties by leveraging the gradients of an E(3)-equivariant graph neural network \citep{satorras2021n} to minimize a composite objective of adiabatic ionization potential, electron affinity, and HOMO-LUMO gap.

In contrast to the experimental setting in \Cref{sec:exp_small_molecules}, this application considers an exhaustively enumerated search space, allowing us to construct clustered data splits in which maximal generalization is required to reach held-out regions with optimal objective values.
Using this setup, we train a diffusion model and different guidance networks on a low-value training set and examine how well they are able to recover molecules from held-out regions with desirable electronic properties. We present a comparison of different methods in~\Cref{fig:materials_results}, confirming that our approach performs significantly better than standard regularization techniques across guidance scales.
We additionally ablate the information content of the context set by removing the most relevant entries, observing that these are a key determinant of our method's success.
Full experimental details and additional results are provided in~\Cref{app:materials}.

\subsection{Discrete Diffusion for Protein Sequences}
\label{sec:exp_proteins}
Furthermore, we explore how well our method extends to categorical diffusion models in the context of protein sequence optimization. In particular, we follow the experimental framework of \citet{gruver2023protein} to apply context-guided diffusion to the property-conditioned infilling of antibody complementarity-determining regions.
In contrast to the applications in~\Cref{sec:exp_small_molecules,sec:exp_materials}, the guidance model of \citet{gruver2023protein} is not a separate network, but rather a regression head of the score model. 
Specifically, a masked language model \citep{austin2021structured,bhargava2021generalization} is used to learn sequence embeddings from which a linear output head and a feed-forward network estimate the scores and properties, respectively.

\begin{figure}[!t]
    \centering
    \includegraphics[width=0.95\linewidth]{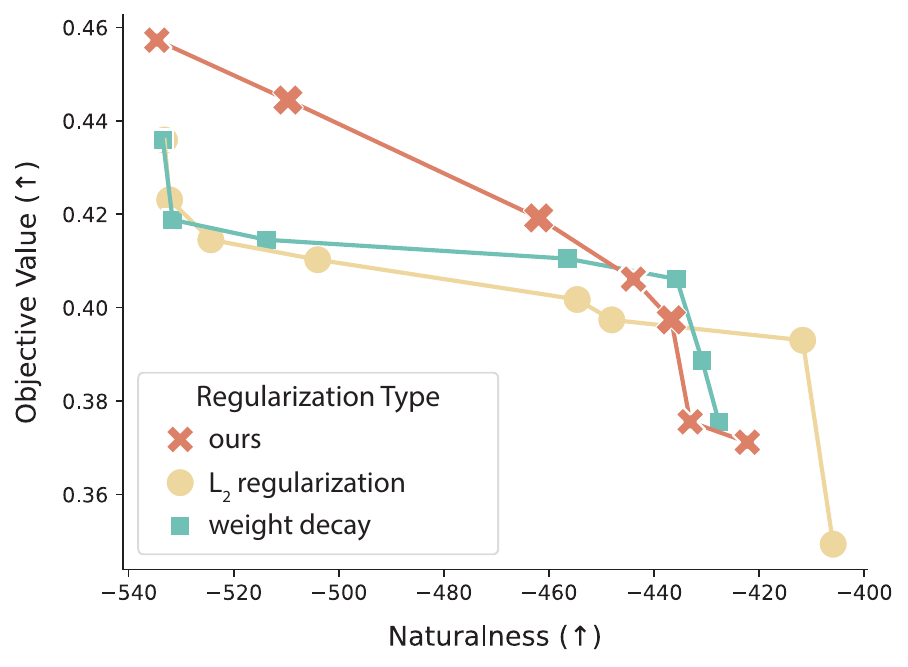}
    \caption{Pareto fronts of samples generated with different regularization schemes, highlighting the trade-off between objective value $(\uparrow)$ and naturalness $(\uparrow)$. As samples move away from the training data and enter an out-of-distribution regime, our method consistently generates sequences with better properties for any given level of naturalness.}
    \label{fig:exp_proteins}
    \vspace*{-6pt}
\end{figure}

At sampling time, \citet{gruver2023protein} generate sequences with a grid of denoising hyperparameters and examine the resulting Pareto front of objective values 
versus ``naturalness'' i.e. how likely a sequence is to be synthesizable, estimated by its likelihood under a protein language model \citep{ferruz2022protgpt2}.
Following this approach, we train different guidance models and use them to generate samples with the same hyperparameter grid as \citet{gruver2023protein}. The corresponding Pareto fronts are presented in~\Cref{fig:exp_proteins}. While standard methods fare well in in-distribution settings, we observe that models trained with our regularizer outperform them as samples progress into an out-of-distribution regime, consistently producing better properties for a given level of naturalness. Full experimental details and additional results are provided in~\Cref{app:proteins}. 

\section{Discussion and Limitations}
\label{sec:discussion}

We conducted a comprehensive empirical evaluation of our method across three distinct molecular and protein design tasks, observing consistent and significant improvements over existing state-of-the-art guided diffusion models. By leveraging unlabeled data to train guidance functions that exhibit high uncertainty and well-behaved gradients on out-of-distribution context points, our context-guided models enabled the more reliable generation of novel molecules with improved properties. These gains were most pronounced when generalization from scarce and biased training data was required to generate compounds from unseen, high-value subsets of chemical and protein sequence space. Our results highlight that even relatively simple regularization techniques can yield substantial empirical benefits when combined with relevant context sets.

The proposed method is best suited for domains in which unlabeled data is abundant or can be easily generated.
While it does not incur any computational overhead at sampling time, where hundreds of model evaluations are required to generate a single molecule, it does come with an increased cost during training.
Specifically, computing the regularization term requires two additional forward passes per training iteration to generate the context set embeddings and predictions.
Fortunately, we found this added cost to be moderate in absolute terms (see \Cref{app:runtime}), as guidance models tend to be relatively lightweight. Additionally, we provide a sensitivity study over context batch sizes in~\Cref{tab:context_batch_size_sensitivity}, which suggests that the computational overhead can be further reduced while maintaining favorable empirical performance.

\newpage

Beyond choosing the context batch size, our approach requires the optimization of two additional hyperparameters: the covariance scale $\sigma_t$ and the diagonal offset parameter $\tau_t$. Additionally, we find that the choice of the context set significantly impacts model performance, requiring sufficient domain expertise to ensure the inclusion of informative and relevant unlabeled data points.


\begin{table}[t]
\centering
\caption{Ablation over the context batch size used by our regularizer. Larger context batches generally yield better results but incur higher computational costs. We find that the context batch size can be reduced substantially with only a marginal deterioration of model performance. 
Performance metrics and runtimes are averaged across 5 independent training and sampling runs with different random seeds.}
\vspace*{-7pt}
\label{tab:context_batch_size_sensitivity}
\begin{adjustbox}{max width=\columnwidth}
\begin{tabular}{lcccc}

\multicolumn{5}{c}{\textbf{Graph-Structured Diffusion for Small Molecules (\Cref{sec:exp_small_molecules})}} \\
\toprule
Batch Size & Batch Time (s) & \textsc{parp1} ($\uparrow$) & \textsc{fa7} ($\uparrow$) & \textsc{5ht1b} ($\uparrow$) \\
\midrule
128 & $0.09\pms{0.01}$ & $\mathbf{0.66\pms{0.01}}$ & $\mathbf{0.48\pms{0.01}}$ & $\mathbf{0.57\pms{0.01}}$ \\
64 & $0.08\pms{0.00}$ & $\mathbf{0.66\pms{0.01}}$ & $\mathbf{0.48\pms{0.01}}$ & $\mathbf{0.57\pms{0.01}}$ \\
32 & $\mathbf{0.07\pms{0.00}}$ & $\mathbf{0.66\pms{0.01}}$ & $\mathbf{0.48\pms{0.01}}$ & $\mathbf{0.57\pms{0.02}}$ \\
16 & $\mathbf{0.07\pms{0.00}}$ & $0.65\pms{0.02}$ & $0.47\pms{0.02}$ & $0.56\pms{0.03}$ \\
8 & $\mathbf{0.07\pms{0.00}}$ & $0.62\pms{0.03}$ & $0.45\pms{0.03}$ & $0.53\pms{0.03}$ \\
\midrule
\multicolumn{5}{c}{}\\[-7pt]
\multicolumn{5}{c}{\textbf{Equivariant Diffusion For Materials (\Cref{sec:exp_materials})}} \\
\toprule
Batch Size & Batch Time (s) & scale=0 ($\downarrow$) & scale=2 ($\downarrow$) & scale=4 ($\downarrow$)\\
\midrule
128 & $0.28\pms{0.02}$ & $0.45\pms{0.00}$ & $\mathbf{0.39\pms{0.01}}$ & $\mathbf{0.34\pms{0.01}}$ \\
64 & $0.25\pms{0.03}$ & $0.45\pms{0.00}$ & $\mathbf{0.40\pms{0.01}}$ & $\mathbf{0.34\pms{0.02}}$ \\
32 & $0.26\pms{0.01}$ & $0.45\pms{0.00}$ & $0.42\pms{0.01}$ & $0.36\pms{0.01}$ \\
16 & $0.24\pms{0.02}$ & $0.45\pms{0.00}$ & $0.41\pms{0.01}$ & $0.38\pms{0.03}$ \\
8 & $\mathbf{0.23\pms{0.01}}$ & $0.45\pms{0.00}$ & $0.42\pms{0.01}$ & $0.39\pms{0.02}$ \\
\bottomrule
\end{tabular}
\end{adjustbox}
\vspace*{-7pt}
\end{table}

\vspace*{-3pt}
\section{Conclusions}
\label{sec:conclusion}

In this paper, we introduced context-guided diffusion, a simple plug-and-play method that leverages unlabeled data and smoothness constraints to improve the out-of-distribution generalization of guided diffusion models.
We demonstrated the versatility of this approach by applying it to the design of small molecules, materials, and proteins, achieving substantial performance gains in each domain. 
Several promising directions for future work exist, such as encoding more complex out-of-distribution behavior into the guidance model, for instance by reverting to the outputs of physics-based methods rather than high predictive uncertainty.
Another promising avenue is the construction of maximally informative context sets, potentially through active learning strategies that iteratively select unlabeled data points based on their relevance and potential to improve guidance model performance.
Exploring the integration of context-guided diffusion with techniques such as multi-task learning or meta-learning could also enable the development of more versatile guidance functions that leverage knowledge from related domains or tasks.
We believe that exploring this line of research has the potential to lead to more adaptable and robust diffusion models that are able to solve a wider range of challenging real-world problems.




\clearpage


\section*{Impact Statement}

This work introduces a method that is designed to improve the performance of guided diffusion models for inverse molecular and protein design.
Our method aims to accelerate the discovery and development of new medicines and more performant and sustainable materials that could benefit society in a variety of ways \citep{wang2023scientific}.
However, there is also the risk of dual-use or misuse, for example for the generation of hazardous agents \citep{urbina2022dual}.

\section*{Acknowledgements}

We thank the anonymous reviewers for their useful feedback.
LK acknowledges financial support from the University of Oxford’s Clarendon Fund and F. Hoffmann-La Roche AG.
We gratefully acknowledge the Oxford Advanced Research Computing service for providing computing resources and infrastructure.

\bibliography{references}
\bibliographystyle{include/icml/icml2024}

\clearpage

\begin{appendices}

\crefalias{section}{appsec}
\crefalias{subsection}{appsec}
\crefalias{subsubsection}{appsec}

\setcounter{equation}{0}
\renewcommand{\theequation}{\thesection.\arabic{equation}}

\onecolumn


\vspace*{-10pt}
\section*{\Huge Appendix}
\label{sec:appendix}

\vspace*{20pt}
\section*{Table of Contents}
\vspace*{-5pt}
\startcontents[sections]
\printcontents[sections]{l}{1}{\setcounter{tocdepth}{2}}
\vspace*{10pt}

\section*{Reproducibility}

\codebox{All data and code needed to reproduce our experiments can be accessed at}{\begin{center}
    \href{https://github.com/leojklarner/context-guided-diffusion}{\centering\texttt{
    https://github.com/leojklarner/context-guided-diffusion}}
\end{center}}


\clearpage

\section{Guidance Function Parameterization}
\label{app:guidance}

To guide sampling in a diffusion model toward certain desired properties, a {\em guidance model} $f_{t}(\cdot ; \theta)$ is trained on a dataset $\calD$, a parametric {\em guidance function} $p(\ybf \vbar \xbf_{t} ; f_{t}(\cdot ; \theta))$ is defined using the learned guidance model, and then the {\em conditional gradient/score function}
\begin{align}
    \nabla \log p(\ybf \vbar \xbf_{t} ; f_{t}(\cdot ; \theta)) + \nabla \log p(\xbf_{t}) ,
\end{align}
is used to guide generation towards high-density regions $p(\ybf \vbar \xbf_{t} ; f_{t}^{1}(\cdot ; \theta))$.
To provide a general framework that accommodates both classification and regression settings, we let $\Ybf$ denote an ``optimality'' random variable indicating the optimality (under some domain-specific optimality function) of a given input $\xbf_{t}$.

For discrete desirable properties, the guidance model $f_{t}(\cdot ; \theta)$ (denoting the logits for a given number of classes) is trained on a classification task, and the guidance function $p(\ybf \vbar \xbf_{t} ; f_{t}(\cdot ; \theta))$ for a specific class $\ybf$ is defined as a categorical likelihood function evaluated at $\ybf$, making $-\log p(\ybf \vbar \xbf_{t} ; f_{t}(\cdot ; \theta))$ a cross-entropy loss.
The score function then becomes the class probability of the desired class under the learned model:
\begin{align}
    \nabla \log p(\ybf \vbar \xbf_{t} ; f_{t}^{1}(\cdot ; \theta))
    =
    \nabla \log \mathrm{softmax}(f_{t}^{1}(\xbf \vbar \theta))_{c} .
\end{align}
For continuous desirable properties, the guidance model (with a mean and a variance head) is trained on a regression task, and the guidance function $p(\ybf \vbar \xbf_{t} ; f_{t}(\cdot ; \theta))$ is defined as a Bernoulli likelihood function with
\begin{align}
    p(\ybf = 1 \vbar \bx_{t}, C ; f_{t}(\cdot ; \theta))
    =
    \exp{(f_{t}^{1}(\xbf_{t} ; \theta) - C}) ,
\end{align}
giving the probability of optimality as a function of the mean-head output $f_{t}^{1}(\xbf_{t} ; \theta)$, and $C$ representing a domain-specific ``goal value''.
Importantly, this formulation requires that $f_{t}^{1}(\xbf_{t} ; \theta) < C$ so that increases in $f_{t}^{1}(\xbf_{t} ; \theta))$ indicate desirable properties.

\subsection{Probabilistic Interpretation}
\label{sec:bayesian_interpretation}

Letting $\ybf$ be an ``optimality'' random variable indicating whether predictions $f_{t}(\Hat{\xbf}_{t} ; \theta)$ are optimal for achieving a desired behavior (e.g., high predictive uncertainty on data points far away from the training data), we can define a Bernoulli observation model where
\begin{align}
    p(\ybf = 1 \vbar \theta, t ; f_{t}, p_{\hat{\Xbf}_{t}})
    =
    \exp{(-R(\theta, f_{t}, t, p_{\hat{\Xbf}_{t}}))} ,
\end{align}
denotes the probability of optimality as a function of the regularizer.
That is, the smaller $R(\theta, f_{t}, t, p_{\hat{\Xbf}_{t}})$, the higher the probability that a prediction $f_t(\Hat{\xbf}_{t} ; \theta)$ is optimal.

Following the result in \citet{rudner2023function}, we note that specifying a data-dependent regularizer of the form shown in \Cref{eq:regularizer} corresponds to specifying a data-driven prior distribution over neural network parameters
\begin{align*}
    p(\theta \vbar t, \ybf=1 ; f_{t}, p_{\hat{\Xbf}_{t}})
    =
    \frac{p(\ybf = 1 \vbar \theta, t ; f_{t}, p_{\hat{\Xbf}_{t}}) \, p(\theta)}{p(\ybf = 1 \vbar t ; f_{t}, p_{\hat{\Xbf}_{t}})}.
\end{align*}
Minimizing the objective function in \Cref{eq:fseb_loss} corresponds to finding a variational approximation to the posterior implied by this data-driven prior and the observed data.
That is, minimizing the objective function in \Cref{eq:fseb_loss}
(approximately) corresponds to finding a variational distribution $q_{\Theta} = \mathcal{N}(\theta ; \theta', \sigma^2)$ (with fixed and very small $\sigma^2$) that solves the variational problem\vspace*{-1pt}
\begin{align}
    \min_{q_{\Theta}} \mathbb{E}_{p_{\mathcal{T}}} \left[\mathbb{D}_{\mathrm{KL}}(q_{\Theta} \,||\, p_{\Theta | \mathcal{D}_{\mathcal{T}}, \ybf, \mathcal{T}})\right] ,
\end{align}
allowing us to interpret the learned guidance model through the lens of Bayesian inference.

\clearpage

\section{Experimental Details}
\label{app:exp_details}

This section provides additional information and experimental results complementing the main text. \Cref{app:toy_example} details the regression experiments from \Cref{sec:method} and contrasts our approach with standard $L_2$ regularization, illustrated in~\cref{fig:reg_illustration}. \Cref{app:small_molecules} presents our extension of \citet{lee2023exploring}'s work on graph-structured diffusion models for small molecule generation (\Cref{sec:exp_small_molecules}). \Cref{app:materials} presents our extension of \citet{weiss2023guided}'s experiments on equivariant diffusion models for generating novel polycyclic aromatic systems (\Cref{sec:exp_materials}). Finally, \Cref{app:proteins} presents our extension of \citet{gruver2023protein}'s approach to optimizing discrete protein sequences using categorical diffusion models (\Cref{sec:exp_proteins}).

All code was written in \citep{python} and can be accessed at \href{https://github.com/leojklarner/context-guided-diffusion}{https://github.com/leojklarner/context-guided-diffusion}. A range of core scientific computing libraries were used for data preparation and analysis, including \textsc{NumPy} \citep{numpy}, \textsc{SciPy} \citep{scipy}, \textsc{Pandas} \citep{pandas1}, \textsc{Matplotlib} \citep{matplotlib}, \textsc{Seaborn} \citep{seaborn}, \textsc{Scikit-learn} \citep{scikit-learn} and \textsc{rdkit} \citep{landrum2013rdkit}. All deep learning models were implemented in \textsc{PyTorch} \citep{paszke2019pytorch}. 

\subsection{Illustrating the Behaviour of Guided Diffusion Models on the Swiss Roll Dataset}
\label{app:toy_example}

\begin{figure}[H]
    \centering
    \includegraphics[width=11cm]{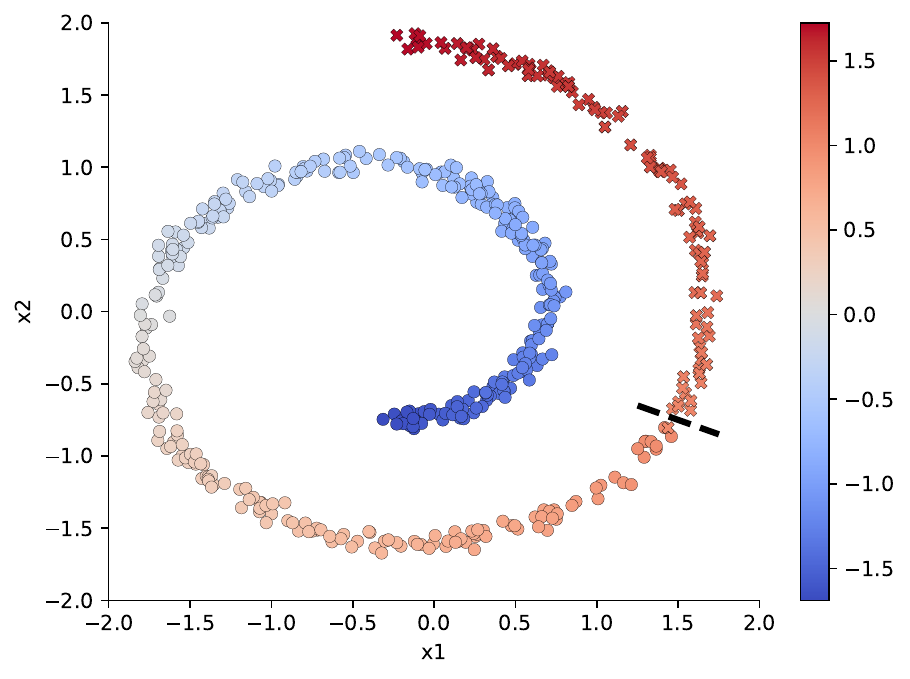}
    \caption{A visualization of the Swiss roll dataset used to train different guidance models. It consists of 500 data points generated with the corresponding function in the \textsc{Scikit-learn} library \citep{scikit-learn} using a noise parameter of 0.3. Both covariates $\Xbf_t\in\R^2$ and labels $\ybf\in\R$ have been normalized to a mean of zero and a standard deviation of one. The color-coded regression labels are derived from the data points' sequential ordering along the manifold's arc length. A label threshold of $\ybf=1$, denoted by a dashed line, shows the training-validation split: the training data (circles) contains data points with lower label values, while the validation data (crosses) consists of data points with higher labels. This split is designed to evaluate the ability of a guided diffusion model trained on a subset of lower-value data to extrapolate to and generate data points from higher-value, out-of-distribution regions.}
    \label{fig:swiss_roll_data}
\end{figure}

\paragraph{Dataset.}In order to illustrate the impact of our regularizer on the learning dynamics of a guidance model, particularly when only trained on a constrained and less diverse subset of the input domain, we conduct a comparison with standard parameter-space regularization schemes, namely explicit $L_2$ regularization. This comparison is performed on the Swiss roll dataset, as described by \citet{marsland2011machine} and implemented in the \textsc{Scikit-learn} library \citep{scikit-learn}. The input data $\Xbf_t\in\R^2$ and regression labels $\ybf\in\R$ are depicted in~\Cref{fig:swiss_roll_data}. A threshold of $\ybf=1$ is used to split the data into training and validation sets. The training set encompasses data points with lower label values, while the validation set consists of data points with higher labels. This split is designed to evaluate the capability of a guided diffusion model trained on a subset of lower-value data to extrapolate to and generate data points from higher-value, out-of-distribution regions.

\paragraph{Diffusion Model Training.}

\begin{figure}[H]
\centering
  \begin{tabular}{c}
    \subfloat[
    Predictions and samples from a context-guided model]{\includegraphics[width=0.8\linewidth]{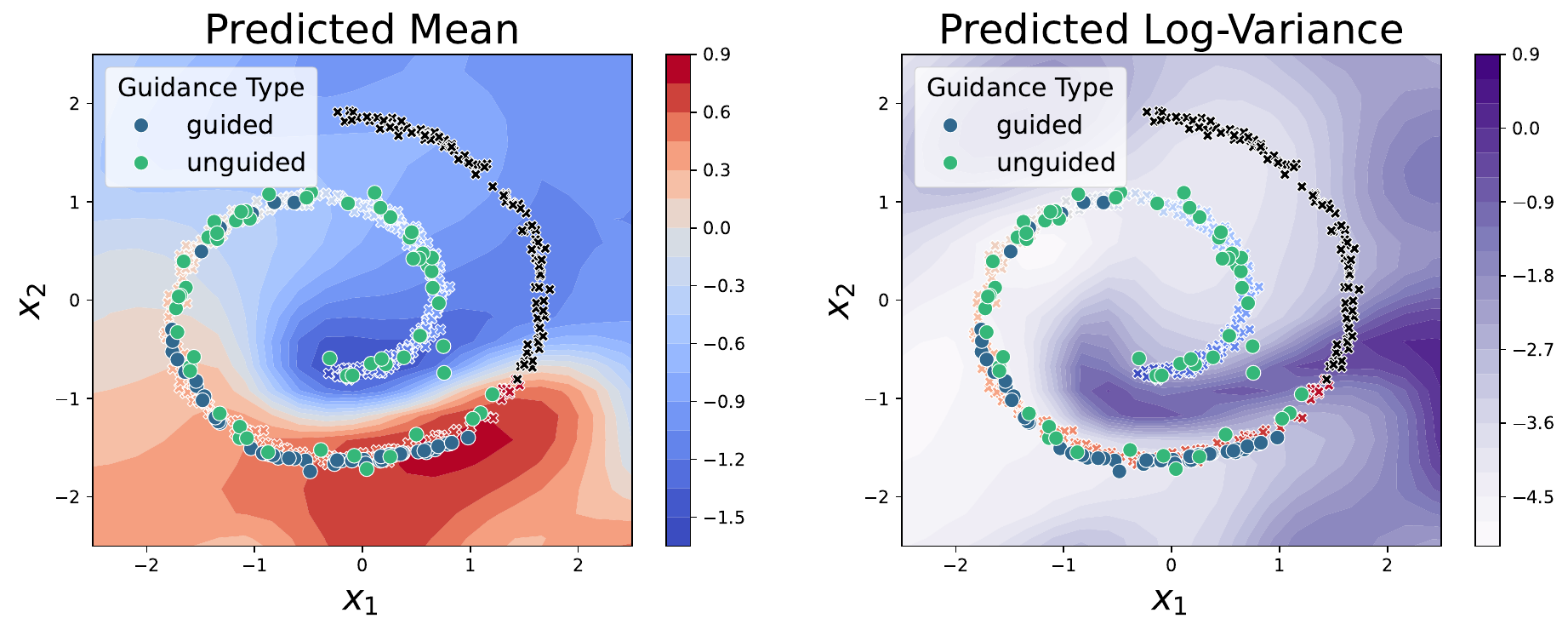}} \\
    \subfloat[Predictions and samples from a context-guided guidance model.]{\includegraphics[width=0.8\linewidth]{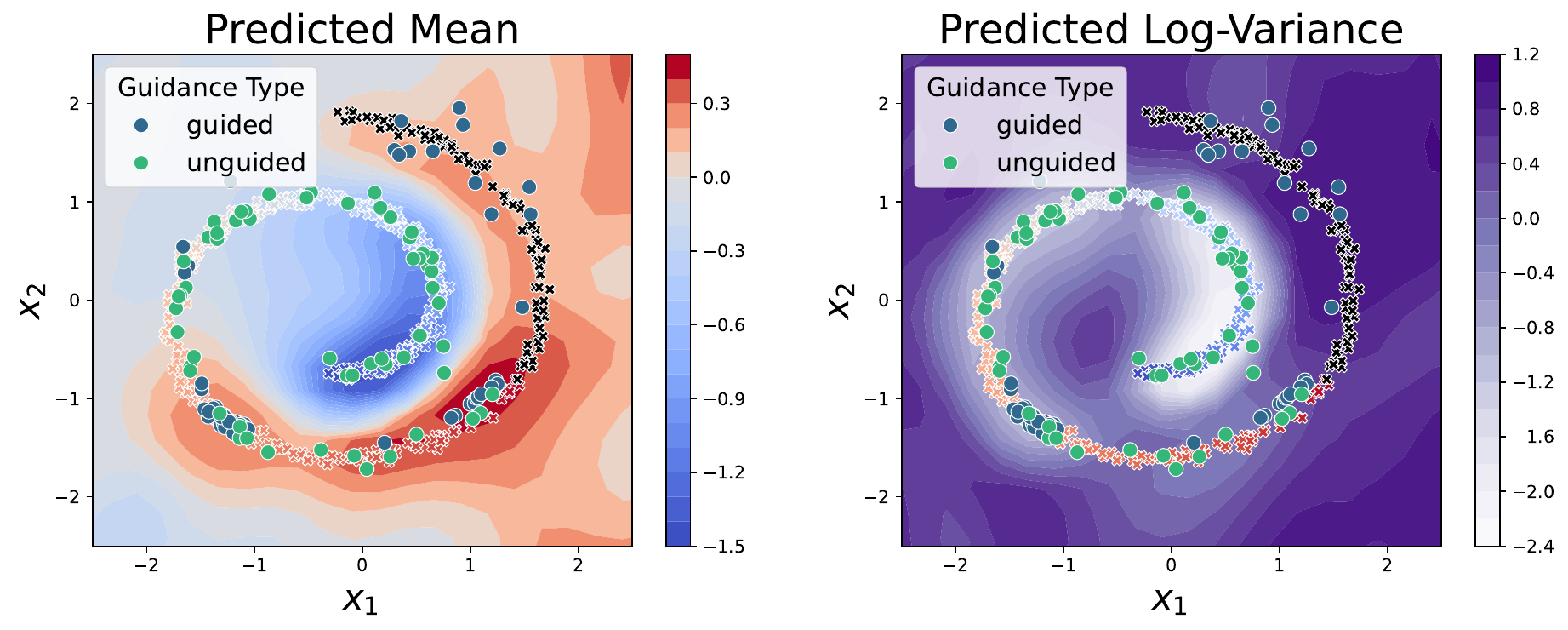}}
  \end{tabular}
\caption{Comparative analysis of predictions, uncertainty estimates, and samples generated by guidance models trained with $L_2$ regularization and context-guidance schemes. The training and validation sets, (see \Cref{fig:swiss_roll_data}), are depicted by color-coded and black crosses, respectively. We display \num{50} samples generated by both unconditional and conditional models, depicted as green and blue circles, respectively. The samples are generated with the same random seed and no cherry-picking. The left panel shows the model's mean predictions across a broader input space \mbox{$[-2.5, 2.5]^2\subset\R^2$}, with warmer colors (red) indicating higher predicted values. The right panel shows the model's log-variance estimates across the same interval, serving as a measure of uncertainty, with darker regions indicating more uncertain predictons. We observe that both $L_2$-regularized and context-guided models fit the training data well. However, without additional information about the input domain, the predictions of the $L_2$-regularized model exhibit poor generalization and miscalibrated confidence estimates when evaluated in new, unseen regions. In contrast, our context-guided approach is able to use this additional information to steer guidance model training toward functions that seem to capture the underlying structure of the input domain more accurately. More importantly, it also generates significantly better-calibrated uncertainty estimates. This improvement translates into higher-quality samples when using gradients from the context-guided model for conditional sampling.}
\label{fig:app_toy_example}
\end{figure}

To train unconditional diffusion models, we adapt an existing implementation of the denoising diffusion probabilistic models from \citet{ho2020denoising} to train diffusion models on datasets in $\R^2$ (\href{https://github.com/albarji/toy-diffusion}{https://github.com/albarji/toy-diffusion}). Specifically, we generate \num{100000} data points from the same data-generating process as our labeled training set, i.e. data points from the Swiss roll generator with standardized labels $\ybf<1$. Using a standard cosine beta schedule \citep{nichol2021improved} 
\begin{equation}
    \beta_t = 1 - \frac{\bar{\alpha}_t}{\bar{\alpha}_{t-1}},\;\text{  with  }\;\bar{\alpha}_t=\frac{f(t)}{f(0)}\;\text{  and  }\;f(t)=\cos \left(\frac{t / T+\epsilon}{1+\epsilon} \cdot \frac{\pi}{2}\right)^2
\end{equation}
where $\epsilon=0.008$ and $T=40$, we ensure that the noising process converges to the target distribution $\calN(\mathbf{0}, \mathbf{I})$. The score network $s_\theta:\R^2\times[1,\ldots,T]\to\R^2$ is constructed as a multi-layer perceptron with \num{5} hidden layers of dimension \num{64} and ReLU activation functions \citep{nair2010rectified}. It is trained with the $L_\text{simple}$ noise prediction objective from \citet{ho2020denoising} over \num{100} epochs of stochastic gradient descent. Specifically, we use the \textsc{Adam} optimizer \citep{kingma2014adam} with a batch size of \num{2048} and a learning rate of \num{1e-3} that drops off to \num{1e-5} following a linear learning rate decay schedule. Samples from this unconditional model are shown in~\Cref{fig:app_toy_example}.

\paragraph{Guidance Function Training.}

For the training of guidance models $f_{t}^{j}(\cdot;\theta):\R^2\times[1,\ldots,T]\to\R$, we use smaller multi-layer perceptrons with \num{3} hidden layers of dimension \num{32}, applying $\mathrm{sin}$ activation functions and dropout \citep{srivastava2014dropout} with $p=0.2$ after each layer. These networks are trained to output both the most likely regression label $\bmu_{t}(\xbf_{t} ; \theta)=f_t^{1}(\xbf_{t} ; \theta)$, as well as log-variances $\log\boldsymbol{\sigma}^2_{t}(\xbf_{t} ; \theta)=f_t^{2}(\xbf_{t} ; \theta)$ that serve as an estimator of their predictive uncertainty. Given the regression labels $\ybf$, the models are optimized with respect to a negative log-likelihood loss
$
\calL_{t} = -\log\calN(\ybf;\bmu_{t},\boldsymbol{\sigma_{t}}^2\mathbf{I})
$
over \num{100} epochs of stochastic gradient descent, using the \textsc{Adam} optimizer with a batch size of $128$ and a learning rate of \num{1e-2}.
The corresponding objective $\calL$ with an explicit $L_2$ regularization term is given by
\begin{equation}
    \calL(\theta, \calD_{\mathcal{T}})
\defines
\E_{p_{\mathcal{T}}}\left[ -\log\calN(\ybf;\bmu_{\mathcal{T}},\boldsymbol{\sigma}^2_{\mathcal{T}}\mathbf{I}) \right]
+ 
\frac{1}{2\lambda} ||\theta||_{2}^{2},
\end{equation}
where the regularization strength $\lambda$ is optimized as a hyperparameter. Similarly, we train context-guided models by adding the additional regularization term $R(\theta, f_{t}, t, p_{\Hat{\Xbf}_{t}})$ as introduced in~\Cref{sec:method} and given by

\begin{equation}
    R(\theta, f_{t}, t, p_{\Hat{\Xbf}_{t}})
    =
    \mathbb{E}_{p_{\Hat{\Xbf}_{t}}} \left[
    \sum\nolimits_{j=1}^{2} \left(f_t^{j}(\Hat{\xbf}_{t} ; \theta) - m_{t}^{j}(\Hat{\xbf}_{t})\right)^\top {K}_{t}(\Hat{\xbf}_{t})^{-1} \left(f_t^{j}(\Hat{\xbf}_{t} ; \theta) - m_{t}^{j}(\Hat{\xbf}_{t})\right)
    \right]
    \hspace*{-2pt} ,
\end{equation}

Here, $\Hat{\xbf}_{t}\sim p_{\Hat{\Xbf}_{t}}(\Hat{\xbf}_{t})$ represents a context batch sampled from an out-of-distribution context set comprising \num{10000} points distributed uniformly over \mbox{$[-2.5, 2.5]^2\subset\R^2$}. As detailed in~\Cref{sec:method}, we use a randomly initialized parameter set $\phi$ to generate \num{32}-dimensional context set embeddings $h_t(\Hat{\xbf}_{t}, \phi)\in\R^{M\times 32}$ and use them to construct the covariance matrix 
$$
{K}_{t}(\Hat{\xbf}_{t}) = \sigma_{t} h_t(\Hat{\xbf}_{t}, \phi)h_t(\Hat{\xbf}_{t}, \phi)^T +\tau_{t}\mathbf{I}.
$$
As the model returns mean and log-variance estimates $f_t^{1}(\Hat{\xbf}_{t} ; \theta)$ and $f_t^{2}(\Hat{\xbf}_{t} ; \theta)$, the target function values $m_{t}^{j}(\Hat{\xbf}_{t})$ are given by the training set mean $m^{1}_{t}(\Hat{\xbf}_{t})=-0.38$ and a log-variance hyperparameter $m^{2}_{t}(\Hat{\xbf}_{t})=\boldsymbol{\sigma}_{0}^{2}$, respectively. The latter is set to $\boldsymbol{\sigma}_{0}=0.7$, chosen to induce high predictive uncertainty estimates of $\exp(\boldsymbol{\sigma}^2)\approx 2$ in out-of-distribution regions of the input domain.
The values of the covariance scale $\sigma_t$ and the diagonal offset $\tau_t$ are optimized as hyperparameters. While empirical performance tends to improve when $\sigma_t$ and $\tau_t$ are made time-dependent, we found that in this setting we can already achieve good performance when keeping them constant.
The resulting augmented training objective is given as
\begin{equation}
    \mathcal{L}^\ast(\theta, \calD)
    =
    \E_{p_{\mathcal{T}}}\left[ \calL(\theta, \calD_{\mathcal{T}}, \mathcal{T}) + R(\theta, f_{t}, \mathcal{T}, p_{\Hat{\Xbf}_{\mathcal{T}}}) \right].\hspace*{-3pt}
\end{equation}

The optimal regularization hyperparameters were selected by performing grid search over the hyperparameter grid presented in~\Cref{tab:hyperparameters_toy_example} with respect to the supervised loss $\calL$ evaluated on the held-out validation set.

\begin{table}[H]
\centering
\caption{Hyperparameter search space for regularization schemes of regression models detailed in~\Cref{app:toy_example}.}
\label{tab:hyperparameters_toy_example}
\begin{tabularx}{\textwidth}{c c X}
\toprule
\textbf{Hyperparameter} & \textbf{Description} & \textbf{Search Space} \\
\midrule
$\lambda$ & $L_{2}$ regularization strength &  \num{e-4}, \num{e-3}, \num{e-2}, \num{e-1}, \num{e0}, \num{e1}, \num{e2}, \num{e3}, \num{e4}
\\
$\sigma_t$ & covariance scale & \num{e-5}, \num{e-3}, \num{e-1}, \num{1e0}, \num{e1}, \num{e3}, \num{e5} \\
$\tau_t$ & diagonal offset & \num{e-5}, \num{e-3}, \num{e-1}, \num{1e0}, \num{e1}, \num{e3}, \num{e5} \\
\bottomrule
\end{tabularx}
\end{table}

\paragraph{Results}

A comparative analysis of model performances is provided in~\Cref{fig:app_toy_example}. 
We observe that both $L_2$-regularized and context-guided models fit the training data well. However, without additional information about the input domain, the predictions of the $L_2$-regularized model exhibit poor generalization and miscalibrated confidence estimates when evaluated in new, unseen regions.
In contrast, the context-guded regularizer is able to use this additional information to steer guidance model training toward functions that seem to capture the underlying structure of the input domain more accurately. More importantly, it also generates significantly better-calibrated uncertainty estimates. This improvement translates into higher-quality samples when using gradients from the context-guided model for conditional sampling.

\subsection{Graph-Structured Diffusion for Small Molecules}
\label{app:small_molecules}

\paragraph{Dataset.}

We evaluate all methods on the same dataset as \citet{lee2023exploring}, consisting of \num{250000} molecules sampled uniformly from the \textsc{ZINC} database of commercially available compounds \citep{irwin2012zinc}. Each molecule is annotated with ground-truth labels $\ybf$, defined in terms of\begin{itemize}
    \item the clipped and normalized negative QuickVina docking score $\ybf_\text{vina}$ \citep{alhossary2015fast},
    \item the normalized synthetic accessibility (\textsc{SA}) score $\ybf_\textsc{SA}$ \citep{ertl2003cheminformatics}, and
    \item the quantitative estimate for drug-likeness (\textsc{QED}) score $\ybf_\textsc{QED}$ \citep{bickerton2012quantifying}.
\end{itemize} 
The resulting composite objective (for which larger values indicate better performance) is given by 
\begin{equation}
\label{eq:exp_mood_obj}
    \ybf = \frac{\operatorname{clip}(-\ybf_\text{vina}, 20, 0)}{20}\cdot \frac{10-\ybf_\textsc{SA}}{9}\cdot\ybf_\textsc{QED}\text{, with }\ybf\in[0,1].
\end{equation}
As the QuickVina docking score $\ybf_\text{vina}$ is a function of the specific protein pocket that a molecule is docked into, all data preparation and model training steps described in this section are repeated across five different proteins from the \textsc{DUD-E} dataset \citep{mysinger2012directory} selected by \citet{lee2023exploring}, namely \textsc{parp1} (Poly [ADP-ribose] polymerase-1), \textsc{fa7} (Coagulation factor VII), \textsc{5ht1b} (5-hydroxytryptamine receptor 1B), \textsc{braf} (Serine/threonine-protein kinase B-raf), and \textsc{jak2} (Tyrosine-protein kinase JAK2). 
All compounds are featurized as topological molecular graphs $\mathbf{g}_0$, defined as tuples $\mathbf{g}_0=(\xbf_0, \mathbf{a}_0)$ of node features $\xbf_0\in\R^{M\times F}$ and adjacency matrices $\mathbf{a}_0\in\R^{M\times M}$, where $M=38$ is the maximum number of heavy atoms in the training set and $F=10$ is the dimensionality of node features, denoting a one-hot encoding of the element types {C}, {N}, {O}, {F}, {P}, {S}, {Cl}, {Br}, {I}, and \textsc{None}.
Different bond types are learned by discretizing the continuous adjacency matric $\mathbf{a}_0$ into bins corresponding to certain bonds, namely $a_{i,j}<0.5\to\text{\textsc{None}}$, $0.5<a_{i,j}<1.5\to\text{single}$, $1.5<a_{i,j}<2.5\to\text{double}$, $a_{i,j}>2.5\to\text{triple}$.
Following \citet{lee2023exploring}, we then order the molecules by their labels $\ybf$ and split them evenly into a low-value training set and a high-value test set of \num{125000} data points each, repeated for each of the five protein targets.
This experimental setup allows us to evaluate the ability of guided diffusion models trained on molecules with low-value properties to extrapolate to and generate data points from higher-value, out-of-distribution regions.

\begin{figure}[H]
    \centering
    \includegraphics[width=0.9\linewidth]{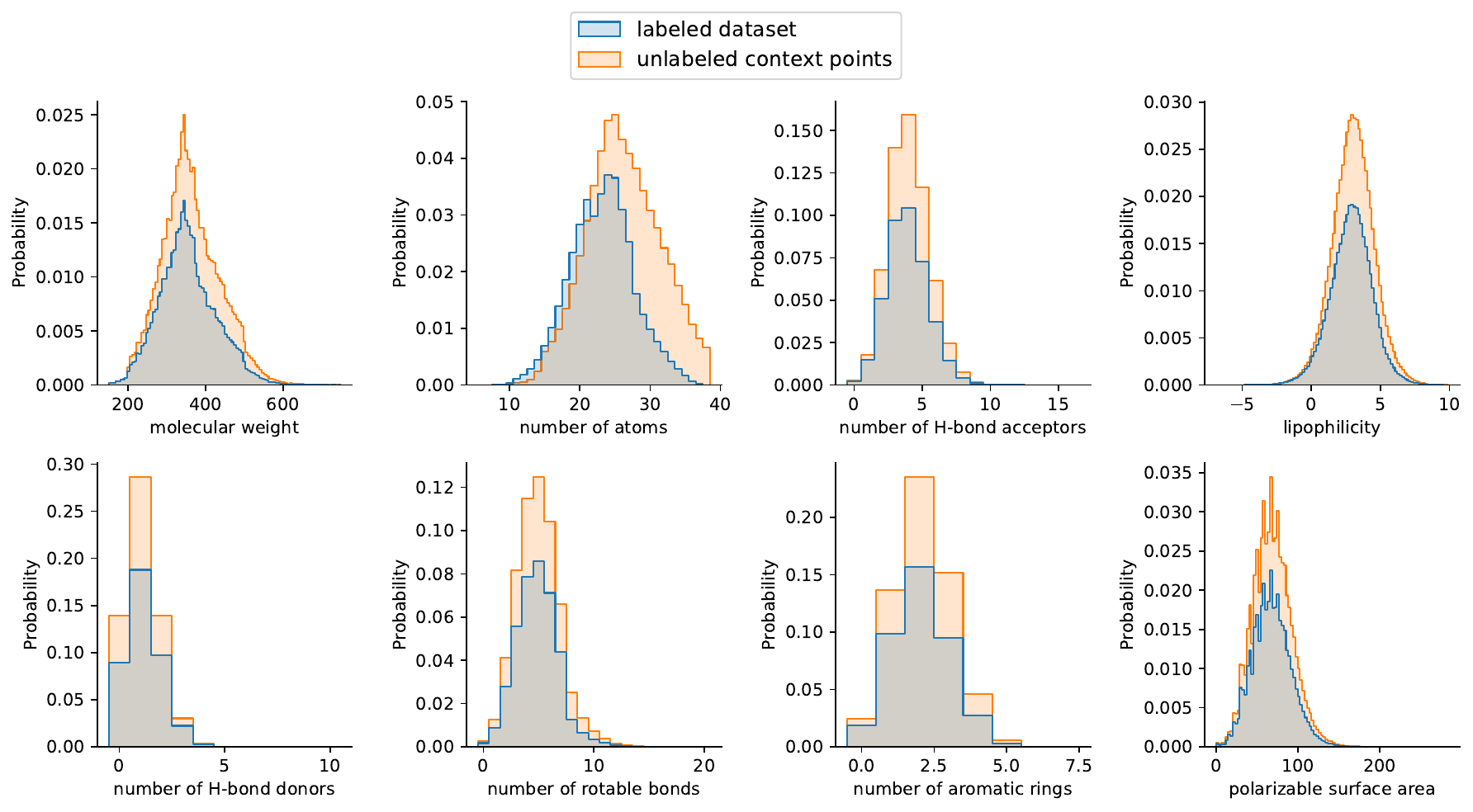}
    \caption{Comparison of the molecular properties of the labeled training and unlabeled context sets, verifying that the latter consists of meaningfully related yet distinct compounds and is suitable for use with our regularization scheme.}
    \label{fig:mood_data_comp}
\end{figure}

To generate a context set $\Hat{\xbf}_\mathcal{C}$ for our context-guided diffusion method, an additional \num{500000} unlabeled compounds were sampled uniformly from the \textsc{ZINC} database and processed in the same fashion as before. A comparison of the training and context set in terms of different molecular properties derived with \textsc{rdkit} \citep{landrum2013rdkit} is presented in~\Cref{fig:mood_data_comp}.

\newpage
\paragraph{Diffusion Models over Graphs.}

The \emph{Molecular Out-Of-distribution Diffusion} (\textsc{MOOD}) framework introduced in \citet{lee2023exploring} leverages diffusion models defined over the space of molecular graphs. Specifically, they employ the \emph{Graph Diffusion via the System of Stochastic differential equations} (\textsc{GDSS}) formalism of \citet{jo2022score}, which we restate in the following. In this framework, a molecular graph $\mathbf{g}_0$ is defined as a tuple $\mathbf{g}_0=(\xbf_0, \mathbf{a}_0)$ of node features $\xbf_0\in\R^{M\times F}$ and adjacency matrices $\mathbf{a}_0\in\R^{M\times M}$, where $M$ is the maximum number of heavy atoms in the training set and $F$ is the dimensionality of the node features. Following the formalism of \citet{song2020score} outlined in~\Cref{sec:background},  the forward diffusion process $\{\mathbf{G}_t=(\Xbf_t, \mathbf{A}_t)\}_{t=0}^T$ is given by
\begin{equation}
    \mathrm{d} \mathbf{G}_t=\mathbf{f}(\mathbf{G}_t, t) \mathrm{~d} t+\beta_{t} \mathrm{~d} \mathbf{B}_t
\end{equation}
and decomposes into a system of stochastic differential equations that can be modeled jointly
\begin{equation}
\begin{aligned}
    \mathrm{d} \mathbf{X}_t&=\mathbf{f}_{\Xbf}(\Xbf_t, t) \mathrm{~d} t+\beta_{\mathbf{X},t} \mathrm{~d} \mathbf{B}_t\\
    \mathrm{d} \mathbf{A}_t&=\mathbf{f}_{\mathbf{A}}(\mathbf{A}_t, t) \mathrm{~d} t+\beta_{\mathbf{A},t} \mathrm{~d} \mathbf{B}_t,
\end{aligned}
\end{equation}
where $\mathbf{f}(\mathbf{G}_t, t)=\left[\mathbf{f}_{\Xbf}(\Xbf_t, t), \mathbf{f}_{\mathbf{A}}(\mathbf{A}_t, t)\right]$ and $\beta_t=\left[\beta_{\mathbf{X},t}, \beta_{\mathbf{A},t}\right]$. The corresponding time reversal $\{\Bar{\mathbf{G}}_t=(\Bar{\Xbf}_t, \Bar{\mathbf{A}}_t)\}_{t\geq0} = \{\mathbf{G}_{T-t}=(\Xbf_{T-t}, \mathbf{A}_{T-t})\}_{t\in[0,T]}$ is given by 
\begin{equation}
\mathrm{d} \Bar{\mathbf{G}}_t=\left\{-\mathbf{f}(\Bar{\mathbf{G}}_t, T-t) + \beta_{T-t}^2\nabla_{\mathbf{G}} \log p_{T-t}\left(\Bar{\mathbf{G}}_t\right) \right\} \mathrm{~d} t+\beta_{T-t} \mathrm{~d} \mathbf{B}_t,
\end{equation}
flowing backward in time and decomposing into a corresponding joint system of \textsc{SDE}s given by
\begin{equation}
\begin{aligned}
\label{eq:app_mood_reverse}
    \mathrm{d} \Bar{\mathbf{X}}_t&=\left\{-\mathbf{f}_{\Xbf}(\Bar{\Xbf}_t, T-t) + \beta_{\mathbf{X},T-t}^2\nabla_{\Xbf} \log p_{T-t}\left(\Bar{\Xbf}_t, \Bar{\mathbf{A}}_t\right) \right\}\mathrm{~d} t+\beta_{\mathbf{X},T-t} \mathrm{~d} \mathbf{B}_t\\
    \mathrm{d} \Bar{\mathbf{A}}_t&=\left\{-\mathbf{f}_{\mathbf{A}}(\Bar{\mathbf{A}}_t, T-t) + \beta_{\mathbf{A},T-t}^2\nabla_{\mathbf{A}} \log p_{T-t}\left(\Bar{\Xbf}_t, \Bar{\mathbf{A}}_t\right) \right\}\mathrm{~d} t+\beta_{\mathbf{A},T-t} \mathrm{~d} \mathbf{B}_t,
\end{aligned}
\end{equation}
where $\nabla_{\Xbf} \log p_t(\cdot)$ and $\nabla_{\mathbf{A}} \log p_t(\cdot)$ that are approximated by separate score networks $s_{\theta_{\Xbf}}(\Xbf_t, \mathbf{A}_t, t)$ and $s_{\theta_{\mathbf{A}}}(\Xbf_t, \mathbf{A}_t, t)$.
To encourage the exploration of novel chemical space, \citet{lee2023exploring} condition the denoising process on a hyperparameter $\gamma\in[0,1]$ that shrinks the unconditional gradients to explore near out-of-distribution regions around the training distribution 
\begin{equation}
\label{eq:app_mood_lambda}
\begin{aligned}
    \nabla_\mathbf{G}\log p_t(\mathbf{G}_t \mid\gamma) &=\nabla_\mathbf{G} \log p_t(\mathbf{G}_t)+\nabla_\mathbf{G}\log p_t(\gamma \mid \boldsymbol{G}_t)\\
    &=\nabla_\mathbf{G} \log p_t(\mathbf{G}_t)+\nabla_\mathbf{G}\log \left(p_t(\mathbf{G}_t)^{-\sqrt{\gamma}}\right) \\
    &= \left(1-\sqrt{\gamma}\right)\nabla_\mathbf{G} \log p_t(\mathbf{G}_t).
\end{aligned}
\end{equation}
In all of our experiments, we follow \citet{lee2023exploring} and set $\sqrt{\gamma}=0.2$. To additionally condition the denoising process on desirable molecular properties, \citet{lee2023exploring} make use of standard guided diffusion as outlined in~\Cref{sec:background}, specified as
\begin{equation}
\begin{aligned}
    \nabla_\mathbf{G}\log p_t(\mathbf{G}_t \mid\gamma,\ybf) &= \nabla_\mathbf{G} \log p_t(\mathbf{G}_t)+\nabla_\mathbf{G}\log p_t(\gamma \mid \boldsymbol{G}_t)+\nabla_\mathbf{G}\log p_t(\ybf\mid \mathbf{G}_t)\\
    &= \left(1-\sqrt{\gamma}\right)\nabla_\mathbf{G} \log p_t(\mathbf{G}_t) + \nabla_\mathbf{G}\log p_t(\ybf\mid \mathbf{G}_t).
\end{aligned}
\end{equation}
Here, a regression model $f_{t}^{j}(\mathbf{g}_t, \theta)$ is used to approximate the conditional distribution $p_t(\ybf\mid \mathbf{g}_t)\approx\operatorname{exp}(\alpha_t f_{t}^{1}(\mathbf{g}_t, \theta))/Z_t$, where $\alpha_t$ is an adaptive guidance scale based on a hyperparameter $r_{\mathbf{G},0}$ and given by
\begin{equation}
\label{eq:app_mood_guidance}
    \alpha_{\mathbf{G},t}=r_{\mathbf{G},t}\frac{\|s_\theta(\mathbf{G}_t, t)\|_2}{\nabla_\mathbf{G}\log \|\beta_\theta(\mathbf{G}_t, t)\|_2}
    \;\text{  with  }\;r_{\mathbf{G},t} = 0.1^t r_{\mathbf{G},0}
\end{equation}
Combining the decomposition into a system of joint \textsc{SDE}s from \Cref{eq:app_mood_reverse}, the shrinking of the unconditional gradients from~\Cref{eq:app_mood_lambda} and the guidance model formulation from~\Cref{eq:app_mood_guidance}, the full conditional sample generation process used by \citet{lee2023exploring} is given by
\begin{equation}
\begin{aligned}
\label{eq:app_mood_conditional}
    \mathrm{d} \Bar{\mathbf{X}}_t
    =
    \beta_{\mathbf{X},T-t}^2&\biggl\{\frac{1}{2}\Bar{\Xbf}_t + \left(1-\sqrt{\gamma}\right) s_{\theta_{\Xbf}}\left(\Bar{\Xbf}_t, \Bar{\mathbf{A}}_t, T-t\right) +
    \nabla_{\Xbf}\log \operatorname{exp} \left(\alpha_{\Xbf, T-t} f_{T-t}\left(\Bar{\Xbf}_t, \Bar{\mathbf{A}}_t; \theta\right) \right)\biggr\}\mathrm{d} t+\beta_{\mathbf{X},T-t} \mathrm{d} \mathbf{B}_t\\
    \mathrm{d} \Bar{\mathbf{A}}_t
    =
    \beta_{\mathbf{A},T-t}^2&\left\{\left(1-\sqrt{\gamma}\right) s_{\theta_{\mathbf{A}}}\left(\Bar{\mathbf{X}}_t, \Bar{\mathbf{A}}_t, T-t\right) + \nabla_{\mathbf{A}}\log \operatorname{exp} \left(\alpha_{\mathbf{A}, T-t} f_{T-t}\left(\Bar{\mathbf{X}}_t, \Bar{\mathbf{A}}_t; \theta\right) \right)\right\}\mathrm{d} t+\beta_{\mathbf{A},T-t} \mathrm{d} \mathbf{B}_t\\
\end{aligned}
\end{equation}
where the drift and diffusion coefficients of $\mathrm{d} \Bar{\mathbf{X}}_t$ and $\mathrm{d} \Bar{\mathbf{A}}_t$ are determined by the variance-preserving (\textsc{VP}) and variance-exploding (\textsc{VE}) \textsc{SDE} from \citet{song2020score}, respectively. 

\paragraph{Diffusion Model Training.}

As in \citet{lee2023exploring}, we adopt the pre-trained node feature and adjacency matrix score networks $s_{\theta_{\Xbf}}(\xbf_t, \mathbf{a}_t, t)$ and $s_{\theta_{\mathbf{A}}}(\xbf_t, \mathbf{a}_t, t)$ from \citet{jo2022score} to define our unconditional diffusion model. These networks are built on graph multi-head attention (\textsc{GMH}) architectures \citep{baek2021accurate} to ensure permutation-equivariance. Each network consists of two graph attention layers with four attention heads of hidden dimension 16 and $\mathrm{tanh}$ activation functions, followed by a multi-layer perceptron output head with three layers, each of dimension 82 and $\mathrm{eLU}$ activation functions \citep{clevert2015fast}.
The node feature \textsc{SDE} is based on the variance-preserving SDE from \citet{song2020score}, using a linear $\beta$-schedule ($\beta_{\mathbf{X},0}=0.1$, $\beta_{\mathbf{X},T}=1$). In contrast, the adjacency matrix \textsc{SDE} uses the variance-exploding SDE also from \citet{song2020score}, with an exponential $\beta$-schedule ($\beta_{\mathbf{A},0}=0.2$, $\beta_{\mathbf{A},T}=1$). Both models employ $T=1000$ discretization steps.
It's important to note that the unconditional score networks from \citep{jo2022score} were trained on an i.i.d 85\%-15\% training-validation split of our dataset, meaning that they may have been optimized using part of the high-property validation set also used for training other guidance models. However, this is not an issue as (a) the unconditional diffusion model was trained without access to label infomration, (b) we only use the validation set for hyperparameter optimization, and (c) the vast search space ensures that most generated compounds are neither in the training nor the validation set (see~\Cref{appsec:similarity} for a detailed analysis).

\paragraph{Guidance Model Training.}

For training the guidance models $f_{t}^{j}(\cdot;\theta)$, we use the graph convolutional neural network (\textsc{GNN}) architecture from \citet{lee2023exploring}, consisting of 3 \textsc{GCN} layers with 16 hidden units and tanh activations \citep{kipf2016semi}.
The embeddings from each layer are concatenated and fed into two fully connected heads with tanh and sigmoid activations, respectively, whose outputs of dimension \num{16} are then multiplied and fed into another two-layer \textsc{MLP} with the same dimension and \textsc{ReLU} activation functions \citep{nair2010rectified}.
We train separate networks for each protein target, predicting both the most likely regression label $\bmu_{t}(\mathbf{g}_{t} ; \theta)=f_t^{1}(\mathbf{g}_{t} ; \theta)$, as well as log-variances $\log\boldsymbol{\sigma}^2_{t}(\mathbf{g}_{t} ; \theta)=f_t^{2}(\mathbf{g}_{t} ; \theta)$ that serve as an estimator of their predictive uncertainty.
Given the regression labels $\ybf$, the models are optimized with respect to a negative log-likelihood loss
$
\calL_{t} = -\log\calN(\ybf;\bmu_{t},\boldsymbol{\sigma_{t}}^2\mathbf{I})
$
Following the protocol of \citet{lee2023exploring}, all models are trained for \num{10} epochs of stochastic gradient descent, using the \textsc{Adam} optimizer with a batch size of $1024$ and a learning rate of \num{1e-3} \citep{kingma2014adam}.
The corresponding objective $\calL$ with an explicit $L_2$ regularization term is given by
\begin{equation}
    \calL(\theta, \calD_{\mathcal{T}})
\defines
\E_{p_{\mathcal{T}}}\left[ -\log\calN(\ybf;\bmu_\mathcal{T},\boldsymbol{\sigma}^2_\mathcal{T}\mathbf{I}) \right]
+ 
\frac{1}{2\lambda} ||\theta||_{2}^{2},
\end{equation}
where the regularization strength $\lambda$ is optimized as a hyperparameter. For weight decay regularization, we instead switch to the \textsc{AdamW} optimizer \citep{loshchilov2017decoupled} and optimize the corresponding hyperparameter.
Similarly, we train context-guided models by adding the additional regularization term $R(\theta, f_{t}, t, p_{\Hat{\Xbf}_{t}})$ as introduced in~\Cref{sec:method}:

\begin{equation}
    R(\theta, f_{t}, t, p_{\hat{\Xbf}_{t}})
    =
    \mathbb{E}_{p_{\hat{\Xbf}_{t}}} \left[
    \sum\nolimits_{j=1}^{2} \left(f_t^{j}(\Hat{\mathbf{g}}_{t} ; \theta) - m_{t}^{j}(\Hat{\mathbf{g}}_{t})\right)^\top {K}_{t}(\Hat{\mathbf{g}}_{t})^{-1} \left(f_t^{j}(\Hat{\mathbf{g}}_{t} ; \theta) - m_{t}^{j}(\Hat{\mathbf{g}}_{t})\right)
    \right]
    \hspace*{-2pt} ,
\end{equation}

Here, $\Hat{\mathbf{g}}_t$ is a context batch sampled from an out-of-distribution context set of \num{500000} unlabeled molecules that were randomly sampled from \textsc{ZINC} and processed identically to the training data.
As described in~\Cref{sec:method}, we use a randomly initialized set of parameters $\phi$ to generate \num{16}-dimensional context set embeddings $h_t(\Hat{\mathbf{g}}_{t}, \phi)\in\R^{M\times 16}$ and use them to construct the covariance matrix 
$$
{K}_{t}(\Hat{\mathbf{g}}_{t}) = \sigma_{t} h_t(\Hat{\mathbf{g}}_{t}, \phi)h_t(\Hat{\mathbf{g}}_{t}, \phi)^T +\tau_{t}\mathbf{I}.
$$

Since the model returns mean and log-variance estimates $f_t^{1}(\Hat{\mathbf{g}}_{t} ; \theta)$ and $f_t^{2}(\Hat{\mathbf{g}}_{t} ; \theta)$, the target function values $m_{t}^{j}(\Hat{\mathbf{g}}_{t})$ are given by the training set mean $m^{1}_{t}(\Hat{\mathbf{g}}_{t})=\operatorname{mean}(\ybf)$ and a log-variance hyperparameter $m^{2}_{t}(\Hat{\mathbf{g}}_{t})=\boldsymbol{\sigma}_{0}^{2}$, respectively. The latter is set to $\boldsymbol{\sigma}_{0}=0.7$, chosen to induce high predictive uncertainty estimates of $\exp(\boldsymbol{\sigma}^2)\approx 2$ in out-of-distribution regions of the input domain.
The resulting training objective is
\begin{equation}
    \mathcal{L}^\ast(\theta, \calD_\mathcal{T})
    =
    \E_{p_{\mathcal{T}}}\left[ \calL(\theta, \calD_{\mathcal{T}}, \mathcal{T}) + R(\theta, f_{\mathcal{T}}, \mathcal{T}, p_{\Hat{\Xbf}_{\mathcal{T}}}) \right].\hspace*{-3pt}
\end{equation}
The covariance scale $\sigma_t$ and diagonal offset $\tau_t$ are optimized as hyperparameters.
To avoid overconfident predictions as context points approach the invariant distribution $\mathcal{N}(\mathbf{0},\mathbf{I})$, we increase $\tau_{t}$ with the same schedule as the noise scales $\beta_t$, starting from $\beta_0=\tau_{t}$ and ending at $\beta_T=10\tau_{t}$.
Similarly, as the distinction between in- and out-of-distribution becomes meaningless at larger noise levels, we decrease $\sigma_{t}$ with an inverted schedule starting from $\beta_0=\sigma_{t}$ and ending at $\beta_T=0.1\sigma_{t}$.

The deep ensembles \citep{lakshminarayanan2017simple} comprise 5 independent models regularized with weight decay, which is optimized as a hyperparameter.
The pre-trained model was optimized with a self-supervised denoising objective, which has been shown to be an effective pre-training method \citet{zaidi2023pretraining}. It was trained on all 500,000 unlabeled molecules from the context set for 50 epochs with the AdamW optimizer \citep{loshchilov2017decoupled}, using a batch size of 1024, a learning rate of $10^{-4}$, and a weight decay of $10^{-4}$. After pre-training, this model was fine-tuned on with weight decay.

All regularization hyperparameters were optimized with respect to the supervised loss $\calL$ on the held-out, high-label test set by performing a grid search over the hyperparameter space presented in~\Cref{tab:hyperparaemters_small_molecules}. The best hyperparameter combination was then selected and used to re-train five independent models using different random seeds.

\begin{table}[H]
\centering
\renewcommand{\arraystretch}{1.2}
\caption{Hyperparameter search space for regularization schemes of regression models detailed in~\Cref{app:small_molecules}.}
\label{tab:hyperparaemters_small_molecules}
\begin{tabularx}{\textwidth}{c c X}
\toprule
\textbf{Description} & \textbf{Hyperparameter} & \textbf{Search Space} \\
\midrule
$L_{2}$ regularization & $\lambda$ &  \num{e-4}, \num{e-3}, \num{e-2}, \num{e-1}, \num{e0}, \num{e1}, \num{e2}, \num{e3}, \num{e4} \\
\midrule
Single model with weight decay & $\lambda$ &  \num{e-8}, \num{e-7}, \num{e-6}, \num{e-5}, \num{e-4}, \num{e-3}, \num{e-2}, \num{e-1}, \num{0.5} \\
\midrule
Fine-tuning with weight decay & $\lambda$ &  \num{e-8}, \num{e-7}, \num{e-6}, \num{e-5}, \num{e-4}, \num{e-3}, \num{e-2}, \num{e-1}, \num{0.5} \\
\midrule
Ensemble with weight decay & $\lambda$ &  \num{e-8}, \num{e-7}, \num{e-6}, \num{e-5}, \num{e-4}, \num{e-3}, \num{e-2}, \num{e-1}, \num{0.5} \\
\midrule
\multirow{3}{*}{Context Guidance} & covariance scale $\sigma_t$ & \num{e-5}, \num{e-3}, \num{e-1}, \num{e0}, \num{e1}, \num{e3} \\
 & diagonal offset $\tau_t$ & \num{e-5}, \num{e-3}, \num{e-1}, \num{e0}, \num{e1}, \num{e3}
 \\
 & number of context points & \num{256}, \num{512} \\
\bottomrule
\end{tabularx}
\renewcommand{\arraystretch}{1}
\end{table}

After retraining five independent guidance models with the best regularization hyperparameters for each of the five different proteins, we use the gradients from these models to condition the denoising process. Following \citet{lee2023exploring}, we set the guidance strength hyperparameters to $r_{\mathbf{X},0}=0.5$ and $r_{\mathbf{G},0}=0$, respectively, and sample \num{3000} molecules each. The \textsc{SA} \citep{ertl2003cheminformatics} and \textsc{QED} \citep{bickerton2012quantifying} scores for these new molecules are derived with \textsc{rdkit} \citep{landrum2013rdkit}, while the docking score is derived with a \textsc{QuickVina} script \citep{alhossary2015fast} prepared by \citet{lee2023exploring}.

\paragraph{Additional Experimental Results.}

\begin{figure}[H]
    \centering
    \includegraphics[width=\textwidth]{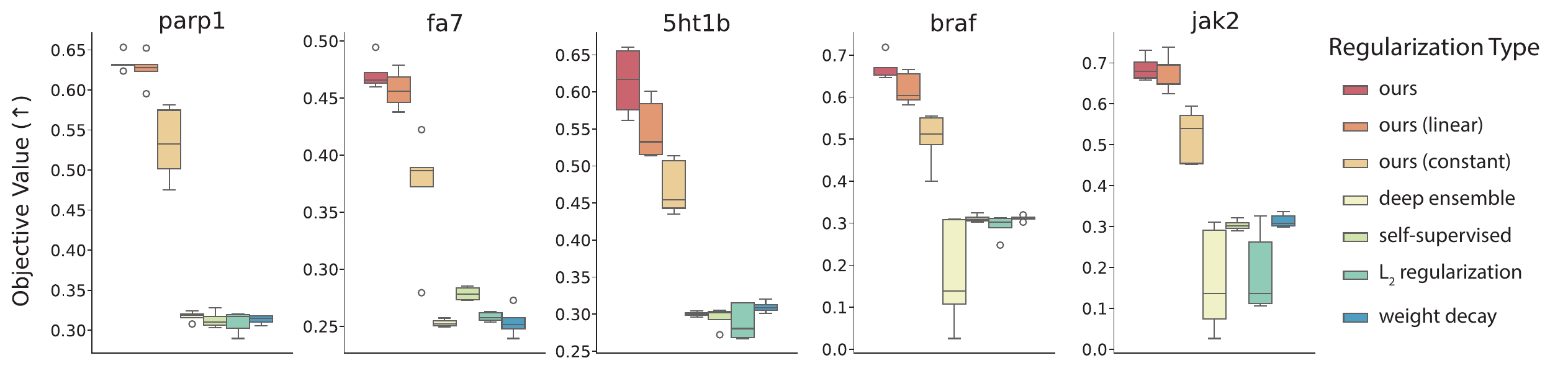}
    \caption{Comparison of the samples generated with different guided diffusion models across five distinct protein targets using the normalized top-5\% docking score of novel and drug-like compounds as in \citet{lee2023exploring}. 
    }
    \label{fig:app_smolecules_main_results}
\end{figure}

\begin{figure}[H]
    \centering
    \includegraphics[width=\textwidth]{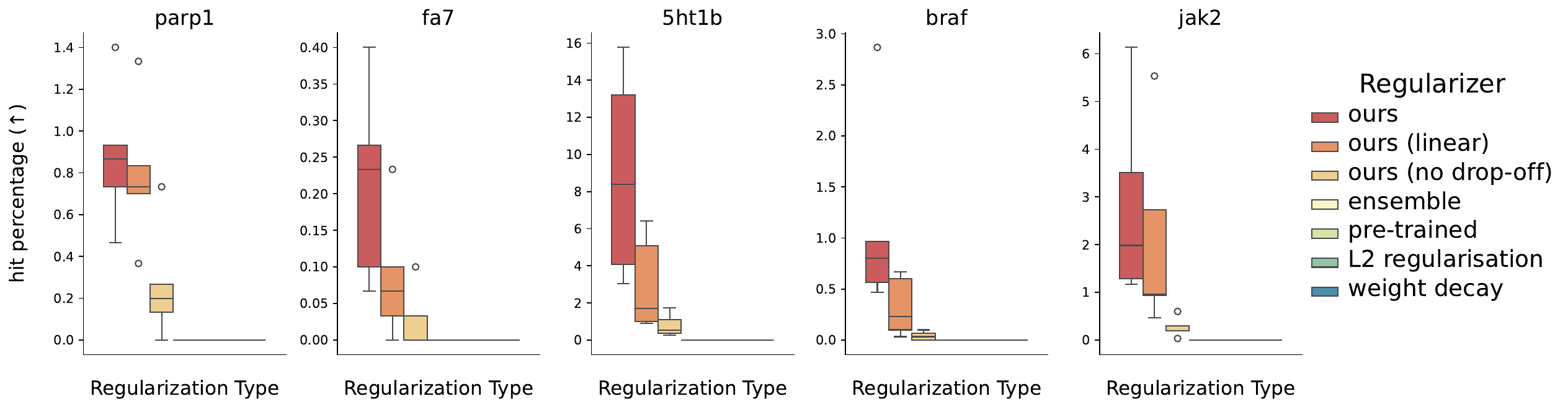}
    \caption{The percentage of small molecules denoted as hits using the thresholds from \citet{lee2023exploring}, given by an unnormalized negative docking score of \num{10} for \textsc{parp1}, \num{8.5} for \textsc{fa7}, \num{8.7845} for \textsc{5ht1b}, \num{9.1} for \textsc{jak2}, and \num{10.3} for \textsc{braf}.
    }
    \label{fig:app_smolecules_hit_percent}
\end{figure}

\begin{figure}[H]
    \centering
    \includegraphics[width=\textwidth]{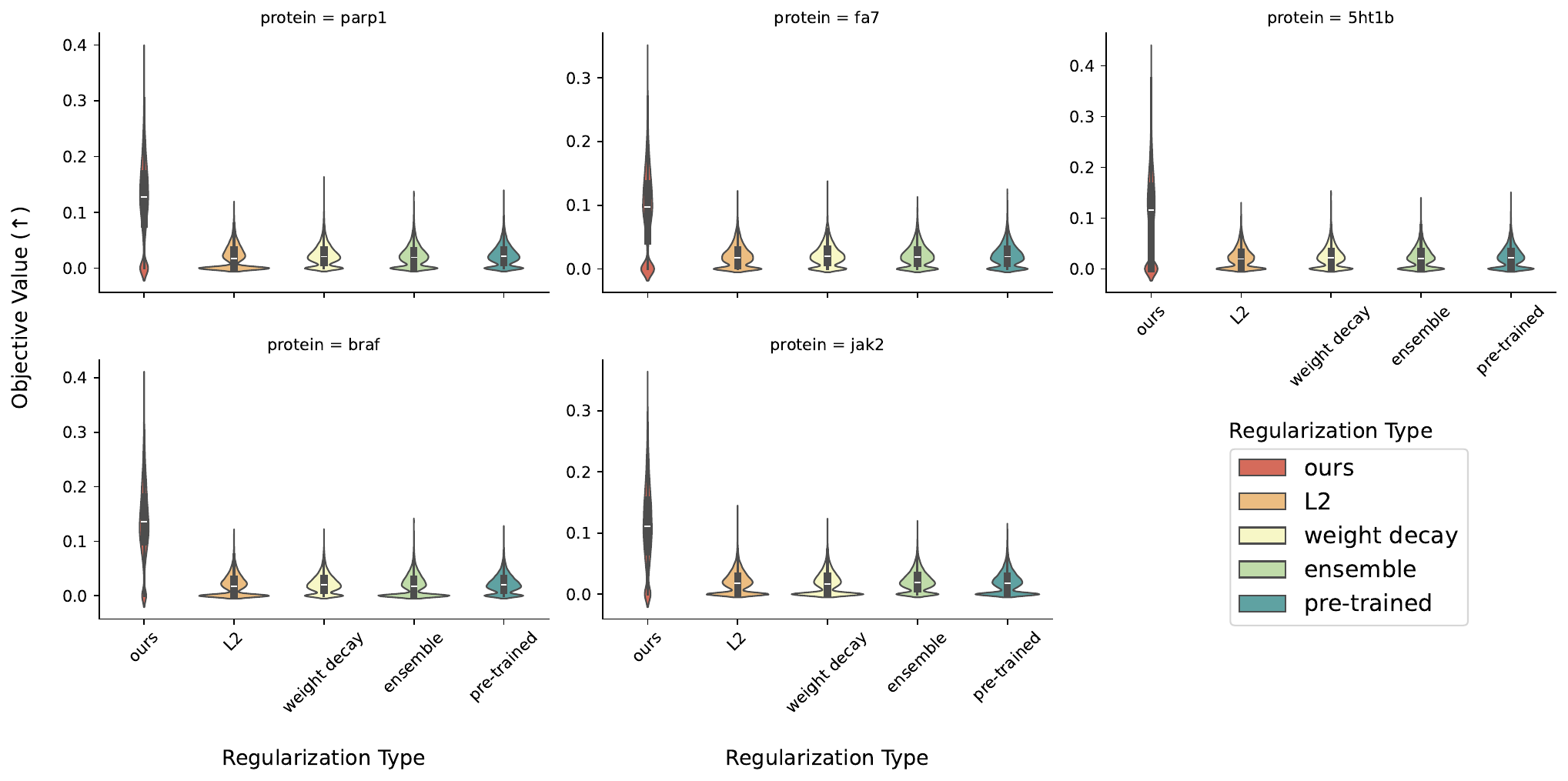}
    \caption{Full distributions of the objective values $\ybf$ of all generated compounds for each protein target across five independent training and sampling runs with different random seeds. These results mirror those in~\Cref{fig:small_molecules_results,fig:app_smolecules_main_results}.}
    \label{fig:app_smolecules_score_full}
\end{figure}

\begin{figure}[H]
    \centering
    \includegraphics[width=\textwidth]{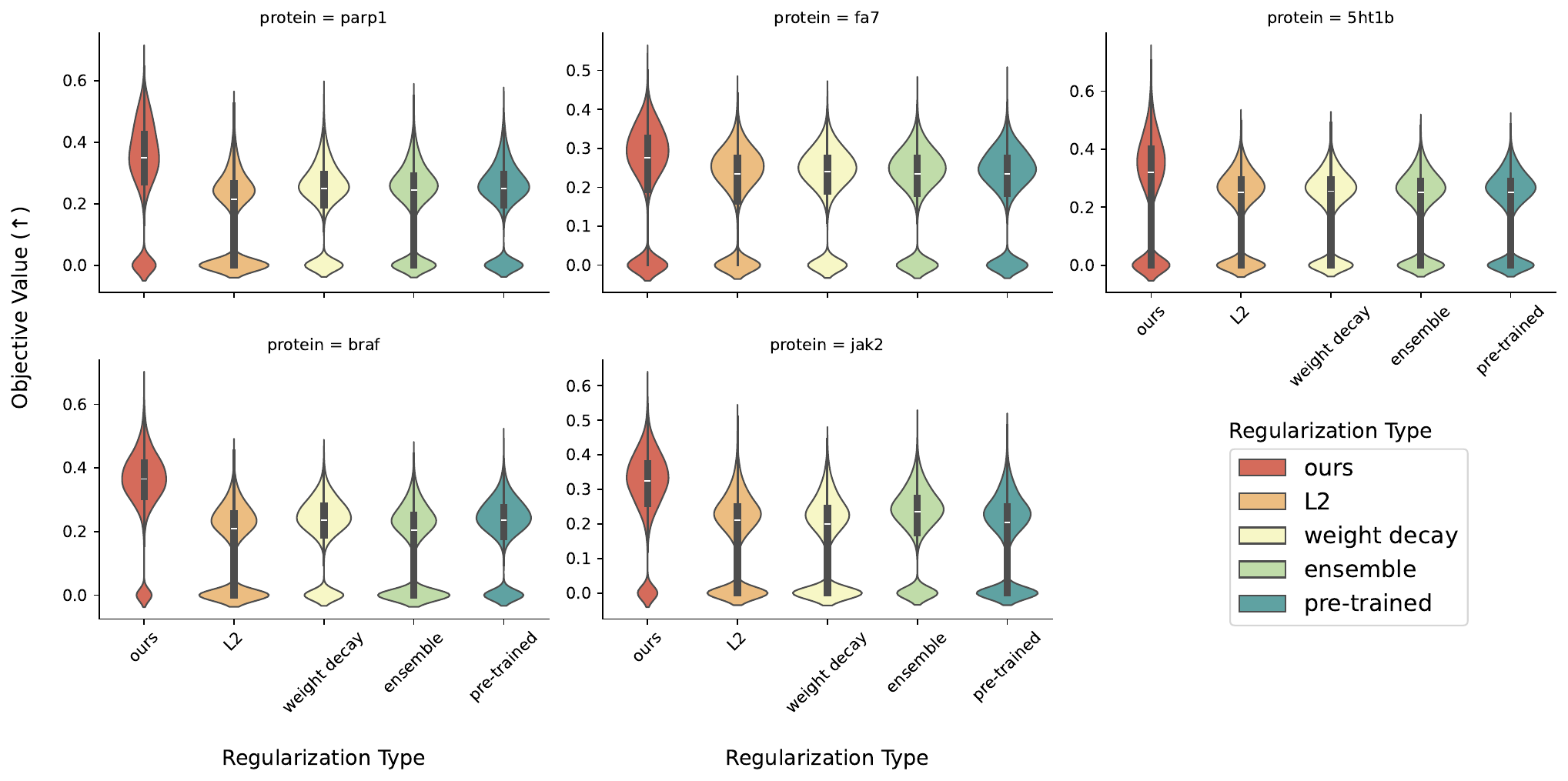}
    \caption{Full distributions of the clipped and normalized negative \textsc{QuickVina} docking score $\y_\text{vina}$ \citep{alhossary2015fast} of all generated compounds for each protein target across five independent training and sampling runs with different random seeds. These results mirror those in~\Cref{fig:small_molecules_results,fig:app_smolecules_main_results}.}
    \label{fig:app_smolecules_protein_score_full}
\end{figure}

\begin{figure}[H]
    \centering
    \includegraphics[width=\textwidth]{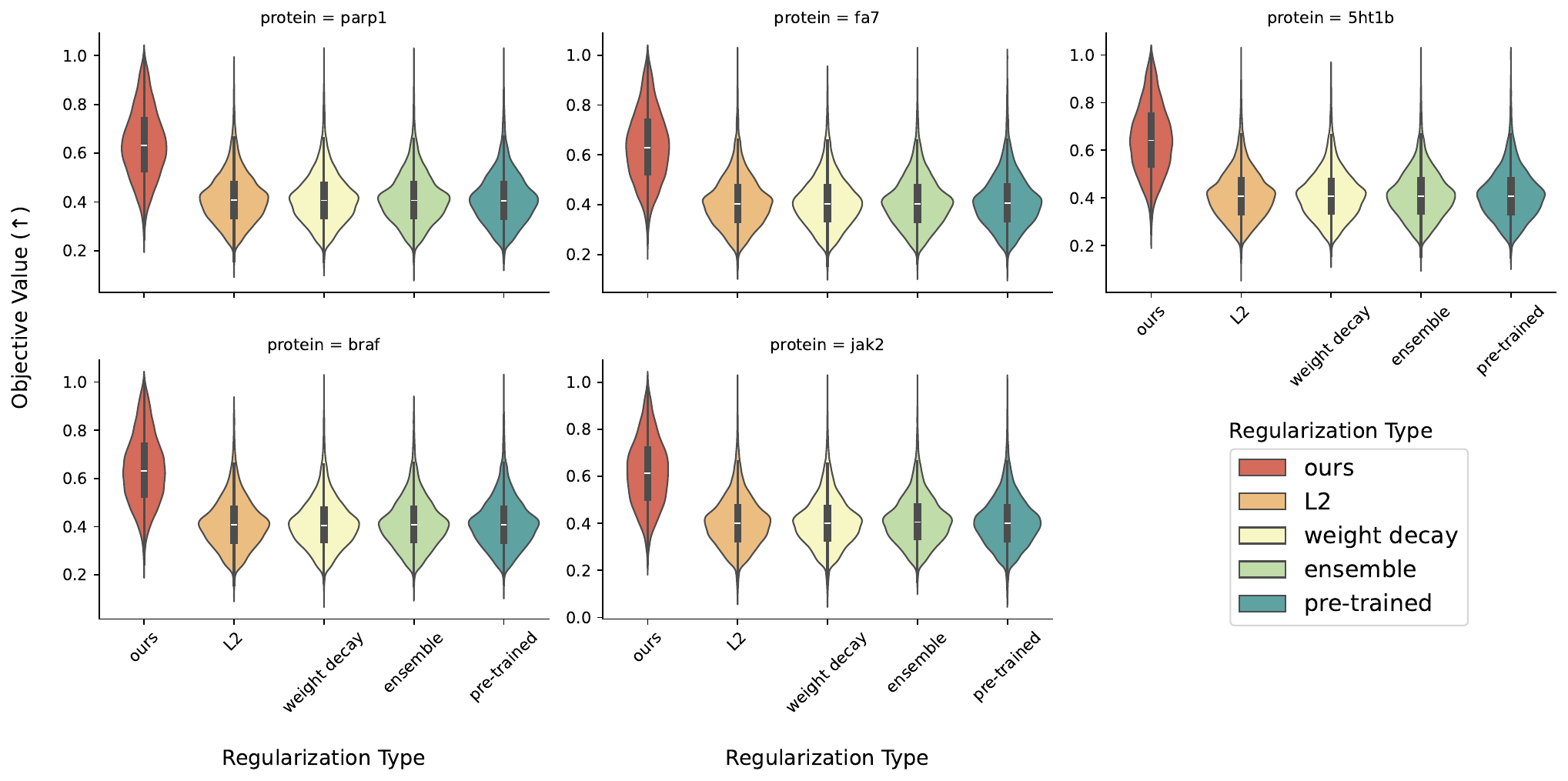}
    \caption{Full distributions of the normalized synthetic accessibility (\textsc{SA}) score $\ybf_\textsc{SA}$ \citep{ertl2003cheminformatics} of all generated compounds for each protein target across five independent training and sampling runs with different random seeds. These results mirror those in~\Cref{fig:small_molecules_results,fig:app_smolecules_main_results}.}
    \label{fig:app_smolecules_sa_score_full}
\end{figure}

\begin{figure}[H]
    \centering
    \includegraphics[width=\textwidth]{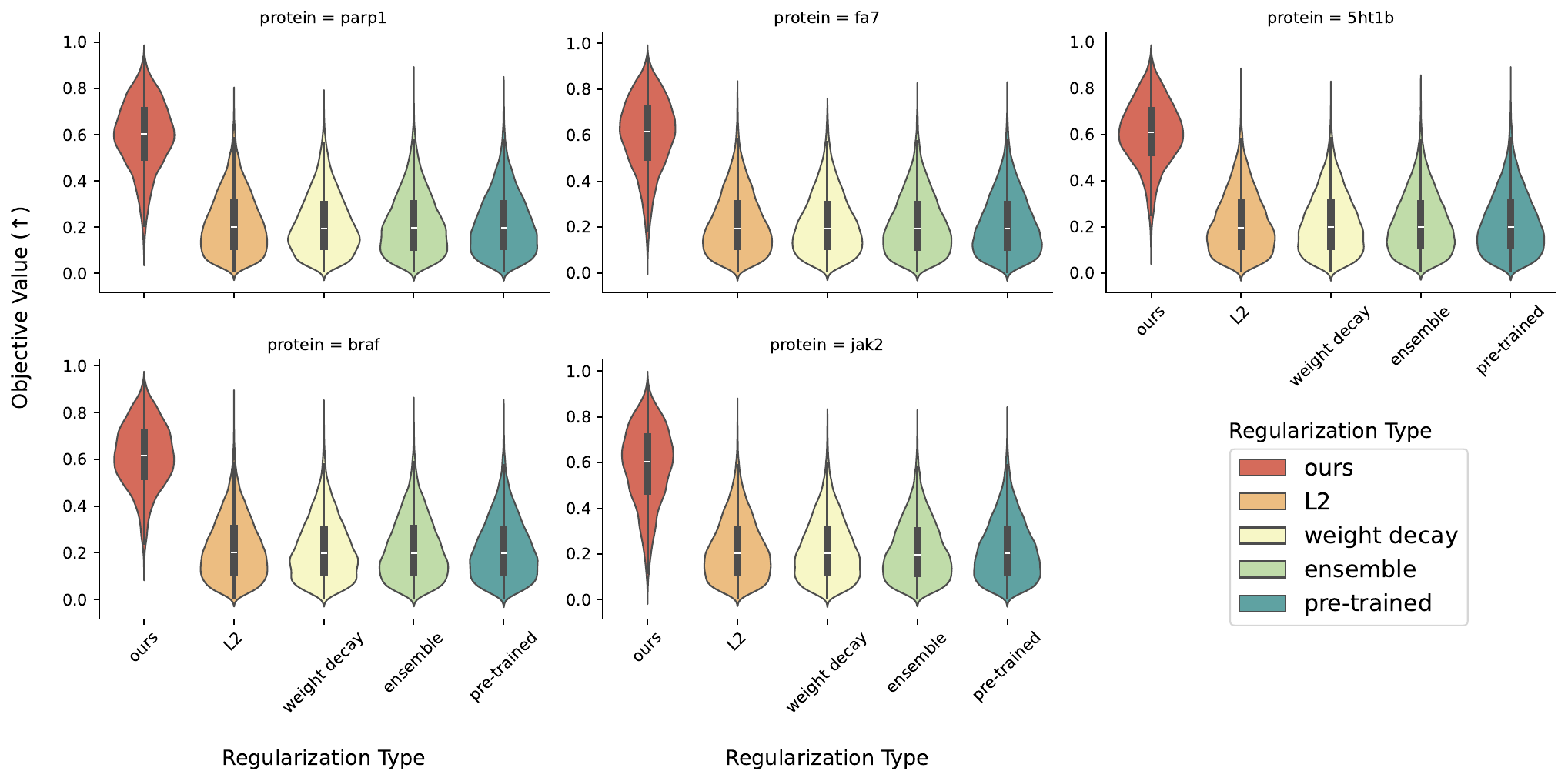}
    \caption{Full distributions of the quantitative estimate for drug-likeness (\textsc{QED}) score $\ybf_\textsc{QED}$ of all generated compounds for each protein target across five independent training and sampling runs with different random seeds. These results mirror those in~\Cref{fig:small_molecules_results,fig:app_smolecules_main_results}.}
    \label{fig:app_smolecules_qed_score_full}
\end{figure}

\begin{figure}[H]
    \centering
    \includegraphics[width=\textwidth]{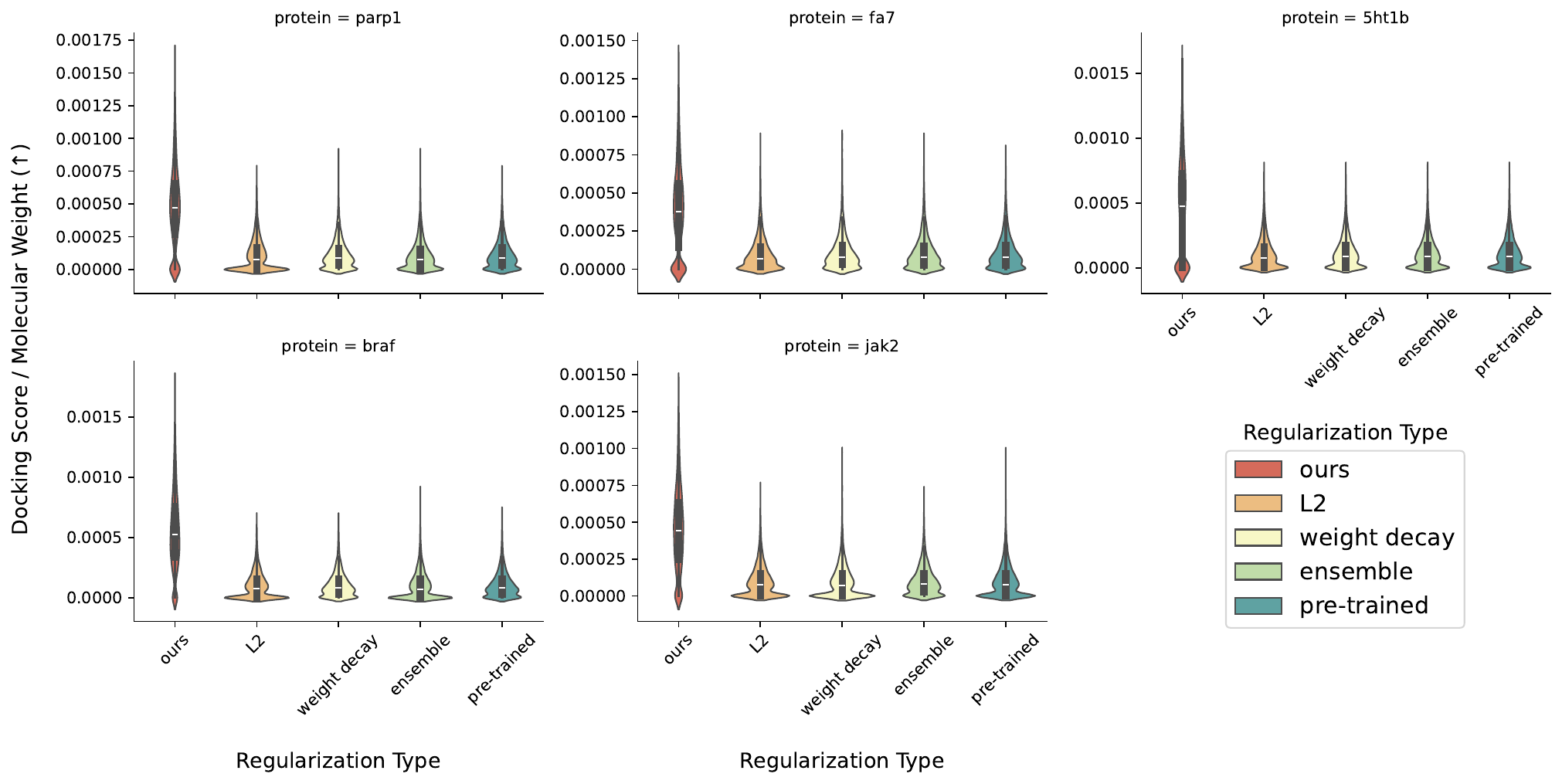}
    \caption{Full distributions of the objective values $\ybf$ of all generated compounds normalized by their molecular weight for each protein target across five independent training and sampling runs with different random seeds. These results mirror those in~\Cref{fig:small_molecules_results,fig:app_smolecules_main_results}.}
    \label{fig:app_smolecules_protein_score_weight}
\end{figure}

\begin{figure}[H]
    \centering
    \includegraphics[width=\textwidth]{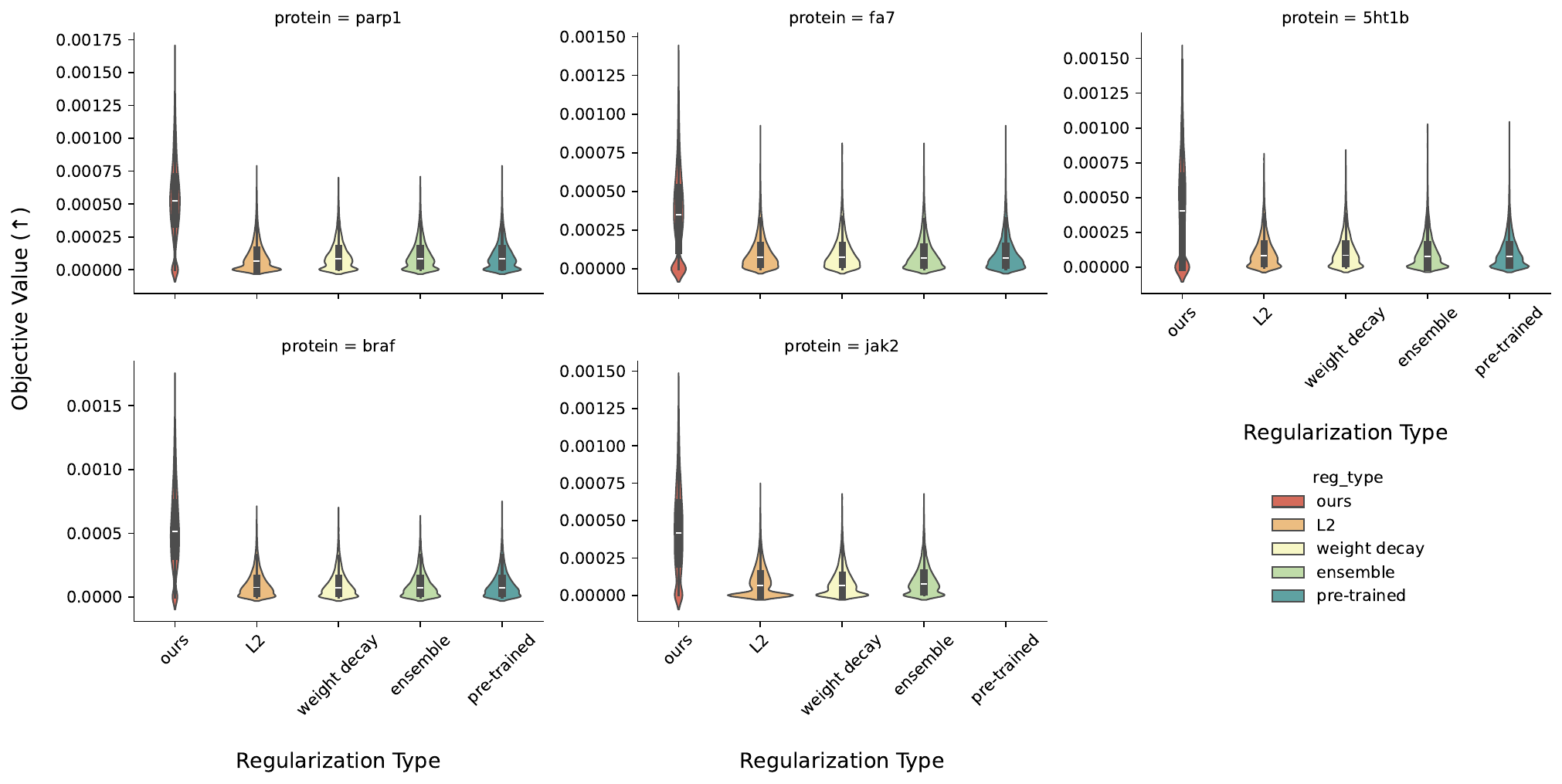}
    \caption{Full distributions of the objective values $\ybf$ of all generated compounds when models are trained on noised labels $\Bar{y}=\ybf \cdot \epsilon$ with added noise $\epsilon\sim\calN(\mathbf{0}, \mathbf{I})$, across five independent training and sampling runs with different random seeds. These results mirror those in~\Cref{fig:small_molecules_results,fig:app_smolecules_main_results}.}
    \label{fig:app_smolecules_protein_score_noised}
\end{figure}

\begin{figure}[H]
    \centering
    \includegraphics[width=\textwidth]{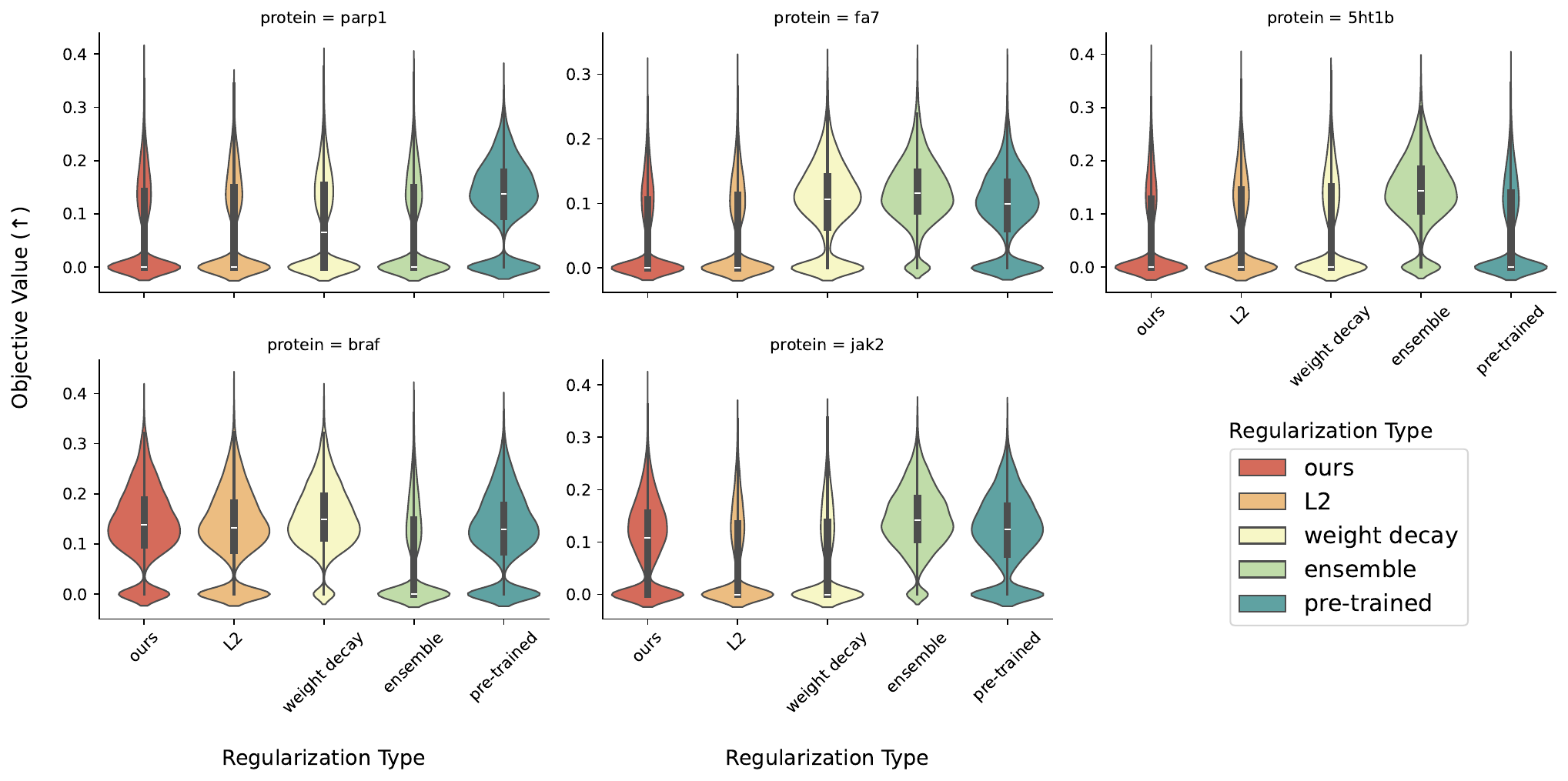}
    \caption{Full distributions of the objective values $\ybf$ of all generated compounds when guidance models that are trained on an 85\%-15\% i.i.d. random split, across five independent training and sampling runs with different random seeds. As expected, the deep ensembles and pre-trained methods perform particularly well in this setting. Interestingly, the performance of our domain-aware guidance method stays approximately the same between the random and label-split evaluation settings.}
    \label{fig:app_smolecules_protein_score_random}
\end{figure}

We repeat the heasline results from~\Cref{sec:exp_small_molecules} in~\Cref{fig:app_smolecules_main_results}; following \citep{lee2023exploring}, we present the top 5\% of normalized docking scores of compounds that are drug-like ($\ybf_\textsc{SA} < 5$ and $\ybf_\textsc{QED} > 0.5$) and novel (Tanimoto similarity of $<0.4$ to the training set \citep{bajusz2015tanimoto}).
We also show the percentage of different compounds that would be classified as hits with the thresholds defined in \citet{lee2023exploring} in~\Cref{fig:app_smolecules_hit_percent}. In addition to the main metrics from \citet{lee2023exploring}, we also present the full distributions of:
\begin{itemize}
    \item The objective $\ybf = \frac{\operatorname{clip}(-\ybf_\text{vina}, 20, 0)}{20}\cdot \frac{10-\ybf_\textsc{SA}}{9}\cdot\ybf_\textsc{QED}$ in~\Cref{fig:app_smolecules_score_full}.
    \item The clipped and normalized negative \textsc{QuickVina} score $\frac{\operatorname{clip}(-\ybf_\text{vina}, 20, 0)}{20}$ in~\Cref{fig:app_smolecules_protein_score_full}.
    \item The normalized synthetic accessibility (\textsc{SA}) score $\frac{10-\ybf_\textsc{SA}}{9}$ in~\Cref{fig:app_smolecules_sa_score_full}.
    \item The quantitative estimate of drug-likeness (\textsc{QED}) score $\mathbf{y}_\textsc{QED}$ in~\Cref{fig:app_smolecules_qed_score_full}.
\end{itemize}
Furthermore, we normalize the scores of the generated compounds by their molecular weight and present the corresponding results in~\Cref{fig:app_smolecules_protein_score_weight}, ensuring the performance differences are not caused by the bias of the docking score toward larger molecules. Finally, we show the same models trained on noisy labels $\Bar{\mathbf{y}} = \mathbf{y} \cdot \epsilon$ where $\epsilon \sim \mathcal{N}(\mathbf{0},\mathbf{I})$ in~\Cref{fig:app_smolecules_protein_score_noised}. Finally, the full distribution of models trained on an 85\%-15\% i.i.d. split is presented in~\Cref{fig:app_smolecules_protein_score_random}. 

\newpage

\paragraph{Comparison to Unsupervised Domain Adaptation Techniques.}

In addition to the models presented above, we also compare our method to guidance functions that are regularized with three robust and well-established unsupervised domain adaptation techniques, namely:
\begin{itemize}
\item \textsc{DeepCoral}: a method that aligns the feature distributions of the last-layer guidance model embeddings across the labeled data and context set through an auxiliary loss term \citep{sun2016deep}.
\item  Domain-Adversarial Neural Networks (\textsc{DANN}): a gradient-reversal approach that aligns the labeled data and context set embeddings by subtracting the gradients of a linear domain classifier from those of the guidance model \citep{ganin2016domain}.
\item \textsc{Domain Confusion}: a method that tries to learn domain-invariant representations by maximizing the predictive entropy of an adversarially trained domain classifier \citep{tzeng2015simultaneous}.
\end{itemize}
We adapted the \textsc{DeepCoral} and \textsc{DANN} implementations from the DrugOOD library \citep{ji2023drugood} and implemented the \textsc{Domain Confusion} models from scratch, closely following the description in \citet{tzeng2015simultaneous}. All models were evaluated with the same training, hyperparameter selection, and sample generation protocols as in~\Cref{app:small_molecules}, using the domain adaptation hyperparameter search spaces presented in~\Cref{tab:hyperparaemters_domain_adaptation}.

\begin{table}[H]
\centering
\renewcommand{\arraystretch}{1.2}
\caption{Hyperparameter search space for the unsupervised domain adaptation algorithms.}
\label{tab:hyperparaemters_domain_adaptation}
\begin{tabularx}{\textwidth}{c c X}
\toprule
\textbf{Method} & \textbf{Hyperparameter} & \textbf{Search Space} \\
\midrule
\textsc{DeepCoral} & loss weight $\lambda$ &  \num{e-2}, \num{5e-2}, \num{e-1}, \num{5e-1}, \num{e0}, \num{5e0}, \num{e1}, \num{5e1}, \num{e2} \\
\midrule
\textsc{DANN} & reversal factor $\alpha$ &  \num{e-2}, \num{5e-3}, \num{e-3}, \num{5e-4}, \num{e-4}, \num{5e-5} \\
\midrule
\textsc{Domain Confusion} & loss weight $\lambda$ &  \num{e-2}, \num{5e-2}, \num{e-1}, \num{5e-1}, \num{e0}, \num{5e0}, \num{e1}, \num{5e1}, \num{e2} \\
\bottomrule
\end{tabularx}
\renewcommand{\arraystretch}{1}
\end{table}

The resulting performance metrics for the first three protein targets from~\citet{lee2023exploring} (\textsc{parp1}, \textsc{fa7}, and \textsc{5ht1b}) are presented in~\Cref{tab:domain_adaptation_results}. While these methods consistently outperform the guidance models trained with standard approaches such as $L_2$ regularization and weight decay, they still perform considerably worse than our approach. One possible explanation for this performance gap is that the \textsc{DeepCoral}, \textsc{DANN}, and \textsc{Domain Confusion} methods only act on the embedding space of a guidance model without considering the behavior of its predictions. In contrast, our context-guided approach directly regularizes the model's predictions, encouraging it to learn functions that revert to an uninformative prior in regions that are far from the training data. This allows our method to better capture the underlying structure of the data and improves its robustness in out-of-distribution settings.

\begin{table}[H]
\centering
\caption{Performance comparison of our method against guidance models regularized with unsupervised domain adaptation techniques (\textsc{DeepCoral}, \textsc{DANN}, and \textsc{Domain Confusion}) and standard regularization approaches (\textsc{Weight Decay}) on the first three protein targets from~\citet{lee2023exploring} (\textsc{parp1}, \textsc{fa7}, and \textsc{5ht1b}). Our context-guided regularization scheme outperforms all domain adaptation techniques while maintaining comparable computational efficiency. Results are reported as the mean and standard deviation across five independent training and sampling runs with different random seeds. The best results for each target are shown in bold.}
\label{tab:domain_adaptation_results}
\renewcommand{\arraystretch}{1.2}
\begin{tabular}{lcccc}
\toprule
Regularizer & Average batch time (s) & \textsc{parp1} ($\uparrow$) & \textsc{fa7} ($\uparrow$) & \textsc{5ht1b} ($\uparrow$) \\
\midrule
Ours & $0.08\pms{0.00}$ & $\mathbf{0.64\pms{0.01}}$ & $\mathbf{0.47\pms{0.00}}$ & $\mathbf{0.60\pms{0.03}}$ \\
\textsc{DeepCoral} & $0.08\pms{0.00}$ & $0.36\pms{0.01}$ & $0.32\pms{0.00}$ & $0.35\pms{0.04}$ \\
\textsc{DANN} & $0.07\pms{0.00}$ & $0.37\pms{0.07}$ & $0.27\pms{0.03}$ & $0.34\pms{0.07}$ \\
\textsc{Domain Confusion} & $0.08\pms{0.00}$ & $0.40\pms{0.01}$ & $0.30\pms{0.01}$ & $0.38\pms{0.01}$ \\
\textsc{Weight Decay} & $0.02\pms{0.00}$ & $0.31\pms{0.00}$ & $0.26\pms{0.01}$ & $0.31\pms{0.01}$ \\
\bottomrule
\end{tabular}
\renewcommand{\arraystretch}{1}
\end{table}

\paragraph{Context Set Sensitivity Studies.}

As the context set provides our method with structured information about a broader subset of the input domain, we found that its size and composition are key factors in determining model performance. To investigate this relationship, we carried out a series of sensitivity studies for the fist three protein targets from \citet{lee2023exploring} (\textsc{parp1}, \textsc{fa7}, and \textsc{5ht1b}). All experiments were performed with the same experimental setup, hyperparameter search space, and evaluation protocol as described in~\Cref{app:small_molecules}, differing only in the choice of context set.
The full results are presented in~\Cref{tab:app_context_set_ablation}

First, we compare the performance of a model trained with the full context set (100\%) to models trained with randomly sampled subsets of size 10\% and 1\%, respectively. Mirroring the results from~\Cref{sec:exp_materials}, we observe a strong deterioration in performance across these settings, with the performance of the models trained with the smallest context set (1\%) reverting to that of models trained with standard weight decay.

To explore whether an even larger and more diverse context set could lead to further performance gains, we trained an additional set of models using the 665k biologically and pharmaceutically relevant molecules from the \textsc{QMUGS} dataset \citep{isert2022qmugs}, originally curated from the \textsc{ChEMBL} database \citep{gaulton2017chembl}. While we observe some performance improvements, they are relatively marginal, indicating that our original context set was likely already at or past the point of diminishing returns regarding its size.

Finally, we performed an additional sensitivity study for which we selected the 10\% of molecules in the original context set that were either most or least similar to the labeled training data (as measured by their maximum ECFP4-based Tanimoto similarity, calculated with rdkit \citep{landrum2013rdkit}). We note that the average maximum Tanimoto similarity between these context sets and the training data is still relatively low, at $0.42\pm0.06$ and $0.31\pm0.04$, respectively. We observe that using a much smaller context set with near-OOD molecules (most similar 10\%) is able to recover the performance of using the full context set (100\%), while relying on less relevant molecules (least similar 10\%) leads to worse performance than a random subsample (10\%).

\begin{table}[H]
\centering
\caption{Impact of the size and composition of the context set on the performance of our proposed regularizer for the small molecules application presented in~\Cref{sec:exp_small_molecules}. We compare the full context set (100\%) to randomly sampled subsets (10\% and 1\%), as well as subsets of the original context set containing the 10\% of molecules that are either most or least similar to the labeled training data. We additionally train models with an even larger and more diverse context set derived from \textsc{QMUGS} \citep{isert2022qmugs}. Results are reported as the mean and standard deviation across five independent training and sampling runs with different random seeds.}
\label{tab:app_context_set_ablation}
\renewcommand{\arraystretch}{1.2}
\begin{tabular}{lccc}
\toprule
Regularizer & parp1 ($\uparrow$) & fa7 ($\uparrow$) & 5ht1b ($\uparrow$) \\
\midrule
ours (100\%) & $0.64\pms{0.01}$ & $0.47\pms{0.00}$ & $0.60\pms{0.03}$ \\
ours (10\%) & $0.54\pms{0.01}$ & $0.45\pms{0.00}$ & $0.44\pms{0.04}$ \\
ours (1\%) & $0.37\pms{0.07}$ & $0.27\pms{0.03}$ & $0.30\pms{0.07}$ \\
ours (QMUGS) & $0.69\pms{0.02}$ & $0.47\pms{0.00}$ & $0.62\pms{0.02}$ \\
ours (most similar 10\%) & $0.66\pms{0.01}$ & $0.45\pms{0.01}$ & $0.59\pms{0.01}$ \\
ours (least similar 10\%) & $0.51\pms{0.02}$ & $0.32\pms{0.04}$ & $0.42\pms{0.01}$ \\
weight decay & $0.31\pms{0.00}$ & $0.26\pms{0.01}$ & $0.31\pms{0.01}$ \\
\bottomrule
\end{tabular}
\renewcommand{\arraystretch}{1}
\end{table}








\newpage
\subsection{Equivariant Diffusion for Materials.}
\label{app:materials}

\paragraph{Dataset.}

We build on the experimental setup from \citet{weiss2023guided} and use a dataset of all 34071 cata-condensed polybenzenoid hydrocarbons (cc-PBHs) with up to 11 benzene rings \citep{wahab2022compas}.
In contrast to the setup in~\Cref{sec:exp_small_molecules} and~\Cref{app:small_molecules}, which uses a subset of the billions of compounds from the \textsc{ZINC} database \citep{tingle2023zinc} to represent the broader chemical space of up to \num{1e60} drug-like small molecules \citep{bohacek1996art,ertl2003cheminformatics}, this application operates in a closed and fully enumerated input space. This allows us to construct data splits that precisely control how much generalization is required to reach held-out regions with optimal objective values.

\begin{figure}[H]
    \centering
    \begin{minipage}[c]{0.49\linewidth}
        \includegraphics[height=7.6cm]{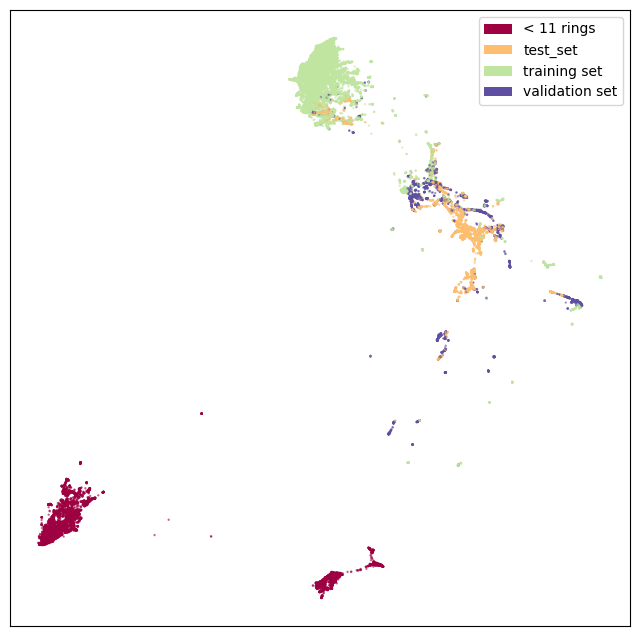}
    \end{minipage}
    \hfill 
    \begin{minipage}[c]{0.49\linewidth}
        \includegraphics[height=7.75cm]{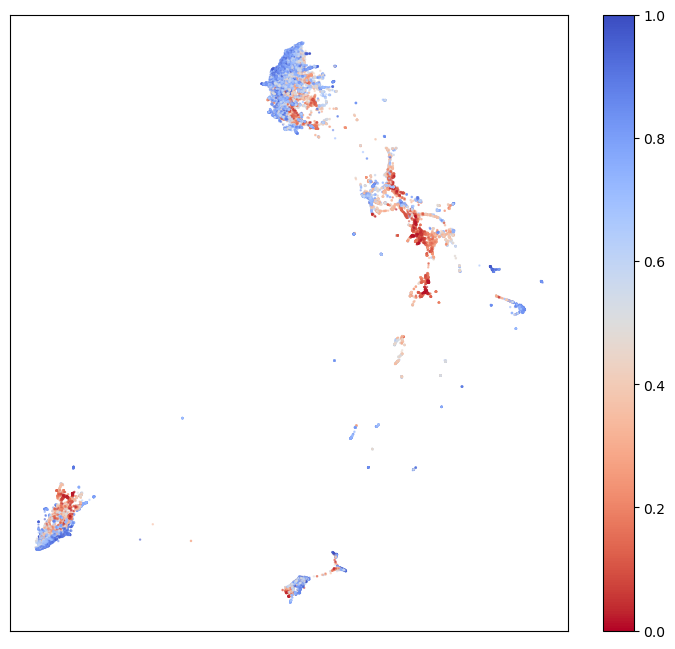}
    \end{minipage}
\caption{\textsc{UMAP} dimensionality reduction plots \citep{mcinnes2018umap} of the cc-PBH dataset using extended reduced graph fingerprints \citep{stiefl2006erg} and the Jaccard-Tanimoto distance. \textbf{Left}: Points are color-coded according to whether they belong to the out-of-distribution context set with fewer than \num{11} rings or the training, validation, and test set. \textbf{Right:} The same points, color-coded according to the composite objective used by \citet{weiss2023guided} and defined as $\ybf = 3\ybf_\textsc{HLG} + \ybf_\textsc{IP} - \ybf_\textsc{EA}$. Lower values indicate more desirable properties and are, for $R=11$, concentrated in the cluster used as a held-out test set.}
\label{fig:app_materials_umaps}
\end{figure}

Starting from the GFN2-xTB optimized ground-state cc-PBH structures, we adopt the graph-of-rings representation from \citet{weiss2023guided}, representing a compound with $R$ rings as a point cloud of ring centroids $\xbf_t \in \mathbb{R}^{R\times 3}$. While computing electronic properties requires 3D representations, much of the combinatorial structure of this data persists in 2D topological graphs. We thus generate extended reduced graph fingerprints \citep{stiefl2006erg} in \textsc{rdkit} (using a fuzziness increment of \num{0} and a maximum path length of \num{50}), which correspond to a 2D equivalent of the 3D graph-of-rings framework. These fingerprints are binarized and compared via the Jaccard-Tanimoto kernel \citep{bajusz2015tanimoto} to measure pairwise similarities. This enables us to cluster the dataset and derive maximally distinct training, validation, test, and context sets (\Cref{fig:app_materials_umaps}). 
Since the framework of \citet{weiss2023guided} only generates compounds with a predetermined number of rings $R$ and $75\%$ of the dataset consists of molecules with $R=11$, we split the data as follows: All compounds with $R<11$ are used as part of an out-of-distribution context set. The remaining compounds are split into 3 groups using spectral clustering (scikit-learn) \citep{scikit-learn}. The largest cluster of \num{16973} compounds is used as the training set. The second largest cluster of \num{4538} compounds is used as the held-out test set. The smallest cluster of \num{3883} compounds is used as the validation set for hyperparameter selection and early stopping. When training models with our context-guided regularization scheme, we use either all molecules with $R<11$ as a \emph{reduced} context set or all compounds not in the training set as a \emph{full} context set

\begin{figure}[H]
\includegraphics[width=\linewidth]{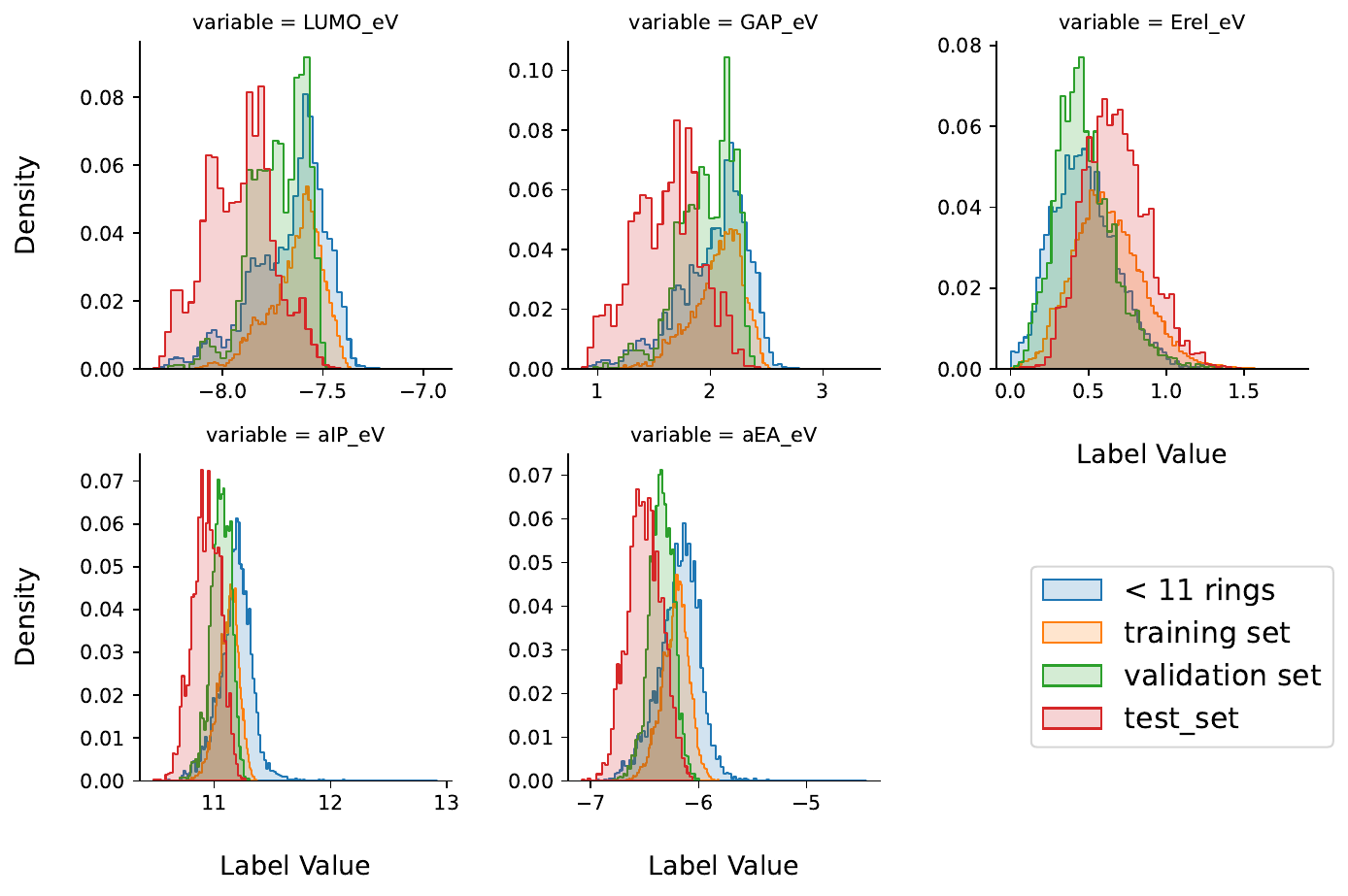}
\caption{The full electronic property distributions for each data split.}
\label{fig:app_materials_data_props}
\end{figure}

Each molecule in the dataset used by \citet{weiss2023guided} is annotated with five electronic properties: LUMO energy ($\ybf_\text{LUMO}$), HOMO-LUMO gap ($\ybf_\text{HLG}$), relative energy ($\ybf_\text{REL}$), adiabatic ionization potential ($\ybf_\text{IP}$), and adiabatic electron affinity ($\ybf_\text{EA}$) \citep{wahab2022compas}. The full property distributions for each data split are shown in~\Cref{fig:app_materials_data_props}. Conveniently, the held-out test set cluster contains compounds with significantly better $\ybf_\text{HLG}$, $\ybf_\text{IP}$, and $\ybf_\text{EA}$ values than the training set, which are the properties used to define the composite objective in \citet{weiss2023guided}, as is also apparent in~\Cref{fig:app_materials_umaps}.

\paragraph{Diffusion Model Training.}

Following \citet{weiss2023guided}, we train an E(3)-equivariant diffusion model (\textsc{EDM}; \citet{hoogeboom2022equivariant}) on the graph-of-rings-derived point clouds $\xbf_t\in\R^{R\times 3}$. Both the forward noising and reverse denoising process outlined in~\Cref{sec:background} are made equivariant to permutations, translations, rotations, and reflections by using an E(3)-equivariant graph neural network (\textsc{E(3)-GNN}) \citep{satorras2021n} as the score model and redefining a translation-invariant noising process in the linear subspace of $\sum_i\xbf_i=0$ \citep{hoogeboom2022equivariant}.

We optimized the resulting equivariant diffusion model on the \num{16973} compounds in the training set using the same hyperparameter settings as in \citet{weiss2023guided}. Specifically, the score network $s_\theta(\xbf_t, t)$ is given by a 9-layer E(3)-GNN with \num{192} hidden units and $\mathrm{tanh}$ activations. It is trained for \num{1000} epochs with the \textsc{Adam} optimizer \citep{kingma2014adam} using a learning rate of \num{1e-3}, a batch size of \num{256}, and a small weight decay hyperparameter of \num{e-12}.
The diffusion process is given by a cosine $\beta_t$ schedule \citep{nichol2021improved} with $T=1000$ time-steps, and $\beta_0=0$ and $\beta_T=1$.
Due to the reduced problem space of the graph-of-rings representation, we find that the vast majority of generated point clouds ($\geq 99\%$, as in \citet{weiss2023guided}) correspond to valid molecules and can be parsed by \textsc{rdkit} \citep{landrum2013rdkit}.

\paragraph{Guidance Model Training.}

Similar to the score network, the guidance model uses a 12-layer \textsc{E(3)-GNN} architecture with \num{192} hidden units and $\mathrm{tanh}$  activations, matching the model hyperparameters from \citet{weiss2023guided}.
These models are trained using the \textsc{Adam} optimizer \citep{kingma2014adam} for up to \num{1000} epochs with a learning rate of \num{6e-4} and batch size of \num{256}, stopping if the validation set performance deteriorates for more than \num{10} consecutive epochs.

As in \citet{weiss2023guided}, we train these models as multi-property prediction networks, outputting both the regression mean $\bmu_{t}(\Hat{\xbf}_{t} ; \theta)=f_t^{1}(\Hat{\xbf}_{t} ; \theta)$ and log-variance $\log\boldsymbol{\sigma}^2_{t}(\Hat{\xbf}_{t} ; \theta)=f_t^{2}(\Hat{\xbf}_{t} ; \theta)$ for all five properties $\ybf_\text{HLG}$, $\ybf_\text{IP}$, $\ybf_\text{LUMO}$, 
$\ybf_\text{REL}$, and $\ybf_\text{EA}$. Given the true regression labels $\mathbf{y}$, the models are trained by minimizing to a negative log-likelihood loss across these five properties
$
\calL_{t} = -\log\calN(\ybf;\bmu_{t},\boldsymbol{\sigma_{t}}^2\mathbf{I}).
$
\newpage
The corresponding objective $\calL$ with an explicit $L_2$-regularization term is given by
\begin{equation}
    \calL(\theta, \calD_{\mathcal{T}})
\defines
\E_{p_{\mathcal{T}}}\left[ -\log\calN(\ybf;\bmu_\mathcal{T},\boldsymbol{\sigma}^2_\mathcal{T}\mathbf{I}) \right]
+ 
\frac{1}{2\lambda} ||\theta||_{2}^{2},
\end{equation}
where the regularization strength $\lambda$ is optimized as a hyperparameter. Similarly, models are regularized with weight decay by switching to the \textsc{AdamW} optimizer \citep{loshchilov2017decoupled} and optimizing the corresponding hyperparameter.
Context-guided models are trained with the regularization term $R(\theta, f_{t}, t, p_{\Hat{\Xbf}_{t}})$ as introduced in~\Cref{sec:method}
\begin{equation}
    R(\theta, f_{t}, t, p_{\Hat{\Xbf}_{t}})
    =
    \mathbb{E}_{p_{\Hat{\Xbf}_{t}}} \left[
    \sum\nolimits_{j=1}^{2} \left(f_t^{j}(\Hat{\xbf}_{t} ; \theta) - m_{t}^{j}(\Hat{\xbf}_{t})\right)^\top {K}_{t}(\Hat{\xbf}_{t})^{-1} \left(f_t^{j}(\Hat{\xbf}_{t} ; \theta) - m_{t}^{j}(\Hat{\xbf}_{t})\right)
    \right]
    \hspace*{-2pt} ,
\end{equation}
where $\Hat{\xbf}_{t}\sim p_{\Hat{\Xbf}_{t}}(\Hat{\xbf}_{t})$ is a context batch sampled from either the \emph{full} context set covering all compounds not in the training set, or the \emph{reduced} context set containing only compounds with fewer than \num{11} rings (see~\Cref{fig:app_materials_umaps}).
As described in~\Cref{sec:method}, we use a randomly initialized parameter set $\phi$ to generate \num{196}-dimensional context set embeddings $h_t(\Hat{\xbf}_{t}, \phi)\in\R^{M\times 196}$ from the \textsc{E(3)-GNN}'s last layer and use them to construct the covariance matrix 
$$
{K}_{t}(\Hat{\xbf}_{t}) = \sigma_{t} h_t(\Hat{\xbf}_{t}, \phi)h_t(\Hat{\xbf}_{t}, \phi)^T +\tau_{t}\mathbf{I}.
$$

Since the model predicts means and log-variances $f_t^{1}(\Hat{\xbf}_{t} ; \theta)$ and $f_t^{2}(\Hat{\xbf}_{t} ; \theta)$, the target function values $m_{t}^{j}(\Hat{\xbf}_{t})$ are given by the training set mean and a log-variance hyperparameter $m^{2}_{t}(\Hat{\xbf}_{t})=\boldsymbol{\sigma}_{0}^{2}$, respectively. The latter is set to $\boldsymbol{\sigma}_{0}=1$, chosen to induce high predictive uncertainty estimates of $\exp(\boldsymbol{\sigma}^2)\approx e$ in out-of-distribution regions of the input domain.
The resulting training objective is
\begin{equation}
    \mathcal{L}^\ast(\theta, \calD)
    =
    \E_{p_{\mathcal{T}}}\left[ \calL(\theta, \calD_{\mathcal{T}}, \mathcal{T}) + R(\theta, f_{\mathcal{T}}, \mathcal{T}, p_{\Hat{\Xbf}_{\mathcal{T}}}) \right].\hspace*{-3pt}
\end{equation}
The covariance scale $\sigma_t$ and diagonal offset $\tau_t$ are optimized as hyperparameters. To avoid overconfident predictions as context points approach the invariant distribution $\mathcal{N}(\mathbf{0},\mathbf{I})$, we increase $\tau_{t}$ with the same schedule as the noise scales $\beta_t$, starting from $\beta_0=\tau_{t}$ and ending at $\beta_T=10\tau_{t}$. Similarly, as the distinction between in- and out-of-distribution becomes meaningless at larger noise levels, we decrease $\sigma_{t}$ with an inverted schedule starting from $\beta_0=\sigma_{t}$ and ending at $\beta_T=0.1\sigma_{t}$.
The regularization hyperparameters are optimized via grid search on the held-out validation set, using the supervised loss $\calL$ as the performance metric. The grid search was over the hyperparameter space shown in ~\Cref{tab:hyperparaemters_materials}. The best hyperparameter combination was selected and used to re-train ten independent models with different random seeds.

\begin{table}[H]
\centering
\renewcommand{\arraystretch}{1.}
\caption{Hyperparameter search space for regularization schemes of regression models detailed in~\Cref{app:materials}.}
\label{tab:hyperparaemters_materials}
\begin{tabularx}{\textwidth}{c c X}
\toprule
\textbf{Description} & \textbf{Hyperparameter} & \textbf{Search Space} \\
\midrule
$L_{2}$ regularization & $\lambda$ &  \num{e-3}, \num{e-2}, \num{e-1}, \num{e0}, \num{e1}, \num{e2}, \num{e-3} \\
\midrule
Weight decay & $\lambda$ &  \num{e-8}, \num{e-7}, \num{e-6}, \num{e-5}, \num{e-4}, \num{e-3}, \num{e-2}, \num{e-1}, \num{0.5} \\
\midrule
\multirow{3}{*}{Context Guidance} & covariance scale $\sigma_t$ & \num{e-2}, \num{e-1}, \num{e0}, \num{e1}, \num{e-2}
 \\
 & diagonal offset $\tau_t$ & \num{e-2}, \num{e-1}, \num{e0}, \num{e1}, \num{e-2} \\
 & number of context points & \num{16}, \num{64}, \num{256}\\
\bottomrule
\end{tabularx}
\renewcommand{\arraystretch}{1}
\vspace{-1.5em}
\end{table}

After retraining ten independent guidance models for each regularization type, their predictions for the HOMO-LUMO gap ($\ybf_\text{HLG}$), adiabatic ionization potential ($\ybf_\text{IP}$), and adiabatic electron affinity ($\ybf_\text{EA}$) properties were used to define the composite objective $\ybf = 3\ybf_\textsc{HLG} + \ybf_\textsc{IP} - \ybf_\textsc{EA}$. The resulting gradients were then used to guide the generation of \num{512} samples each, using an additional guidance scale $s$
$$
\nabla_{\Xbf} \log p_t(\xbf_t\mid\ybf) = \nabla_{\Xbf} \log p_t(\xbf_t) + \nabla_{\Xbf} \log p_t(\ybf\mid\xbf_t)^s = s_\theta(\xbf_t, t) + s\cdot\nabla_{\Xbf} \log f_{t}^{1}(\xbf_t;\theta)
$$

\paragraph{Additional Experimental Results.}

\begin{figure}[H]
    \centering
    \includegraphics[width=\linewidth]{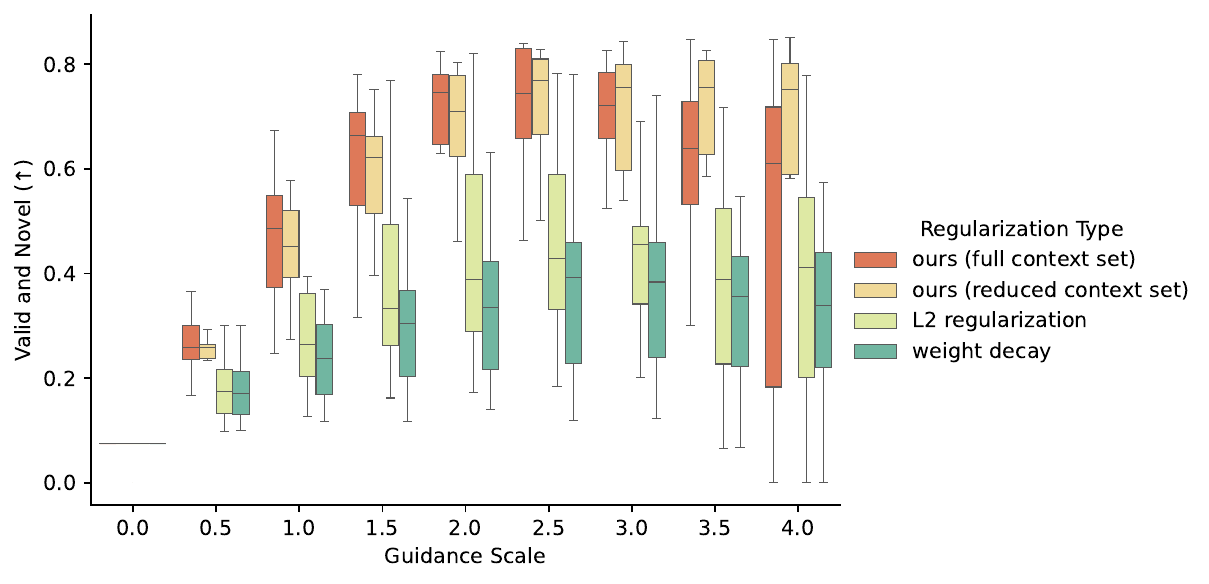}
    \caption{Performance comparison of guidance models with different regularization schemes across ten independent training and sampling runs. The proportion of compounds that are both valid and novel (i.e. not in the training set) is computed for each run and aggregated across random seeds.}
    \label{fig:app_mat_valid_and_novel}
\end{figure}

\begin{figure}[H]
    \centering
    \includegraphics[width=\linewidth]{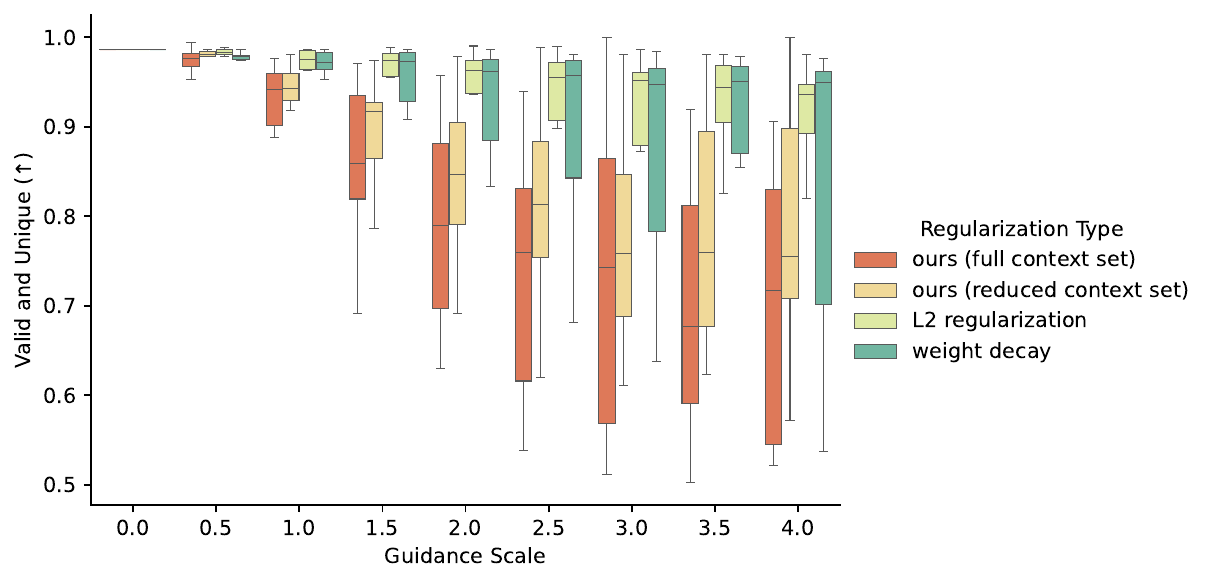}
    \caption{Performance comparison of guidance models with different regularization schemes across ten independent training and sampling runs. The proportion of compounds that are both valid and unique (i.e. only generated once) is computed for each run and aggregated across random seeds.}
    \label{fig:app_mat_valid_and_unique}
\end{figure}

In addition to the objective values in~\Cref{fig:materials_results}, we also calculate the proportion of generated samples that are (1) valid and novel (i.e. not in the training set) in~\Cref{fig:app_mat_valid_and_novel} and (2) valid and unique (i.e. only generated once) in~\Cref{fig:app_mat_valid_and_unique}.
Together with~\Cref{fig:materials_results}, the results suggest that as guidance strength increases, our method generates more novel molecules with better objective values, yet less diversity. This may be explained by stronger guidance signals causing the denoising process to converge to the same high-value subsets of the input domain.

\newpage 

\subsection{Discrete Diffusion for Protein Sequences.}
\label{app:proteins}

\paragraph{Dataset.}

We use the same antibody sequence dataset as \citet{gruver2023protein}, consisting of approximately \num{100000} paired heavy and light chain sequences from the \textsc{OAS} and \textsc{SAbDab} databases \citep{olsen2022observed, dunbar2014sabdab} aligned with \textsc{ANARCI} \citep{dunbar2016anarci} and padded to length \num{300}. 
We focus on the synthetic-label applications presented by \citet{gruver2023protein}, namely the single-objective optimization of a protein's solvent-accessible surface area (\textsc{SASA}) and percentage of beta sheets. In the following experiments we optimize the latter, using code from \citet{gruver2023protein} and the \textsc{BioPython} library \citep{cock2009biopython} to derive the labels $\ybf$.
Similar to~\Cref{app:small_molecules,app:materials}, we split the data into a low-property training set and a high-property validation set to optimize for hyperparameters that perform well in label-shifted, out-of-distribution settings. Additionally, a random subsample of a third of the dataset is split off as a held-out unlabeled context set for our context-guided method, ordering the remaining two-thirds by their labels $\ybf$ and splitting them in half to generate the low-value training and high-value validation sets. 

\paragraph{Diffusion and Guidance Model Training.}

In contrast to ~\Cref{sec:exp_small_molecules,sec:exp_materials}, the guidance model of \citet{gruver2023protein} is not a standalone model, but rather a regression head of the diffusion model's score network, trained jointly with the denoising objective. Specifically, a masked language model \citep{austin2021structured} learns sequence embeddings, from which two MLPs estimate the score and properties. 
Following \citet{gruver2023protein}, we use a \textsc{BERT}-based transformer backbone \citep{bhargava2021generalization,devlin2018bert} with \num{4} layers and \num{512} hidden units. The score head is a single linear layer, while the property prediction head is a feedforward network with one \num{512}-unit hidden layer and $\mathrm{tanh}$ activations.
This model is trained for \num{500} epochs with \textsc{AdamW} \citep{loshchilov2017decoupled}, using a batch size of \num{128} and a learning rate of \num{5e-4} with a \num{10}-step linear warmup schedule.
The noising process is given by a discrete formulation from \citet{gruver2023protein}, using cosine $\beta_t$ schedule \citep{nichol2021improved} with $T=1000$ time-steps, and $\beta_0=0$ and $\beta_T=1$. 

We train the regression head to estimate both the regression mean $\bmu_{t}(\Hat{\xbf}_{t} ; \theta)=f_t^{1}(\Hat{\xbf}_{t} ; \theta)$ and log-variance $\log\boldsymbol{\sigma}^2_{t}(\Hat{\xbf}_{t} ; \theta)=f_t^{2}(\Hat{\xbf}_{t} ; \theta)$. Given the true regression labels $\mathbf{y}$, the property prediction model is optimized with respect to the corresponding negative log-likelihood loss:
$$
\calL = -\log\calN(\ybf;\bmu_{\mathcal{T}},\boldsymbol{\sigma_{\mathcal{T}}}^2\mathbf{I})
$$
Following \citep{gruver2023protein}, we take one gradient step with respect to $\calL$ for every \num{5} gradient steps with respect to the score matching objective, logging the validation loss every \num{5} epochs and loading the best checkpoint for sample generation.
We train $L_2$-regularized models by augmenting the supervised objective $\mathcal{L}$ with an explicit $L_{2}$ penalty
\begin{equation}
    \calL(\theta, \calD_{\mathcal{T}})
\defines
\E_{p_{\mathcal{T}}}\left[ -\log\calN(\ybf;\bmu_\mathcal{T},\boldsymbol{\sigma}^2_\mathcal{T}\mathbf{I}) \right]
+ 
\frac{1}{2\lambda} ||\theta||_{2}^{2},
\end{equation}
where the regularization strength $\lambda$ is optimized as a hyperparameter. Similarly, models are regularized with weight decay by switching to the \textsc{AdamW} optimizer \citep{loshchilov2017decoupled} and optimizing the corresponding hyperparameter.

Context-guided models are trained by adding the additional regularization term $R(\theta, f_{t}, t, p_{\Hat{\Xbf}_{t}})$ given by
\begin{equation}
    R(\theta, f_{t}, t, p_{\Hat{\Xbf}_{t}})
    =
    \mathbb{E}_{p_{\Hat{\Xbf}_{t}}} \left[
    \sum\nolimits_{j=1}^{2} \left(f_t^{j}(\Hat{\xbf}_{t} ; \theta) - m_{t}^{j}(\Hat{\xbf}_{t})\right)^\top {K}_{t}(\Hat{\xbf}_{t})^{-1} \left(f_t^{j}(\Hat{\xbf}_{t} ; \theta) - m_{t}^{j}(\Hat{\xbf}_{t})\right)
    \right]
    \hspace*{-2pt} ,
\end{equation}
where $\Hat{\xbf}_{t}\sim p_{\Hat{\Xbf}_{t}}(\Hat{\xbf}_{t})$ is a context batch sampled from the unlabeled context set split off as detailed before.
As described in~\Cref{sec:method}, we use a randomly initialized parameter set $\phi$ to generate \num{512}-dimensional context set embeddings $h_t(\Hat{\xbf}_{t}, \phi)\in\R^{M\times 512}$ from the last layer of the regression head and use them to construct the covariance matrix 
$$
{K}_{t}(\Hat{\xbf}_{t}) = \sigma_{t} h_t(\Hat{\xbf}_{t}, \phi)h_t(\Hat{\xbf}_{t}, \phi)^T +\tau_{t}\mathbf{I}.
$$
Since the model predicts mean and log-variance estimates $f_t^{1}(\Hat{\xbf}_{t} ; \theta)$ and $f_t^{2}(\Hat{\xbf}_{t} ; \theta)$, the target function values $m_{t}^{j}(\Hat{\xbf}_{t})$ are given by the training set mean and a log-variance hyperparameter $m^{2}_{t}(\Hat{\xbf}_{t})=\boldsymbol{\sigma}_{0}^{2}$, respectively. The latter is set to $\boldsymbol{\sigma}_{0}=1$, chosen to induce high predictive uncertainty estimates of $\exp(\boldsymbol{\sigma}^2)\approx e$ in out-of-distribution regions of the input domain.
The resulting training objective is
\begin{equation}
    \mathcal{L}^\ast(\theta, \calD)
    =
    \E_{p_{\mathcal{T}}}\left[ \calL(\theta, \calD_{\mathcal{T}}, \mathcal{T}) + R(\theta, f_{\mathcal{T}}, \mathcal{T}, p_{\Hat{\Xbf}_{\mathcal{T}}}) \right].\hspace*{-3pt}
\end{equation}
The covariance scale $\sigma_t$ and diagonal offset $\tau_t$ are optimized as hyperparameters. To avoid overconfident predictions as context points approach the invariant distribution $\mathcal{N}(\mathbf{0},\mathbf{I})$, we increase $\tau_{t}$ with the same schedule as the noise scales $\beta_t$, starting from $\beta_0=\tau_{t}$ and ending at $\beta_T=10\tau_{t}$. Similarly, as the distinction between in- and out-of-distribution becomes meaningless at larger noise levels, we decrease $\sigma_{t}$ with an inverted schedule starting from $\beta_0=\sigma_{t}$ and ending at $\beta_T=0.1\sigma_{t}$.


At sampling time, \citet{gruver2023protein} generate sequences using a grid of denoising hyperparameters and examine the resulting Pareto front of objective values versus ``naturalness'' i.e. how likely a sequence is to be synthesizable, estimated by its likelihood under a protein language model \citep{ferruz2022protgpt2}. We use same hyperparameter grid (presented in~\Cref{tab:hyperparaemters_proteins}), generating samples with five independently trained models with different random seeds.

\begin{table}[H]
\centering
\renewcommand{\arraystretch}{1}
\caption{Hyperparameter search space for regularization schemes of regression models detailed in~\Cref{app:proteins}.}
\label{tab:hyperparaemters_proteins}
\begin{tabularx}{\textwidth}{c c X}
\toprule
\textbf{Hyperparameter} & \textbf{Description} & \textbf{Sampling Range} \\
\midrule
$\lambda$ & strength of naturalness regularization &  \num{e1}, \num{e0}, \num{e-1}, \num{e-2}, \num{e-3}\\
\midrule
$\eta$ & step size of Langevin dynamics update &  \num{1.0}\\
\midrule
$K$ & number of steps used to update embedding  &  \num{10}\\
\midrule
guidance layer & which embedding to use for update & first \\
\midrule
return best & return samples with best guidance model predictions &  true \\
\bottomrule
\end{tabularx}
\renewcommand{\arraystretch}{1}
\end{table}

\paragraph{Additional Experimental Results.}

\begin{figure}[H]
    \centering
    \includegraphics[width=0.8\linewidth]{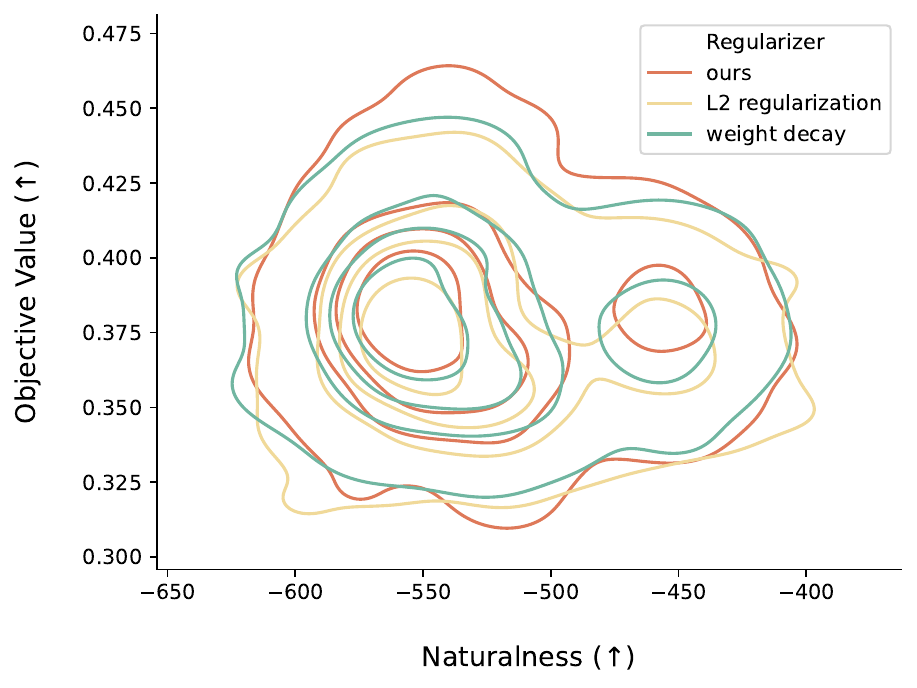}
    \caption{Kernel density estimates (\textsc{KDE}) of all samples generated with different regularization schemes and sampling hyperparameters, using the standard \textsc{KDE} parameters of the \textsc{Seaborn} plotting library \citep{seaborn} and five density isocontours each. 
    Similar to the Pareto front, this highlights the trade-off between objective value and naturalness. As samples move away from the training data and enter an out-of-distribution regime, our method consistently generates sequences with better properties at the same level of naturalness.}
    \label{fig:app_proteins_kde}
\end{figure}

In addition to the Pareto front of the generated samples shown in~\Cref{fig:exp_proteins}, we follow \citet{gruver2023protein} and also visualize the full distribution of generated samples by plotting their kernel density estimates. The results are presented in~\Cref{fig:app_proteins_kde}, mirroring the conclusions drawn from the analysis of the Pareto front.

\subsection{Runtime Comparison.}
\label{app:runtime}

\begin{figure}[H]
    \centering
    \includegraphics[width=0.85\linewidth]{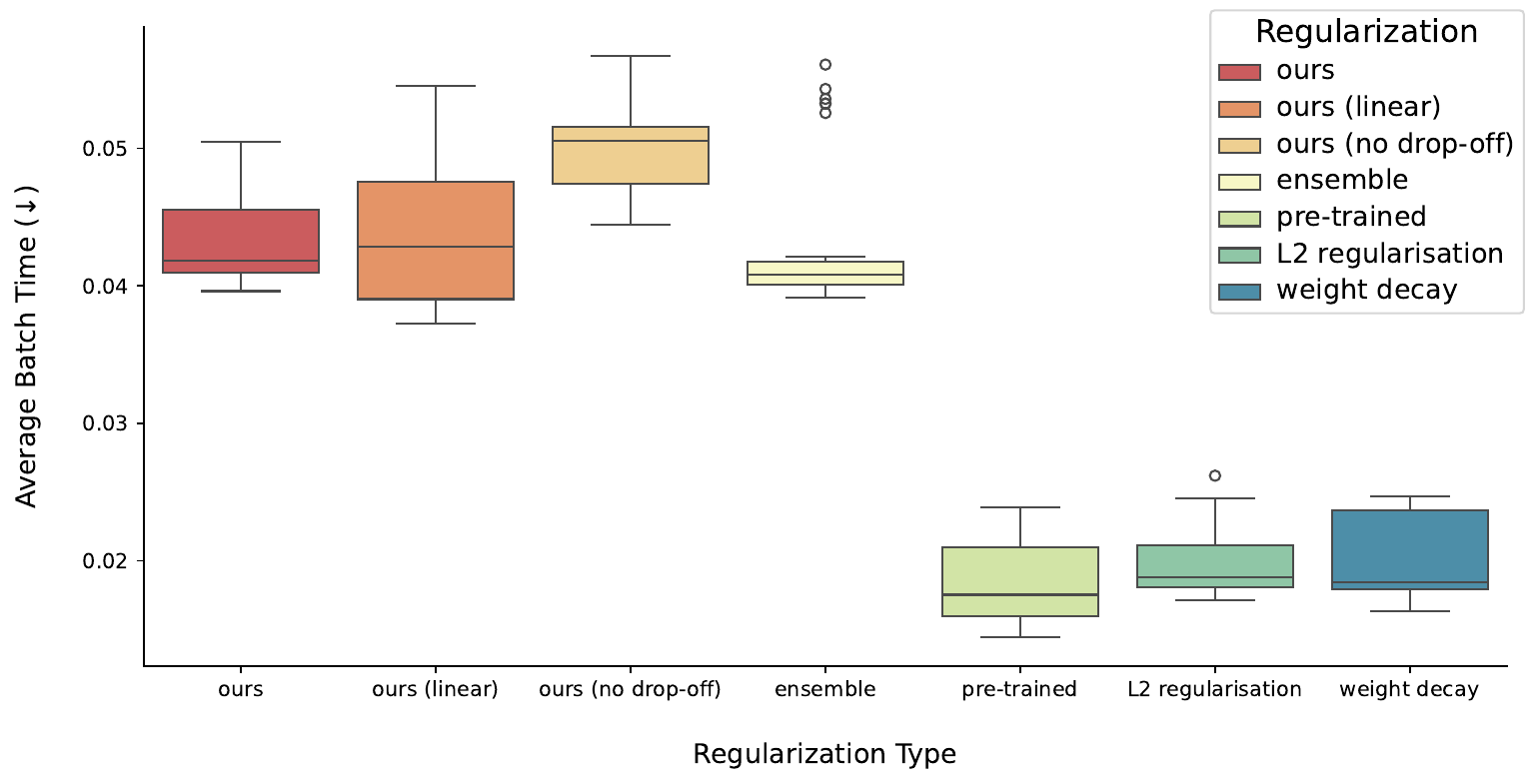}
    \caption{Average batch times of models from~\Cref{sec:exp_small_molecules} trained with an identical setup and on the same \textsc{GPU}s.}
    \label{fig:app_timing_molecules}
\end{figure}

\begin{figure}[H]
    \centering
    \includegraphics[width=0.46\linewidth]{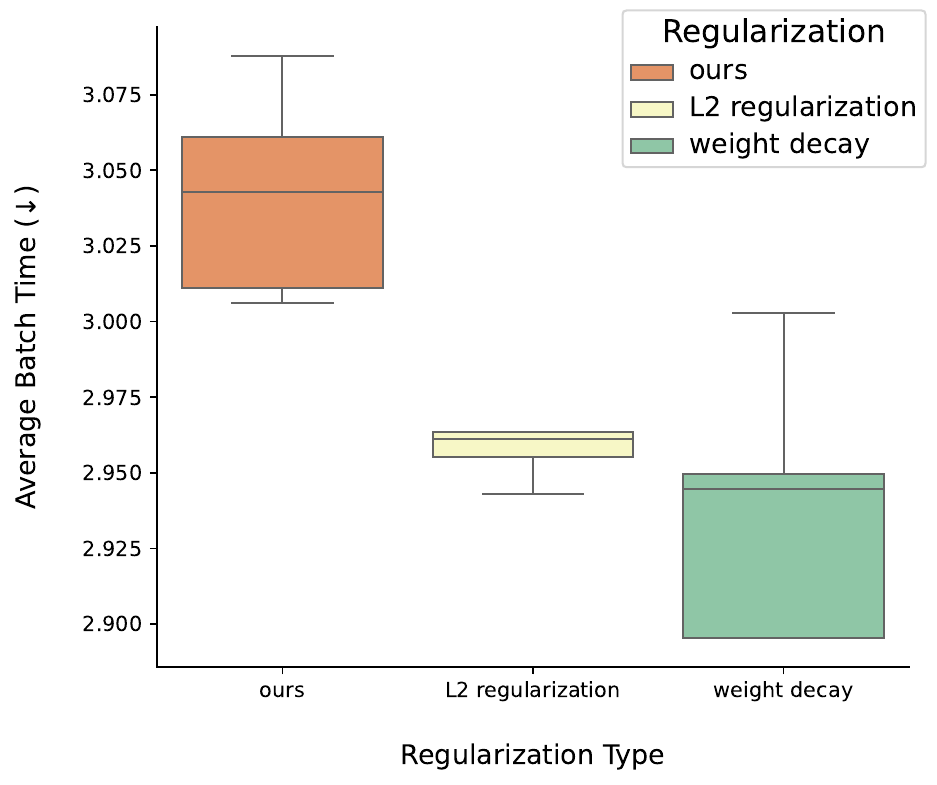}
    \caption{Average batch times of models from~\Cref{sec:exp_proteins} trained with an identical setup and on the same \textsc{GPU}s.}
    \label{fig:app_timing_proteins}
\end{figure}

We quantify computational cost as the average execution time per mini-batch iteration, including data loading, model forward passes, backpropagation, and parameter updates. \Cref{fig:app_timing_molecules} compares regularization schemes for the application in~\Cref{sec:exp_small_molecules} across 5 model training runs. \Cref{fig:app_timing_proteins} shows the same comparison for the experiments in~\Cref{sec:exp_proteins}. 
All models were trained with an identical setup and on the same \textsc{NVIDIA A100 GPU}s, differing only in the regularizer.

When training a standard guidance model, as in \citet{lee2023exploring}, we observe that the added computational cost at training time is roughly equivalent to training a deep ensemble with $M=5$ models (\cref{fig:app_timing_molecules}). However, we note that our method does not induce any additional cost at inference, i.e., sampling time. 
When training the score and guidance models jointly and only updating the latter every \num{5} steps, as in \citet{gruver2023protein}, we observe that the increase in computational cost becomes negligible (\Cref{fig:app_timing_proteins}).

\subsection{Similarity of the Generated Samples to the Training and Validation Sets.}
\label{appsec:similarity}

As our experimental setups select regularization hyperparameters based on an out-of-distribution validation set with more desirable properties than the training data, it is important to verify that the models do not simply generate molecules from the high-value validation set. 
The results are presented below and demonstrate that guidance model trained with our regularizer do not simple generate molecules from the validation set, but are instead learning to sample molecules with improved properties from novel subsets of the input domain.

\subsubsection{Graph-Structured Diffusion for Small Molecules}

To investigate this, we computed the maximum ECFP4-based \citep{rogers2010extended} Jaccard-Tanimoto similarity of the molecules generated by our model in \cref{sec:exp_small_molecules} to any compound in the training and validation sets, respectively. The results are provided in~\cref{tab:app_sim_val_small_molecules} and show that the similarities are consistently low and comparable across the training and validation sets (a threshold of 0.7 is often used to indicate molecules as closely related).

\begin{table}[H]
\centering
\renewcommand{\arraystretch}{1.2}
\caption{Maximum extended-connectivity fingerprint (ECFP4; \citet{rogers2010extended})-based Tanimoto similarity~\citep{bajusz2015tanimoto} of the small molecules generated in \cref{sec:exp_small_molecules} to compounds in the training and validation sets. Means and standard deviations are reported across five independent training and sampling runs with different random seeds.}
\label{tab:small_mol_similarity}
\label{tab:app_sim_val_small_molecules}
\begin{tabular}{lccccc}
\toprule
Set & \textsc{parp1} & \textsc{fa7} & \textsc{5ht1b} & \textsc{braf} & \textsc{jak2} \\
\midrule
Training Set & $0.33\pm0.09$ & $0.35\pm0.08$ & $0.35\pm0.08$ & $0.37\pm0.07$ & $0.36\pm0.08$ \\
\hline
Validation Set & $0.33\pm0.09$ & $0.36\pm0.09$ & $0.37\pm0.09$ & $0.39\pm0.10$ & $0.38\pm0.09$ \\
\bottomrule
\end{tabular}

\renewcommand{\arraystretch}{1}
\end{table}

\subsubsection{Equivariant Diffusion For Materials}

Similarly, we investigated the distribution of generated molecules across the training, validation, and test sets of the materials design experiments from~\cref{sec:exp_materials}. As this experimental setup operates in an exhaustively enumerated search space, we compare the proportions of generated samples across all sets with both unguided and weight decay-regularized baselines. We use a guidance scale of 4, as this is the setting in which both guided diffusion models perform best (see \cref{fig:materials_results}). We observe that our model generates a similar proportion of molecules from the medium-value validation set as the weight decay-regularized baseline, while producing substantially more compounds from the completely held-out high-property test set.

\begin{table}[H]
\centering
\renewcommand{\arraystretch}{1.2}
\caption{Distribution of generated molecules across the training, validation, and test sets for the materials science experiments in~\cref{sec:exp_materials}. Means and standard deviations are reported across ten independent training and sampling runs with different random seeds.}
\label{tab:yourlabel}
\begin{tabular}{lccc}
\toprule
Model & ratio in training set & ratio in validation set & ratio in test set \\
\midrule
Unguided & $0.91\pm0.01$ & $0.05\pm0.00$ & $0.04\pm0.00$ \\
Weight Decay & $0.49\pm0.07$ & $0.32\pm0.05$ & $0.19\pm0.03$ \\
Ours & $0.19\pm0.03$ & $0.39\pm0.05$ & $0.46\pm0.06$ \\
\bottomrule
\end{tabular}
\renewcommand{\arraystretch}{1}
\end{table}

\end{appendices}

\end{document}